\documentclass[aps,prb,twocolumn,superscriptaddress,showpacs]{revtex4}
\usepackage{graphicx} 
\newcommand{\be}{\begin{equation}}
\newcommand{\ee}{\end{equation}}
\newcommand{\bea}{\begin{eqnarray}}
\newcommand{\eea}{\end{eqnarray}}

\begin{document}

\title{Magnetic Exchange Interactions in ${\rm\bf BaMn_2As_2}$: A Case Study of the $J_1$-$J_2$-$J_c$ Heisenberg Model}

\author{ D. C. Johnston}
\author{R. J. McQueeney}
\affiliation{Ames Laboratory, Iowa State University, Ames, Iowa 50011, USA}
\affiliation{Department of Physics and Astronomy, Iowa State University, Ames, Iowa 50011, USA}
\author{B. Lake}
\affiliation{Hahn-Meitner-Institut, Glienicker Str.~100, D-14109 Berlin,Germany}
\affiliation{Institut f\"{u}r Festk\"{o}rperphysik, Technische
Universit\"{a}t Berlin, Hardenbergstr.~36, D-10623 Berlin, Germany}
\author{A. Honecker}
\affiliation{Institut f\"ur Theoretische Physik, Universit\"at
G\"ottingen, D-37077 G\"ottingen, Germany}
\author{M.~E.~Zhitomirsky}
\affiliation{Service de Physique Statistique, Magn\'etisme et Supraconductivit\'e, UMR-E9001 CEA-INAC/UJF, 17 rue des Martyrs, 38054 Grenoble Cedex 9, France}
\affiliation{Max-Planck-Institut f\"ur Physik Komplexer Systeme, N\"othnitzer Stra{\ss}e~38, D-01187 Dresden, Germany}
\author{R. Nath}
\altaffiliation{Present address: Indian Institute of Science Education \& Research Trivandrum, CET Campus, Trivandrum-695016, Kerala, India.}
\affiliation{Ames Laboratory, Iowa State University, Ames, Iowa 50011, USA}
\author{Y. Furukawa}
\affiliation{Ames Laboratory, Iowa State University, Ames, Iowa 50011, USA}
\affiliation{Department of Physics and Astronomy, Iowa State University,
Ames, Iowa 50011, USA}
\author{V. P. Antropov}
\affiliation{Ames Laboratory, Iowa State University, Ames, Iowa 50011, USA}
\author{Yogesh Singh}
\altaffiliation{Present address: Indian Institute of Science Education \& Research Mohali, MGSIPAP Complex, Sector 26, Chandigarh 160019, India.}
\affiliation{Ames Laboratory, Iowa State University, Ames, Iowa 50011, USA}

\date{\today}

\begin{abstract}

${\rm BaMn_2As_2}$ is unique among Ba$T_2$As$_2$ compounds crystallizing in the body-centered-tetragonal ${\rm ThCr_2Si_2}$ structure, which contain stacked square lattices of $3d$ transition metal $T$ atoms, since it has an insulating large-moment ($3.9~\mu_{\rm B}$/Mn) G-type (checkerboard) antiferromagnetic AF ground state.  We report measurements of the anisotropic magnetic susceptibility $\chi$ versus temperature $T$ from 300 to 1000~K of single crystals of ${\rm BaMn_2As_2}$, and magnetic inelastic neutron scattering measurements at 8~K and $^{75}$As NMR measurements from 4 to 300~K of polycrystalline samples.  The N\'eel temperature determined from the $\chi(T)$ measurements is $T_{\rm N} = 618$(3)~K\@.  The measurements are analyzed using the $J_1$-$J_2$-$J_c$ Heisenberg model for the stacked square lattice, where $J_1$ and $J_2$ are respectively the nearest-neighbor (NN) and next-nearest-neighbor intraplane exchange interactions and $J_c$ is the NN interplane interaction.  Linear spin wave theory for G-type AF ordering and classical and quantum Monte Carlo simulations and molecular field theory calculations of $\chi(T)$ and of the magnetic heat capacity $C_{\rm mag}(T)$ are presented versus $J_1$, $J_2$ and $J_c$.  We also obtain band theoretical estimates of the exchange couplings in ${\rm BaMn_2As_2}$.  From analyses of our $\chi(T)$, NMR, neutron scattering, and previously published heat capacity data for ${\rm BaMn_2As_2}$ on the basis of the above theories for the $J_1$-$J_2$-$J_c$ Heisenberg model and our band-theoretical results, our best estimates of the exchange constants in ${\rm BaMn_2As_2}$ are $J_1 \approx 13$~meV, $J_2/J_1 \approx 0.3$ and $J_c/J_1 \approx 0.1$, which are all antiferromagnetic.  From our classical Monte Carlo simulations of the G-type AF ordering transition, these exchange parameters predict $T_{\rm N}\approx 640$~K for spin $S = 5/2$, in close agreement with experiment.  Using spin wave theory, we also utilize these exchange constants to estimate the suppression of the ordered moment due to quantum fluctuations for comparison with the observed value and again obtain $S = 5/2$ for the Mn spin.  

\end{abstract}


\maketitle

\section{Introduction}

The observations of superconductivity up to 56~K in several classes of Fe-based superconductors in 2008 (Refs.~\onlinecite{Kamihara2008,Wang2008, rotter2008a, Johnston2010}) have reinvigorated the high-$T_{\rm c}$ field following the discovery of high-$T_{\rm c}$ superconductivity in the layered cuprates 25 years ago.\cite{Bednorz1986, Johnston1997}  Interestingly, the Fe atoms have the same layered square lattice structure as the Cu atoms do.  Even though the maximum $T_{\rm c}$ of the Fe-based materials is far below the maximum $T_{\rm c}$ of 164~K for the cuprates,\cite{Gao1994} the Fe-based materials have generated much interest because the superconductivity appears to be caused by a magnetic mechanism\cite{Johnston2010} as also appears to be the case in the cuprates.  One of the many motivations for carrying out detailed measurements on the Fe-based materials is to see if these studies can clarify the superconducting mechanism in the high-$T_{\rm c}$ cuprates for which a clear consensus has not yet been reached despite 25 years of intensive research.

Many studies of the magnetic properties of the Fe-based superconductors have been carried out.\cite{Johnston2010, Lumsden2010}  For the FeAs-based materials such as Ba$_{1-x}$K$_x{\rm Fe_2As_2}$ with the body-centered-tetragonal ${\rm ThCr_2Si_2}$ structure, the magnetic susceptibility $\chi$ increases approximately linearly with increasing temperature above $T_{\rm c}$ or above the N\'eel temperature $T_{\rm N}$ of the nonsuperconducting parent compounds up to at least 700~K.\cite{wang2008, GMZhang2008}  In a model of local magnetic moments on a square lattice with strong antiferromagnetic (AF) Heisenberg interactions, this type of behavior is explained as being due to the measurement temperature ($T$) range being on the low-$T$ side of a broad maximum in $\chi(T)$ at higher temperatures.\cite{Johnston1997}  On the other hand, many magnetic measurements of the FeAs-based superconductors have been explained in terms of itinerant magnetism models, and indeed the consensus is pointing in this direction, although this view is not universal.\cite{Johnston2010}

\begin{figure}
\includegraphics [width=1.75in]{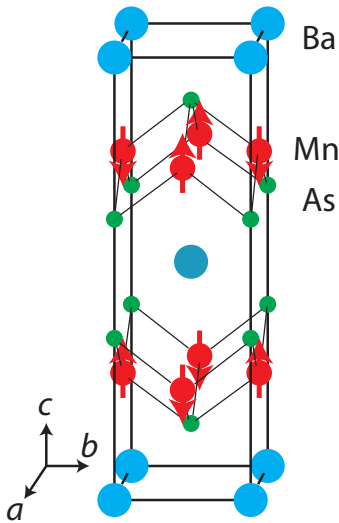}
\caption{(Color online) Crystal and magnetic structures of ${\rm BaMn_2As_2}$.\cite{YSingh2009}  The crystal structure is body-centered-tetragonal ${\rm ThCr_2Si_2}$-type in which the Mn atoms form a square lattice within the $ab$-plane, the axes of which are rotated by 45$^\circ$ with respect to the {\bf a} and {\bf b} unit cell axes.  The Mn atoms in adjacent layers are directly above or below each other along the $c$-axis.  The magnetic structure is a G-type antiferromagnetic structure in which the magnetic moments of nearest-neighbor spins are antiparallel both in the $ab$-plane and along the $c$-axis.  The ordered moment at $T = 0$ is $3.9(1)~\mu_{\rm B}$/Mn.\cite{YSingh2009} }
\label{Fig:BaMn2As2_Xtal_Mag_struct}
\end{figure}

In this context it is very useful to have a benchmark compound with the same ${\rm ThCr_2Si_2}$-type structure and similar composition as many of the Fe-based superconductors, but for which a local moment model must be used to explain the magnetic properties.  Such a compound is ${\rm BaMn_2As_2}$ because it has an insulating ground state.\cite{singh2009, an2009}  The crystal structure of ${\rm BaMn_2As_2}$ is shown in Fig.~\ref{Fig:BaMn2As2_Xtal_Mag_struct}.\cite{YSingh2009}  It is a small-band-gap semiconductor\cite{singh2009, an2009} with an activation energy of 30~meV.\cite{singh2009}  The electronic structure calculations of An et al.\cite{an2009} for the predicted conventional G-type (checkerboard) AF state give a small band gap of 0.1--0.2~eV, qualitatively consistent with the experimental value of the activation energy\cite{singh2009} that is expected to be a lower limit to half the band gap.  The anisotropic $\chi$ of single crystals was previously measured at $T \leq 400$~K.\cite{singh2009} These data indicate that the compound is in a collinear AF state at these temperatures, with the ordered moment direction along the $c$-axis, and with a $T_{\rm N}$ significantly above 400~K\@.  From subsequent magnetic neutron powder diffraction measurements, the N\'eel temperature was determined to be $T_{\rm N} = 625$(1)~K and the AF structure was found to be a conventional G-type (checkerboard) structure in all three directions as shown in Fig.~\ref{Fig:BaMn2As2_Xtal_Mag_struct}, with an ordered moment direction along the $c$-axis in agreement with the $\chi(T)$ data, and with an ordered moment $\mu = 3.9(1)~\mu_{\rm B}$/Mn at 10~K, where $\mu_{\rm B}$ is the Bohr magneton.\cite{YSingh2009}  These characteristics are radically different from those of the similar FeAs-based metallic parent compound ${\rm BaFe_2As_2}$ with the same room temperature crystal structure.   ${\rm BaFe_2As_2}$ has a much smaller ordered moment $\mu \approx 0.9~\mu_{\rm B}$/Fe and much smaller $T_{\rm N} = 137$~K than ${\rm BaMn_2As_2}$, the structure of ${\rm BaFe_2As_2}$ distorts to orthorhombic symmetry below a temperature $T_{\rm S} \approx T_{\rm N}$ instead of remaining tetragonal as in ${\rm BaMn_2As_2}$, the ordered moment direction is in the $ab$-plane instead of along the $c$-axis, and the in-plane AF structure is a stripe structure (see the bottom panel of Fig.~\ref{Fig:Magnetic_Structures} below) instead of G-type.\cite{Johnston2010}  These large differences between the magnetic properties of  ${\rm BaMn_2As_2}$ and  ${\rm BaFe_2As_2}$ evidently arise because  ${\rm BaMn_2As_2}$ is a local moment antiferromagnet whereas  ${\rm BaFe_2As_2}$ is an itinerant antiferromagnet.

An intriguing aspect of the in-plane electrical resistivity $\rho(T)$ data for ${\rm BaMn_2As_2}$ single crystals is that above $\sim 100$~K the slope of the resistivity versus temperature changes from negative (semiconductor-like) to positive (metal-like).\cite{singh2009, an2009}  The $\rho(T)$ of a material can be written in an effective single carrier model as
\be
\rho(T)=\frac{1}{e\,n(T)\mu(T)},
\label{Eq:rho}
\ee
where $e$ is the magnitude of the electron charge, and $n(T)$ and $\mu(T)$ are respectively the effective conduction carrier density and the effective carrier mobility, respectively.  Thus, a positive temperature coefficient of resistivity can be obtained for a band semiconductor if the increase in carrier concentration with increasing temperature is slower than the decrease in mobility with increasing temperature.  Our $^{75}$As NMR measurements in Sec.~XI were in fact initially motivated in order to address this issue.  As stated in that section, we found no evidence for a Korringa contribution to the $^{75}$As nuclear spin-lattice relaxation rate that would have indicated metallic behavior, and indeed we could interpret the data from 50 to 300~K in terms of a local moment insulator model.  Furthermore, there is no evidence from the previously published neutron diffraction,\cite{YSingh2009} resistivity or heat capacity\cite{singh2009, an2009} measurements for any phase transition from a band insulator at low temperatures to a metal at high temperatures.  Thus in the absence of experimental data to the contrary, our present interpretation of the positive temperature coefficient of resistivity above $\sim 100$~K is as discussed below Eq.~(\ref{Eq:rho}) above.  

A related Mn-based compound is ${\rm Sr_2Mn_3As_2O_2}$ which consists of ${\rm Mn_2As_2}$ layers that are the same as in ${\rm BaMn_2As_2}$, alternating along the $c$-axis with MnO$_2$ layers with the same structure as the CuO$_2$ layers in the layered cuprate superconductor parent compounds.\cite{Brock1996}  Due to geometric frustration effects, the Mn spins in the MnO$_2$ layers do not show any obvious long-range magnetic ordering for $T \geq 4$~K, but the Mn spins in the ${\rm Mn_2As_2}$ layers show long-range G-type AF ordering below $T_{\rm N} = 340$~K with a low-temperature ordered moment $\mu = 3.50(4)~\mu_{\rm B}$/Mn.\cite{Brock1996, Nath2010}  Thus, in both ${\rm BaMn_2As_2}$ and ${\rm Sr_2Mn_3As_2O_2}$, the Mn spins in the ${\rm Mn_2As_2}$ layers exhibit the same G-type AF structure and significant reductions in the ordered moments from the value $\mu = gS\mu_{\rm B} = 5~\mu_{\rm B}$/Mn that would be expected for the high-spin $S=5/2$ $d^5$ Mn$^{+2}$ ion with spectroscopic splitting factor $g=2$.

The main goal of the present work was to determine the magnitudes of the exchange interactions in the fiducial compound ${\rm BaMn_2As_2}$ and their signs, i.e., AF or ferromagnetic (FM).  Experimentally, we extend the single-crystal anisotropic $\chi(T)$ measurements from 300 to 1000~K, significantly above $T_{\rm N}$.  We also report inelastic magnetic neutron scattering measurements at 8~K and $^{75}$As NMR measurements from 4 to 300~K on polycrystalline samples.  We analyze these data using the $J_1$-$J_2$-$J_c$ Heisenberg stacked square spin lattice model for which we develop extensive theory.  This model has also been investigated recently by other groups.\cite{Schmalfuss2006, Viana2008, Nunes2010, Yao2010, Holt2011, Majumdar2011, Rojas2011, Stanek2011}  We calculate the spin wave dispersion relations for this model.  We report classical and quantum Monte Carlo simulations and molecular field theory calculations of $\chi(T)$ and the magnetic heat capacity $C_{\rm mag}(T)$.  We extract the values of $J_1$, $J_2$ and $J_c$ by fitting our experimental data for ${\rm BaMn_2As_2}$ by these theoretical predictions for the $J_1$-$J_2$-$J_c$ model.  From our classical Monte Carlo simulations of the heat capacity of coupled layers, we derive a formula for $T_{\rm N}$ versus the exchange parameters which yields a $T_{\rm N}$ very close to experiment from the independently-derived exchange constants, which indicates that the spin on the Mn ions is 5/2.  We also utilize these exchange constants to determine from spin wave theory the suppression of the ordered moment due to quantum fluctuations for comparison with the observed value, and again arrive at the estimate of $S = 5/2$ for the spin of the Mn$^{+2}$ ions when the additional expected suppression of the ordered moment due to hybridization and/or to charge and/or magnetic moment amplitude fluctuations, which arise from both on-site and intersite interactions, are taken into account.  Finally, we report band-theoretical calculations of $J_1$, $J_2$ and $J_c$ for ${\rm BaMn_2As_2}$.

The applicability of the local moment Heisenberg model to a specific compound depends on the degree of variation of atomic magnetic moments and interatomic exchange parameters found from electronic structure calculations for the relevant magnetic ordering configurations. Such variations are usually found to be small in magnetic insulators. In the case of ${\rm BaMn_2As_2}$, our band theory analysis in Sec.~\ref{Sec:BandTheory} indicates that insulating character is conserved for both the N\'eel and stripe antiferromagnetic structures, as observed, with a tiny metallicity appearing in the ferromagnetic case.  As noted above, An et al.\ previously estimated that the band gap is 0.1--0.2~eV from electronic structure calculations for G-type AF order in ${\rm BaMn_2As_2}$.\cite{an2009}  Moreover, our direct calculations of the atomic magnetic moment and exchange couplings for different spin configurations demonstrate that the ordered moment variations do not exceed 10--12\%, while the exchange coupling variation is only about 5\%. The largest change appears for the ferromagnetic state which due to its high energy is expected to contribute little to thermodynamic properties. Finally, our determination of a self-consistent set of antiferromagnetic exchange coupling parameters in ${\rm BaMn_2As_2}$ from both static and dynamic experiments confirm the validity of our analyses in terms of the local moment Heisenberg model.

The remainder of the paper is organized as follows.  The experimental details for the sample preparation and characterization of ${\rm BaMn_2As_2}$ and for the various measurements are given in Sec.~\ref{SecExpDetails}. The $J_1$-$J_2$-$J_c$ Heisenberg model is introduced and defined in Sec.~\ref{Sec:J1J2J3Model}.  The inelastic neutron scattering measurements of polycrystalline ${\rm BaMn_2As_2}$ and the analysis of these data are presented in Sec.~\ref{Sec:INS}.  This includes the presentation of  spin wave theory for the $J_{1}$-$J_{2}$-$J_{c}$ model of the G-type antiferromagnet in Sec.~\ref{SWT} that is used to fit the neutron data and to later obtain an estimate of the spin wave contribution to the heat capacity at low temperatures in Sec.~\ref{Sec:SWTCmag} and to analyze the $^{75}$As nuclear spin-lattice relaxation NMR data below $T_{\rm N}$ in Sec.~\ref{Sec:NSLR}.  The high-temperature anisotropic magnetic susceptibility measurements of single crystals of ${\rm BaMn_2As_2}$ are presented in Sec.~\ref{SecExpResults}.  The predictions of molecular field theory (MFT) and related topics for the $J_1$-$J_2$-$J_c$ Heisenberg model are given in Secs.~\ref{Sec:MFT}, \ref{Sec:MFTJ1J2J3} and the Appendices.  Comparisons of the MFT predictions with our experimental $\chi(T)$, $C_{\rm p}(T)$ and ordered moment $\mu(T)$ data for ${\rm BaMn_2As_2}$ are given in Sec.~\ref{Sec:MFT vs Expt}.  In this section we also calculate the spin wave contribution to the heat capacity at low temperatures assuming a negligible anisotropy gap in the spin wave spectrum and compare this contribution with the experimental heat capacity data at low temperatures.  Classical and quantum Monte Carlo simulations of $T_{\rm N}$, $\chi(T)$ and $C_{\rm mag}(T)$ in the $J_1$-$J_2$-$J_c$ Heisenberg model are presented in Sec.~\ref{SecThy} and comparisons with the experimental $T_{\rm N}$ and $\chi(T>T_{\rm N})$ data are carried out in Sec.~\ref{SecThyExpFit}. The NMR measurements and analysis are given in Sec.~\ref{Sec:NMR}, and our band-theoretical estimates of the exchange couplings in ${\rm BaMn_2As_2}$ are presented in Sec.~\ref{Sec:BandTheory}.  Our spin wave theory results for the suppression of the ordered moment are given in Sec.~\ref{SecSWTMomentSuppr}.  From a comparison with the experimental ordered moment, we infer that the spin on the Mn ions is 5/2.  A summary of our results and of our most reliable values of the $J_1$, $J_2$ and $J_c$ exchange constants and of the spin value derived for the Mn ions in ${\rm BaMn_2As_2}$ is given in Sec.~\ref{Sec:Summary}.

\section{\label{SecExpDetails} Experimental Details}

A 25-g polycrystalline sample of BaMn$_2$As$_2$ was prepared by solid state synthesis for inelastic neutron scattering (INS) measurements.  Stoichiometric amounts of Ba dendritic pieces (Aldrich, 99.9\%), Mn powder (Alfa Aesar, 99.9\%), and As chunks (Alfa Aesar, 99.9\%) were ground and mixed together in a He-filled
glovebox, pelletized, placed in a 50-mL Al$_2$O$_3$ crucible with a lid and sealed in a quartz tube under a 0.5~atm pressure of Ar gas (99.999\%).  The tube was placed in a box furnace and heated at a rate of 50~$^\circ$C/h to 575~$^\circ$C and kept there for 24~h.  The furnace was then heated at 100~$^\circ$C/h to 800~$^\circ$C and kept there for 48~h before cooling to room temperature by turning off the furnace. The quartz tube was opened inside the glovebox and the product was ground and mixed thoroughly and pelletized again. The pellet was placed in the same crucible and sealed again in a quartz tube. The quartz tube was heated in the box furnace at 100~$^\circ$C/h to 850~$^\circ$C and kept there for 24~h and then heated at 100~$^\circ$C/h to 900~$^\circ$C and kept there for 24~h, followed by furnace-cooling at $\sim300~^\circ$C/h to room temperature. The resulting product was ground and pelletized and the above
heat treatment was repeated. The resulting product was characterized by x-ray powder diffraction and
the majority phase ($\approx 83$\%) was found to be the desired ThCr$_2$Si$_2$ structure. The major impurity phase was identified to be tetragonal ${\rm Ba_2Mn_3As_2O}$.  ${\rm BaMn_3As_2O}$ is an insulator, shows low-dimensional magnetic behavior with a broad maximum in $\chi(T)$ at 100~K and antiferromagnetic ordering at $\approx 75$~K.\cite{Brechtel1979, Brock1996a}  From x-ray diffraction measurements the weight fraction of this impurity phase in the INS sample was estimated to be $\approx 17$\%.  The INS spectra at 8~K and at 100~K (not shown) exhibited no noticeable differences. ÊSince 100~K is well above the purported ordering temperature of the impurity phase, this eliminates any concern for serious contamination of the magnetic INS data by this phase.

About 20~g of the above material was used for INS measurements. About 5~g of the polycrystalline material was used to grow single crystals.  About 3~g of polycrystalline BaMn$_2$As$_2$ and 20~g of Sn flux (Alfa Aesar, 99.999\%) were placed in an Al$_2$O$_3$ crucible and sealed in a quartz tube. The crucible was heated at 250~$^\circ$C/hr to 1000~$^\circ$C and kept there for 24~h and then cooled at 5~$^\circ$C/h to 575~$^\circ$C
and kept there for 5~h at which point the Sn flux was centrifuged off to give isolated single crystals of typical dimension ${\rm 5\times 5 \times 0.2~mm^3}$.  Energy-dispersive x-ray (EDX) measurements using a Jeol scanning electron microscope confirmed the composition of the crystals to be BaMn$_2$As$_2$.

For the INS measurements, the powder sample of mass $\approx 20$~g was characterized for phase purity by x-ray powder diffraction as discussed above. The INS measurements were performed on the Pharos spectrometer at the Lujan Center of Los Alamos National Laboratory. Pharos is a direct geometry time-of-flight spectrometer and measures the scattered intensity over a wide range of energy transfers $\hbar\omega$ and angles between 1 and 140$^\circ$ allowing determination of the scattered intensity $S(Q,\omega)$ over large ranges of momentum transfer $\hbar Q$ and $\hbar\omega$.  The powder sample was packed in a flat aluminum can oriented at 135$^\circ$ to the incident neutron beam, and INS spectra were measured with incident energies $E_{\rm i}$ of 150 and 200 meV\@. The data were measured at a temperature $T = 8$~K, well below the antiferromagnetic ordering temperature of 625~K.\cite{YSingh2009} The time-of-flight data were reduced into $\hbar\omega$ and scattering angle ($2\theta$) histograms and corrections for detector efficiencies, empty can scattering, and instrumental background were performed.

The high-temperature anisotropic $\chi(T)$ measurements of a ${\rm BaMn_2As_2}$ single crystal took place in a physical properties measurement system (PPMS, Quantum Design, Inc.)\ at the Laboratory for Magnetic Measurements at the Helmholtz Zentrum Berlin f\"ur Materialien und Energie.  For these measurements the vibrating sample magnetometer option was used. Data were collected with the magnetic field applied both parallel and perpendicular to the Mn layers. For field pointing within the $ab$-plane a sample of mass 15.31~mg was used. The sample had to be cut for field parallel to $c$ and the sample weight was 12.058~mg. For all measurements a constant magnetic field $H = 3$~T was used while the temperature was varied between 300 and 1000~K\@. To achieve these temperatures an oven set-up provided as an option by Quantum Design was utilized.  The crystal was fixed on a zirconia sample stick containing a wire system that acts as a heating element.  The sample was glued on the stick with heat-resistant cement and wrapped in low emissivity copper foil to minimize the heat leak from the hot region to the surrounding coil set. The measurements took place with heating rates of between 1 and 2~K per minute. The magnetic moment of the empty sample holder, 7.63~mg of cement and of the copper foil was measured separately and subtracted from the data.

The NMR measurements were carried out on a polycrystalline sample using the conventional pulsed NMR technique on $^{75}$As nuclei (nuclear spin $I=3/2$ and gyromagnetic ratio $^{75}\gamma/2\pi=7.2919$~MHz/T) in the temperature range $4 \leq T \leq 300$~K\@. The measurements were done at a radio frequency of about $52$~MHz. Spectra were obtained by sweeping the field at fixed frequency. The $^{75}$As nuclear spin-lattice relaxation rate $1/T_{1}$ was measured by the conventional single saturation pulse method.

\section{\label{Sec:J1J2J3Model} The $J_1$-$J_2$-$J_c$ Heisenberg Model}

\begin{figure}
\includegraphics [width=2.25 in]{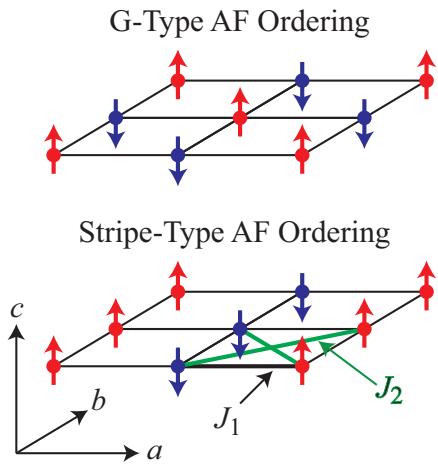}
\caption{(Color online) Collinear commensurate in-plane magnetic structures in the $J_1$-$J_2$-$J_c$ model for the square lattice antiferromagnet.  The top panel shows the G-type (N\'eel or checkerboard) AF structure where nearest-neighbor spins are aligned antiparallel.  The bottom panel shows stripe-type ordering, along with the definitions of the in-plane exchange constants $J_1$ and $J_2$.  A $J_2$ interaction is present for both diagonals of each square.   According to Eqs.~(\ref{Eq:JRestrictions}), the G-type in-plane ordering is favored if $J_2<J_1/2$, whereas the stripe-type ordering is favored if $J_2>J_1/2$. By examining the bottom panel, one sees that the stripe magnetic structure consists of two interpenetrating G-type magnetic structures, each respectively consisting of next-nearest-neighbor spins.}
\label{Fig:Magnetic_Structures}
\end{figure}

A bipartite spin lattice is defined as consisting of two distinct spin sublattices in which a given spin on one sublattice only interacts with nearest-neighbor (NN) spins on the other sublattice.  In the FeAs-based superconductors and parent compounds, when the magnetism is analyzed in a local moment model, the magnetic lattice is found not to be bipartite.\cite{Johnston2010}  In addition to the in-plane ($J_1$) and out-of-plane ($J_c$) NN inter-sublattice interactions, in-plane diagonal next-nearest-neighbor (NNN) intra-sublattice interactions $J_2$ are also present along both diagonals of each square, as shown in Fig.~\ref{Fig:Magnetic_Structures}.  The spin Hamiltonian in the $J_1$-$J_2$-$J_c$ Heisenberg model is
\bea
{\cal H} &=& J_1{\displaystyle\sum\limits_{\rm NN}}\mathbf{S}_{i}\cdot\mathbf{S}_{j}
 + J_{2}{\displaystyle\sum\limits_{\rm NNN}}\mathbf{S}_{i}\cdot\mathbf{S}_{j} \label{Eq:HamilJ1J2Jz}\\*
&&+\ J_{c} {\displaystyle\sum\limits_{c}} \mathbf{S}_{i}\cdot\mathbf{S}_{j} 
 + g\mu_{B}H {\displaystyle\sum\limits_{i}} S^z_{i},\nonumber
\eea
where $\mathbf{S}_{i}$ is the spin operator for the $i$th site, $g$ is the spectroscopic splitting factor ($g$-factor) of the magnetic moments, $\mu_{\rm B}$ is the Bohr magneton and $H$ is the magnitude of the magnetic field which is applied in the $+z$ direction. Throughout this paper, a positive $J$ corresponds to an antiferromagnetic interaction and a negative $J$ to a ferromagnetic interaction.  The indices ${\rm NN}$ and ${\rm NNN}$ indicate sums restricted to distinct spin pairs in a Mn layer, while the index $c$ indicates a sum over distinct ${\rm NN}$ Mn spin pairs along the $c$ axis.  This is the minimal model needed to explain our INS results below for ${\rm BaMn_2As_2}$.  

The classical energies of collinear commensurate ordered spin configurations in this model with $H=0$ are analyzed as discussed in Ref.~\onlinecite{Johnston2010}.  We consider four competing magnetic structures in the $J_1$-$J_2$-$J_c$ model.  One is the simple FM structure.  The other three are two AF stripe structures and the G-type (N\'eel) structure shown in Fig.~\ref{Fig:Magnetic_Structures}.  By definition, the NN spins in alternate layers are aligned antiferromagnetically in the G-type AF ordered state, whereas the stripe state can have either AF or FM spin alignments along the $c$-axis which depend on the sign of $J_c$.  The classical energies of these states for $H = 0$ are\cite{Johnston2010}
\bea
E_{\rm FM} &=& NS^2(2J_1+J_c+2J_2)\nonumber\\*
E_{\rm stripe} &=& NS^2(-2J_2 \pm J_c)\label{Eq:StripevsNeel}\\*
E_{\rm G} &=& NS^2(-2J_1  -J_c +2J_2),\nonumber
\eea
where $N$ is the number of spins $S$ and a factor of 1/2 has been inserted on the right-hand sides to avoid double-counting distinct pairs of spins.  The $\pm$ signs in the expression for the stripe phase arise due to the possibilities of either antiferromagnetic ($-$~sign) or ferromagnetic ($+$~sign) alignment of adjacent spins along the $c$-axis.  From these expressions, the in-plane G-type AF magnetic structure observed in ${\rm BaMn_2As_2}$ is lower in energy than the stripe structure if
\bea
J_1 &>& 0\nonumber\hspace{0.2in}({\rm G\ type\ AF})\\*
J_1 &>& 2J_2\label{Eq:JRestrictions}.
\eea
In order that G-type AF ordering occurs along the $c$-axis, one also requires that 
\be
J_c > 0.\hspace{0.2in}({\rm G\ type\ AF})\nonumber
\label{Eq:Jccondx}
\ee

These results place restrictions on the exchange coupling parameter space that is relevant to the G-type AF ordering observed in ${\rm BaMn_2As_2}$.  Equations~(\ref{Eq:JRestrictions}) and~(\ref{Eq:Jccondx}) require both $J_1$ and $J_c$ to be positive (antiferromagnetic), but $J_2$ can have either sign as long as it satisifies the second of Eqs.~(\ref{Eq:JRestrictions}).  The compound ${\rm BaFe_2As_2}$, on the other hand, has an in-plane stripe AF state at low temperatures (and with the ordered moment in the $ab$-plane instead of along the $c$-axis as in Fig.~\ref{Fig:BaMn2As2_Xtal_Mag_struct} for ${\rm BaMn_2As_2}$),\cite{Johnston2010} which in a local moment model requires $J_1 < 2J_2$ according to Eqs.~(\ref{Eq:StripevsNeel}).  The in-plane stripe phase can be considered to consist of two interpenetrating G-type AF sublattices, where in this case a sublattice consists of all NNN spins of a given spin, and which are connected by an antiferromagnetic interaction $J_2$ (see the bottom panel of Fig.~\ref{Fig:Magnetic_Structures}).

\section{\label{Sec:INS} Inelastic Neutron Scattering (INS) Measurements and Analysis}

\begin{figure}
\centering
\includegraphics[width = 0.46\textwidth]{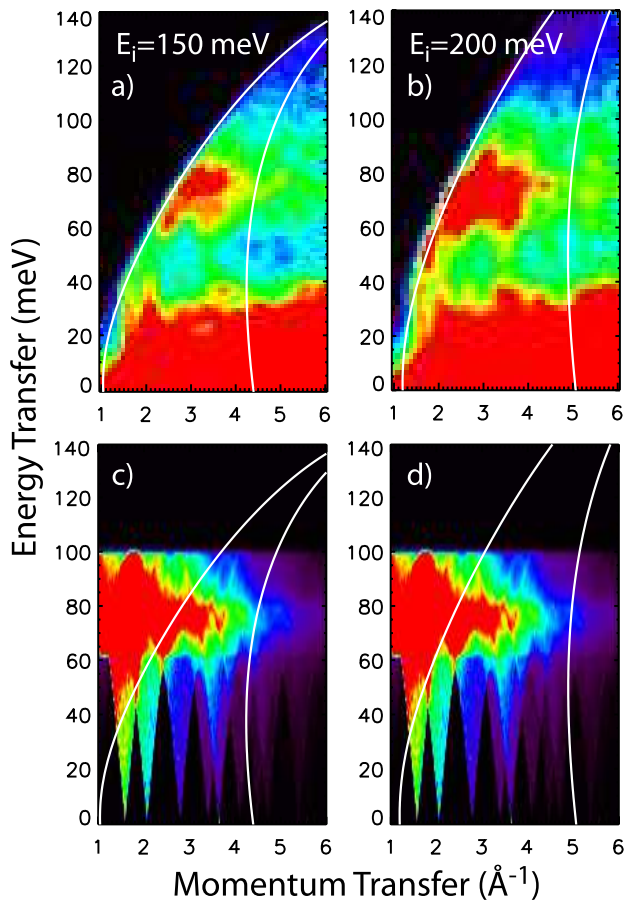}
\caption{(Color online) Inelastic neutron scattering data from a powder sample of BaMn$_{2}$As$_{2}$ as measured on the Pharos spectrometer with incident energies a) $E_{\rm i}=150$~meV and b) $E_{\rm i}=200$~meV\@.  The white lines delineate scattering angles of 7 and 30$^\circ$ where the magnetic scattering was estimated.  Panels c) and d) show calculations of the polycrystalline averaged spin wave scattering using a Heisenberg model.  The calculations in c) and d) are identical, however panel c) shows the trajectories of the angle summation limits for $E_{\rm i}=150$~meV and d) for $E_{\rm i}=200$~meV\@.}
\label{Fig:sqw_images}
\end{figure}

Figures~\ref{Fig:sqw_images}(a) and~\ref{Fig:sqw_images}(b) show images of the INS data taken at the base temperature of 8~K which share similar features at each incident energy.  Unpolarized inelastic neutron scattering contains contributions from both magnetic and phonon scattering. The magnetic scattering intensity falls off with $Q$ (or scattering angle $2\theta$) due to the magnetic form factor, while phonon scattering intensity increases like $Q^{2}$.  One can then observe a large contribution from magnetic scattering between 60 and 80 meV, presumably due to spin wave excitations in the magnetically ordered phase, whose intensity only appears at small $Q$.  On this intensity scale, strong phonon scattering is apparent below approximately 40~meV\@.

\begin{figure}
\centering \includegraphics[width = 1.1\columnwidth]{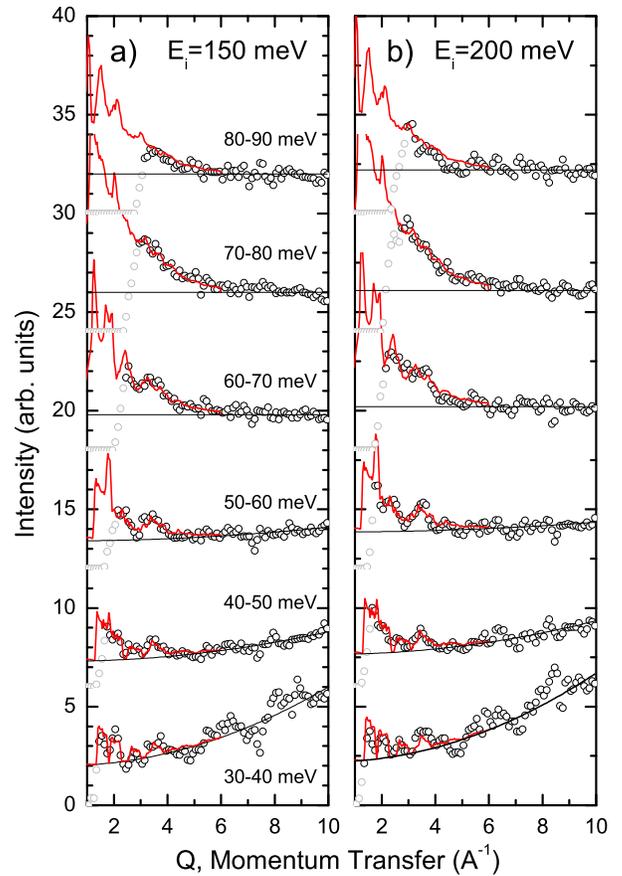}
\caption{(Color online) $Q$ dependence of the inelastic neutron scattering intensity averaged over several energy ranges with incident energies a) $E_{\rm i}=150$~meV and b) $E_{\rm i}=200$~meV\@.  The red lines show calculations of the polycrystalline averaged spin wave scattering using the $J_{1}$-$J_{2}$-$J_{c}$ Heisenberg model.}
\label{Fig:qdep}
\end{figure}

This separation of magnetic and phonon scattering is more clearly shown by plots of the $Q$-dependence of the scattering averaged over different energy ranges, as shown in Fig.~\ref{Fig:qdep}. For an energy range from 30 to 40~meV, the scattering is dominated by a large phonon contribution, whose intensity is proportional to $Q^{2}$, and a large constant background due to multiple scattering and other background contributions.  $Q$-dependent oscillations arise from the powder averaging of the coherent phonon scattering and weak magnetic scattering.  At the higher energy ranges between 60 and 90~meV, the $Q^{2}$ phonon contributions are gone and magnetic scattering appears superimposed on a constant background.  The magnetic scattering intensity falls off with $Q$ as the magnetic form factor for the Mn$^{2+}$ ion and is no longer visible above $\sim 6$~\AA$^{-1}$. Similar to the phonon cross-section, $Q$-dependent oscillations in the magnetic scattering occur due to coherent scattering of spin waves.

\begin{figure}
\centering \includegraphics[width = 1.1\columnwidth]{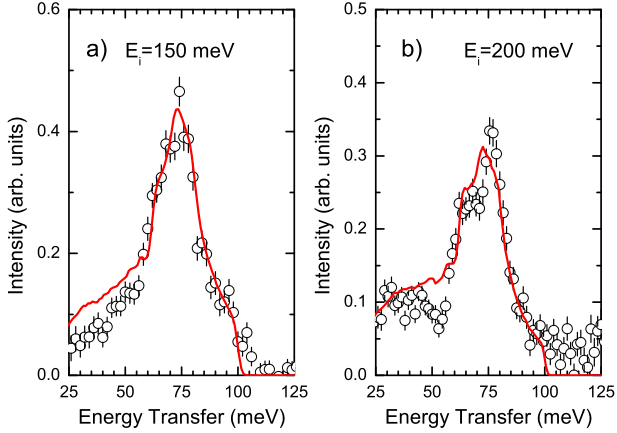}
\caption{(Color online) Energy dependence of the inelastic neutron scattering intensity averaged over a scattering angle range from 7 to 30$^\circ$ with incident energies a) $E_{\rm i}=150$~meV and b) $E_{\rm i}=200$~meV\@.  The magnetic intensity was extracted from the total scattering as described in the text. The red lines show calculations of the polycrystalline averaged spin wave scattering using the $J_{1}$-$J_{2}$-$J_{c}$ Heisenberg model.}
\label{Fig:edep}
\end{figure}

The spin wave spectrum can be obtained by averaging the low-$Q$ (low $2\theta$) data to improve statistics. However, the magnetic scattering, especially below $\sim 50$~meV, must be separated from the phonon scattering and other background contributions.  The pure phonon signal can be estimated from the high-angle spectra, where magnetic scattering is absent.  The magnetic scattering component can then be estimated by subtracting the high angle data (averaged from $2\theta =50$--$90^\circ$) from low angle data (averaged from $2\theta=7$--$30^\circ$ and indicated by the white lines in Fig.~\ref{Fig:sqw_images}) after scaling by a constant factor.  These spectra are shown in Fig.~\ref{Fig:edep} and show a strong and broad magnetic peak at $\sim 70$~meV\@.  At energies below 50~meV, the subtraction of the phonon intensity is subject to error since the phonon intensity may not scale uniformly to low-$Q$ due to coherent scattering effects and also due to the different Debye-Waller factors for each atomic species.  It is difficult to quantify this error without detailed phonon models; however, most of the magnetic scattering occurs above the phonon cutoff.  Thus the errors introduced are only a problem below 50~meV and the isolated magnetic data in this energy range can contain large errors.

\subsection{\label{SWT} Spin Waves in the $J_{1}$-$J_{2}$-$J_{c}$ Heisenbeerg Model for a G-type Antiferromagnet}

\subsubsection{\label{Sec:SWT} Spin Wave Theory}

In order to analyze the $Q$ and $\omega$ dependence of the magnetic spectra, we utilize a model of the spin wave scattering in BaMn$_{2}$As$_{2}$. Spin waves in insulators such as BaMn$_{2}$As$_{2}$ with the ThCr$_{2}$As$_{2}$ structure can be described by the Heisenberg Hamiltonian~(\ref{Eq:HamilJ1J2Jz}) except that here we set the magnetic field $H$ in the last term to zero.

The spin wave dispersions for the G-type AF structure are obtained from a Holstein-Primakoff spin-wave expansion of the Heisenberg model. When the single-ion anisotropy is zero, the dispersions with respect to the body-centered-tetragonal (bct) $I$4/$mmm$ unit cell containing two formula units of BaMn$_{2}$As$_{2}$ are
\bea
\left[\frac{\hbar\omega\left(\mathbf{q}\right)}{2SJ_{1}}\right]^{2} &=& \Big\{2+\frac{J_{c}
}{J_{1}}-\frac{J_{2}}{J_{1}}\left[  2-\cos(q_{x}a)-\cos(q_{y}a)\right]
\Big\}^{2} \nonumber\\*
&&-\ \Big\{  \cos\left[(q_{x}+q_{y})\frac{a}{2}\right]+\cos\left[(q_{x}-q_{y}%
)\frac{a}{2}\right]\nonumber\\*
&&+\ \frac{J_{c}}{J_{1}}\cos\left(\frac{q_{z}c}{2}\right)\Big\}^{2}
\label{Eq:SpnWaveDispersions}
\eea
where $a = 4.15$ and $c =13.41$~\AA\ are the lattice parameters of the bct unit cell at our measurement temperature of 8~K.\cite{YSingh2009}

In the absence of an anisotropy-induced energy gap in the spin-wave spectrum, the long-wavelength spin wave energies are described for an orthogonal (cubic, tetragonal, or orthorhombic) antiferromagnetically ordered spin lattice by the generic dispersion relation
\be
E_{\bf q} = \hbar\omega({\bf q}) = \hbar\sqrt{v_{a}^2q_x^2 + v_{b}^2q_y^2 + v_c^2 q_z^2},
\label{Eq:SWEnergies0}
\ee
where $v_{a}$, $v_{b}$ and $v_c$ are the spin wave velocities (speeds) along the respective axes.  In our case of tetragonal symmetry we have  
\be
E_{\bf q} = \hbar\omega({\bf q}) = \hbar\sqrt{v_{ab}^2(q_x^2 + q_y^2) + v_c^2 q_z^2},
\label{Eq:SWEnergies}
\ee
where $v_{ab}\equiv v_a = v_b$.  For G-type AF ordering of a spin lattice with our bct unit cell, these velocities are derived from  the dispersion relation in Eq.~(\ref{Eq:SpnWaveDispersions}) as
\bea
\hbar v_{ab} &=& 2J_1Sa\, \sqrt{\left(1 - \frac{2J_2}{J_1}\right)\left(1+\frac{J_c}{2J_1}\right)}\label{Eq:SWvels}\\*
\hbar v_{c} &=& \sqrt{2}J_1Sc\, \sqrt{\frac{J_c}{J_1}\left(1+\frac{J_c}{2J_1}\right)}.\nonumber
\eea
From the first of Eqs.~(\ref{Eq:SWvels}) the in-plane spin wave velocity $v_{ab}$ decreases with increasing $J_2$, consistent with expectation since according to Fig.~\ref{Fig:Magnetic_Structures}, a positive (AF) $J_2$ is a frustrating interaction for G-type AF ordering.  Indeed, $v_{ab}$ vanishes when $J_2 = J_1/2$, which is the classical criterion in Eq.~(\ref{Eq:JRestrictions}) for the transition between the G-type and stripe-type in-plane AF ordering arrangements.

In order to make contact with previous spin wave calculations for isotropic and anisotropic primitive orthogonal spin lattices, one can change unit cell variables to those of the primitive tetragonal (pt) spin lattice containing one spin at each lattice point with lattice parameters $a^\prime$ and $c^\prime$.  Referring to the bct structure with lattice parameters $a$ and $c$ in Fig.~\ref{Fig:BaMn2As2_Xtal_Mag_struct}, the pt spin lattice parameters are related to these according to
\bea
a&=&\sqrt{2}\,a^\prime\label{Eq:actoMconvert}\\*
c&=&2\,c^\prime.\nonumber
\eea
Furthermore the pt unit cell is rotated about the $c$-axis by 45$^\circ$ with respect to the bct unit cell, so the pt wave vectors $q^\prime_{x}$, $q^\prime_{y}$ and $q^\prime_{z}$ are related to those with respect to the bct cell $q_x,q_y,q_z$ by
\bea
q_x + q_y &=& \sqrt{2}\,q^\prime_{x}\nonumber\\*
-q_x + q_y &=& \sqrt{2}\,q^\prime_{y}\label{Eq:qtoMconvert}\\*
q_z &=& q^\prime_{z}.\nonumber
\eea
With these conversion expressions, the dispersion relation in Eq.~(\ref{Eq:SpnWaveDispersions}) becomes
\bea
\left[\frac{\hbar\omega\left(\mathbf{q}^\prime\right)}{2SJ_{1}}\right]^{2} &=& \Big\{2+\frac{J_{c}
}{J_{1}}-\frac{2J_{2}}{J_{1}}\Big[  1-\cos(q^\prime_{x}a^\prime)\cos(q^\prime_{y}a^\prime)\Big] \Big\}^{2}\nonumber\\*
&-&\Big[  \cos(q^\prime_x a^\prime)+\cos(q^\prime_y a^\prime) + \frac{J_{c}}{J_{1}}\cos(q^\prime_z c^\prime)\Big]^{2}. \label{Eq:SpnWaveDispersionsMbased}
\eea
Our dispersion relation~(\ref{Eq:SpnWaveDispersionsMbased}) is identical to that in Refs.~\onlinecite{Majumdar2011} and~\onlinecite{Rojas2011} derived from linear spin wave theory for the $J_1$-$J_2$-$J_c$ model.  Also, Eq.~(\ref{Eq:SpnWaveDispersionsMbased})  with $J_2$ set to zero is identical to that in Eq.~(5) of Ref.~\onlinecite{Raczkowski2002} and in Eq.~(3) of Ref.~\onlinecite{mcq08} for the anistropic simple cubic G-type bipartite antiferromagnet.

Using Eqs.~(\ref{Eq:actoMconvert}), for the primitive tetragonal spin lattice the spin wave velocities in Eqs.~(\ref{Eq:SWvels}) become
\bea
\hbar v_{a^\prime b^\prime} &=& 2\sqrt{2}J_1Sa^\prime\, \sqrt{\left(1 - \frac{2J_2}{J_1}\right)\left(1+\frac{J_c}{2J_1}\right)}\nonumber\\*
\hbar v_{c^\prime} &=& 2\sqrt{2}J_1Sc^\prime\, \sqrt{\frac{J_c}{J_1}\left(1+\frac{J_c}{2J_1}\right)}.\label{Eq:SWvelsPrimed}
\eea
In a simple cubic bipartite spin lattice with one spin per lattice point and isotropic interactions with $c^\prime = a^\prime$, $J_c/J_1 = 1$ and $J_2=0$, the spin wave velocity is isotropic with magnitude $\hbar v^\prime = 2\sqrt{3}J_1Sa^\prime$, which is the same as given previously in Table~I of Ref.~\onlinecite{Raczkowski2002} where $a^\prime$ was set to unity and is the standard well-known result when magnetocrystalline anisotropy is negligible.\cite{Kranendonk1958}

\subsubsection{Application of Spin Wave Theory to ${\rm BaMn_2As_2}$}

\begin{figure}
\centering \includegraphics[width = \columnwidth]{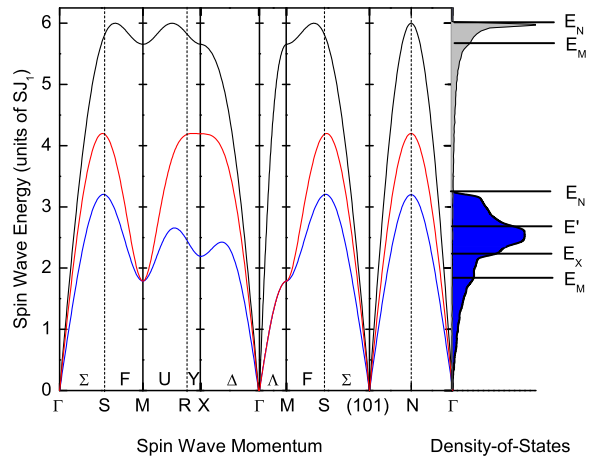}
\caption{(Color online) The spin wave dispersion of the $J_{1}$-$J_{2}$-$J_{c}$ model along various symmetry directions for three different combinations of the exchange parameters: $J_{2}/J_{1}=0$ and $J_{c}/J_{1}=1$ (top black curve); $J_{2}/J_{1}=0$ and $J_{c}/J_{1}=0.1$ (middle red curve); and $J_{2}/J_{1}=0.25$ and $J_{c}/J_{1}=0.1$ (bottom blue curve).  Energies are reported in units of $SJ_{1}$.  The $\Gamma$ point in the Brillouin zone corresponds to wave vector {\bf q} = 0.  The {\bf q} values corresponding to the high-symmetry M, X and N points are given in Table~\ref{tbl1}.  These wave vector directions are written with respect to the lattice translation vectors of the $I$4/$mmm$ chemical unit cell.  The labels $\Sigma$, F, U, Y, $\Delta$ and $\Lambda$ correspond to high symmetry lines in reciprocal space.  In the far right-hand panel, the energy versus the spin wave density-of-states is shown for $J_{2}/J_{1}=0$ and $J_{c}/J_{1}=1$ (top gray region) and $J_{2}/J_{1}=0.25$ and $J_{c}/J_{1}=0.1$ (bottom blue region).}
\label{Fig:dispr}
\end{figure}

Spin wave dispersions using the bct notation in Eq.~(\ref{Eq:SpnWaveDispersions}) are plotted in Fig.~\ref{Fig:dispr} (in units of $SJ_{1}$) for three different combinations of the exchange ratios $J_{2}/J_{1}$ and $J_{c}/J_{1}$.  The notations in Fig.~\ref{Fig:dispr} and Table~\ref{tbl1} for labeling the zone boundary reciprocal space positions are given by Kovalev.\cite{kovalev}  The magnetic excitation wavevector {\bf q} values are in reciprocal lattice units given by $(H,K,L)$ r.l.u.  This is a shorthand for {\bf q} expressed in inverse length units of the bct chemical unit cell according to
\[
{\bf q} = \frac{2\pi H}{a}\,\hat{\bf a} + \frac{2\pi K}{a}\,\hat{\bf b} + \frac{2\pi L}{c}\,\hat{\bf c}.
\]
In the $I$4/$mmm$ bct unit cell notation, the magnetic propagation vector for G-type AF ordering is $\tau_{\rm G} = (1,0,1)$, which gives $\tau_{\rm G} = (0,0,0)$ when translated by a reciprocal lattice vector to the $\Gamma$-point in the Brillouin zone.  This corresponds to the more familiar G-type wave vector $\tau^\prime_{\rm G} = \left(\frac{1}{2},\frac{1}{2},\frac{1}{2}\right)$ in the pt cell according to the transformation ${\bf q}^\prime = \left(\frac{H+K}{2},\frac{H-K}{2},\frac{L}{2}\right)$, where $H,K,L$ are referred to the $I$4/$mmm$ crystallographic unit cell, as shown in Eqs.~(\ref{Eq:qtoMconvert}).

For $J_{2}=0$ and $J_{c}/J_{1}=1$ (the top black curve in Fig.~\ref{Fig:dispr}), the dispersion is that of an isotropic G-type antiferromagnet (similar to LaFeO$_{3}$)\cite{mcq08} with a maximum spin wave energy of $6SJ_{1}$.  For the layered BaMn$_{2}$As$_{2}$ structure, $J_{c}$ is expected to be much weaker than $J_{1}$.  When $J_{2}=0$ and $J_{c}/J_{1}=0.1$ (the middle red curve in Fig.~\ref{Fig:dispr}), the maximum spin wave energy is reduced to $\sim$ $4SJ_{1}$ and the zone boundary $M$-point spin wave at \textbf{q} = (001) is strongly reduced.  If we now turn on antiferromagnetic NNN interactions with $J_{2}/J_{1}=0.25$ (the bottom blue curve in Fig.~\ref{Fig:dispr}), we observe a further softening of the spin wave spectrum, most notably at the $X$-point.  When $J_{2}>J_{1}/2$, the G-type ordering becomes unstable and the stripe AF order is the new ground state with ordering wavevector at the $X$-point as discussed above in Sec.~\ref{Sec:J1J2J3Model}, and our spin wave expressions are no longer applicable.  The spin wave theory for the stripe phase in the Fe-based superconductor parent compounds with the ${\rm ThCr_2As_2}$ structure is reviewed in Ref.~\onlinecite{Johnston2010}.

The powder-averaged spin wave scattering is closely associated with the spin wave density-of-states [SWDOS, $g(\omega)$].  The SWDOS is the distribution of spin wave energies and is determined by the summation over all wavevectors in the Brillouin zone ($\mathbf{q}$),
\be
g(\omega)  ={\displaystyle\sum\limits_{\mathbf{q}}}\,
\delta\left[  \hbar\omega-\hbar\omega\left(  \mathbf{q}\right)  \right].
\label{eqn1}
\ee
The SWDOSs versus energy $\hbar\omega$ are shown on the right-hand side of Fig.~\ref{Fig:dispr}.  It is observed that the SWDOS remains sharply peaked when $J_{2}=0$, and that $J_{2}$ acts to broaden the SWDOS.  Table~\ref{tbl1} indicates the energies of the various extremal features in the SWDOS (van Hove singularities) for ratios $J_{2}/J_{1}$ and $J_{c}/J_{1}$ that are associated with zone boundary spin wave energies.

\begin{table}
\caption{Energies of various van Hove singularities in the spin wave density-of-states of G-type antiferromagnets with NN ($J_{1}$), NNN ($J_{2}$) and interlayer ($J_{c}$) exchange interactions.}
\begin{ruledtabular}
\begin{tabular}{ccc}
\textbf{q} & label & van Hove singularity energy \\
\hline
(001) & $E_{M}$ & 4$SJ_{1}\sqrt{2\frac{J_{c}}{J_{1}}}$\\

($\frac{1}{2},\frac{1}{2}$,0) & $E_{X}$ & $2SJ_{1}\sqrt{\left(  2+\frac{J_{c}}{J_{1}}%
-4\frac{J_{2}}{J_{1}}\right)  ^{2}-\left(  \frac{J_{c}}{J_{1}}\right)  ^{2}}$\\

($\frac{1}{2}$,0,$\frac{1}{2}$) & $E_{N}$
& $2SJ_{1}\left(  2+\frac{J_{c}}{J_{1}}-2\frac{J_{2}}{J_{1}}\right)  $\\

($\frac{1}{2}$,$\frac{1}{2}$,$\frac{1}{2}$) &
$E_{P}$ & $2SJ_{1}\left(
2+\frac{J_{c}}{J_{1}}-4\frac{J_{2}}{J_{1}}\right)$\\

($\frac{3}{4}$,$\frac{1}{4}$,0) &
$E'$ & $2SJ_{1}%
\sqrt{\left(  2+\frac{J_{c}}{J_{1}}-2\frac{J_{2}}{J_{1}}\right)
^{2}-\left(  1-\frac{J_{c}}{J_{1}}\right)  ^{2}}$\\
\end{tabular}
\end{ruledtabular}
\label{tbl1}
\end{table}

\subsection{\label{Sec:NeutIntensity} Calculations of the Scattered Intensity}

When performing an INS experiment on a powder, the resulting INS intensities arise from the averaging of the inelastic scattering structure factor $S(\mathbf{Q},\omega)$ over all orientations of the crystallites.  Despite the orientational averaging, the spectra can show evidence of the spin wave dispersions, especially at low angles (within the first Brillouin zone) and in the vicinity of the first few magnetic Bragg peaks.  Due to the weighting of the spin wave modes by coherent scattering intensities, the $Q$-averaged intensity, $S(\omega)$, as shown in Fig.~\ref{Fig:edep} does not necessarily give the SWDOS\@.  This is only true in the incoherent scattering approximation, which does not apply to the case of scattering from a magnetically ordered system.  Therefore, model calculations of the powder-averaged spin wave intensities are necessary for accurate comparison to the data.

Numerical calculations of the spin waves in the Heisenberg model give not only the dispersion relation $\omega_{n}(\mathbf{q})$ for the $n^{\rm th}$ (degenerate) branch [as shown in Eq.~(\ref{Eq:SpnWaveDispersions})], but also the spin wave eigenvectors, $T_{ni}(\mathbf{q})$, for the $i^{\rm th}$ spin in the magnetic unit cell. The dispersion and associated eigenvectors can be used to calculate the spin wave structure
factor for unpolarized neutron energy loss scattering from a single-crystal sample, $S_{\rm mag}(\mathbf{Q},\omega)$, given by
\bea
S_{\rm mag}(\mathbf{Q},\omega) &=& \frac{1}{2}\left(  \gamma r_{o}\right)
^{2}\left[  1+\frac{(\mathbf{\hat{\mu}\cdot Q)}^{2}}{Q^{2}}\right]  \\*
&&  \times\sum\limits_{n}\Big\vert \sum\limits_{i}F_{i}(\mathbf{Q})\sigma
_{i}\sqrt{S_{i}}T_{ni}(\mathbf{q})e^{-i\mathbf{Q}\cdot\mathbf{d}_{i}%
}\Big\vert ^{2}\nonumber\\*
&&  \times[n(\omega)+1]\,\delta[\omega-\omega_{n}(\mathbf{q})], \nonumber
\eea
where the $i^{\rm th}$ spin with magnitude $S_{i}$ pointed in direction $\mathbf{\hat{\mu}}$ is located at position $\mathbf{d}_{i}$ and $\sigma_{i}=\pm1$ is the direction of the spin relative to the quantization axis
$\mathbf{\hat{\mu}}$ for a collinear spin structure, as shown in the top two rows of Table~\ref{tbl2}.  The vector $\mathbf{q}=\mathbf{Q}-\vec{\tau}$ is the spin wave wavevector in the first Brillouin zone.  Finally, the function $n(\omega)$ is the temperature-dependent Bose factor and $F_{i}(\mathbf{Q})=\frac{1}{2}g_{i}f_{i}(\mathbf{Q})e^{-W_{i}(\mathbf{Q})}$ is a product of the spectroscopic splitting factor ($g$-factor), magnetic form factor, and Debye-Waller factor for the $i^{\rm th}$ spin, respectively. The constant $\left(  \gamma r_{o}\right)  ^{2}=$ 290.6 millibarns allows calculations of the cross-section to be reported in absolute units of [millibarns steradian$^{-1}$ meV$^{-1}$ (formula~unit)$^{-1}%
$].  ForBaMn$_{2}$As$_{2}$, all Mn ions in the magnetic cell are equivalent.  The structure factor can then be written
\bea
S_{\rm mag}(\mathbf{Q},\omega)  &  =&\frac{1}{2}\left(  \gamma r_{o}\right)  ^{2}
SF^{2}(\mathbf{Q})\left[  1+\frac{(\mathbf{\hat{\mu}\cdot Q)}^{2}}{Q^{2}}
\right] \nonumber\\*
& & \times\sum\limits_{n} \Big| \sum\limits_{i}\sigma_{i}T_{ni}%
(\mathbf{q})e^{-i\mathbf{Q}\cdot\mathbf{d}_{i}}\Big| ^{2}\label{eqn5}\\*
& & \times[n(\omega)+1]\, \delta[\omega-\omega_{n}(\mathbf{q})].\nonumber
\eea
In the calculations, we use the isotropic magnetic form factor for Mn found in the International Crystallography Tables \cite{magformfactor} and the Debye-Waller factor is set equal to unity.  The differential magnetic cross section that is measured in the inelastic neutron scattering experiments is proportional to the structure factor.

\begin{table}
\caption{Parameters of the Heisenberg model in Eq.~(\ref{Eq:HamilJ1J2Jz}) for BaMn$_{2}$As$_{2}$ determined from magnetic inelastic neutron scattering measurements at a temperature of 8~K\@.  The Mn positions $i$ in the top two rows refer to the crystallographic $I4/mmm$ unit cell with lattice parameters $a = 4.15$ and $c =13.41$~\AA\ at 8~K\@. The moment direction is along the $c$-axis. The spin $S$ of the Mn atoms is not independently determined from the measurements.  Only the products of $S$ with the exchange constants $J_i$ can be modeled.  All exchange parameters are positive (antiferromagnetic).   Also shown are the low-energy spin wave velocities in the $ab$-plane $v_{ab}$ and along the $c$-axis $v_c$, each multiplied by $\hbar$, calculated from the exchange constants using Eqs.~(\ref{Eq:SWvels}).}
\begin{ruledtabular}
\begin{tabular}{ccc}
$i$ & $\sigma_i$ & $d_i$ \\
\hline
1 & $+1$ & (0,0,0) \\
2 & $-1$ & $\left(\frac{1}{2},\frac{1}{2},0\right)$ \\
\hline\hline
exchange constant & value & value (K) \\
\hline
$SJ_{1}$ & $(33\pm3)$ meV & 380 K \\
$J_{1}\ (S = 2,\ 5/2)$ & 16.5, 13.2 meV & 190, 150 K \\
$SJ_{2}$ & ($9.5\pm1.3$) meV & 110 K \\
$J_{2}\ (S = 2,\ 5/2)$& 4.8, 3.8 meV & 55, 44 K \\
$SJ_{c}$ & ($3.0\pm0.6$) meV & 35 K \\
$J_{c}\ (S=2,\ 5/2)$ & 1.5, 1.2 meV & 18, 14 K \\
$S(2J_1+J_c)$ & $69$~meV & 800 K \\
$2J_1+J_c\ (S = 2,\ 5/2)$ & 18.0, 14.4 meV & 400, 320 K \\
$J_2/J_1$ & $0.29\pm0.05$ \\
$J_c/J_1$ & $0.09\pm0.02$ \\
\hline\hline
spin wave velocity & value\\
 & (meV\,\AA)\\
\hline
$\hbar v_{ab}$ & 180\\
$\hbar v_{c}$ & 190\\
\end{tabular}
\end{ruledtabular}
\label{tbl2}
\end{table}

To compare Heisenberg model spin wave results to the powder INS data, powder-averaging of $S_{\rm mag}(\mathbf{Q},\omega)$ is performed by Monte Carlo integration over 25\,000 $\mathbf{Q}$ vectors lying on a constant-$Q$ sphere, giving the orientationally averaged $S_{\rm mag}(Q,\omega)$ which depends only on the magnitude of $Q$. By a comparison of the total $S(Q\mathbf{,}\omega)$ in Fig.~\ref{Fig:sqw_images}, the $Q$-cuts in Fig.~\ref{Fig:qdep}, and the energy spectra in Fig.~\ref{Fig:edep}, we arrive at the following parameters; $SJ_{1}= 33$~meV, $SJ_{2}= 9.5$~meV ($J_{2}/J_{1}=0.29$), and $SJ_{c}= 3$~meV ($J_{c}/J_{1}=0.09$), as summarized in Table~\ref{tbl2}.  Figures~\ref{Fig:sqw_images}(c) and~\ref{Fig:sqw_images}(d) show that calculations of $S_{\rm mag}(Q,\omega)$ at 8~K using these parameters compare well to the corresponding data in Figs.~\ref{Fig:sqw_images}(a) and~\ref{Fig:sqw_images}(b) and show clearly the coherent scattering of the powder-averaged spin waves. The most obvious coherent scattering feature is the necking down of acoustic spin waves in the vicinity of allowed magnetic Bragg reflections.  The first two observed magnetic Bragg peaks are at {\bf Q} = (101) and (103).  Additional coherent scattering features can also be seen for zone boundary spin waves, where intensities tend to peak in between the allowed magnetic Bragg peaks. Figure~\ref{Fig:sqw_images} enforces the general agreement of the Heisenberg model calculations of the spin wave intensity with neutron scattering measurements.

More quantitative estimates of the agreement of the calculated spin waves and the data are shown in Figs.~\ref{Fig:qdep} and~\ref{Fig:edep}. The calculations can be summed over scattering angles in order to compare the equivalent angle-summed data, as shown in Fig.~\ref{Fig:edep}.  The success of the Heisenberg model in estimating the measured spin wave intensities is better observed by plotting constant energy $Q$-cuts, as shown in Fig.~\ref{Fig:qdep}.  The plots show $Q$ oscillations of the experimental magnetic spin wave scattering above a background due mainly to phonon scattering and background/multiple scattering.  A constant background and incoherent phonon scattering intensity (proportional to $Q^{2}$) are added to the calculated spin wave scattering in order to compare to the measured data.  The agreement confirms the adequacy of the parameters.

The low-energy spin wave velocities in the $ab$-plane $v_{ab}$ and along the $c$-axis $v_c$ calculated from the exchange constants in Table~\ref{tbl2} using Eqs.~(\ref{Eq:SWvels}) are shown in Table~\ref{tbl2} for our measurement temperature of 8~K\@.  Remarkably, in spite of the layered nature of the spin lattice, the $ab$-plane and $c$-axis spin wave velocities are seen to have nearly the same value $\hbar v_{ab}\approx \hbar v_c \approx  180$--190~meV\,\AA.  For comparison, the spin wave velocities in the $A{\rm Fe_2As_2}$ compounds are in the ranges $\hbar v_{ab} \approx 280$--570~meV\,\AA\ and $\hbar v_c \approx 57$--280~meV\,\AA.\cite{Johnston2010}

\section{\label{SecExpResults} Magnetic Susceptibility Measurements}

\begin{figure}
\includegraphics [width=3.2in]{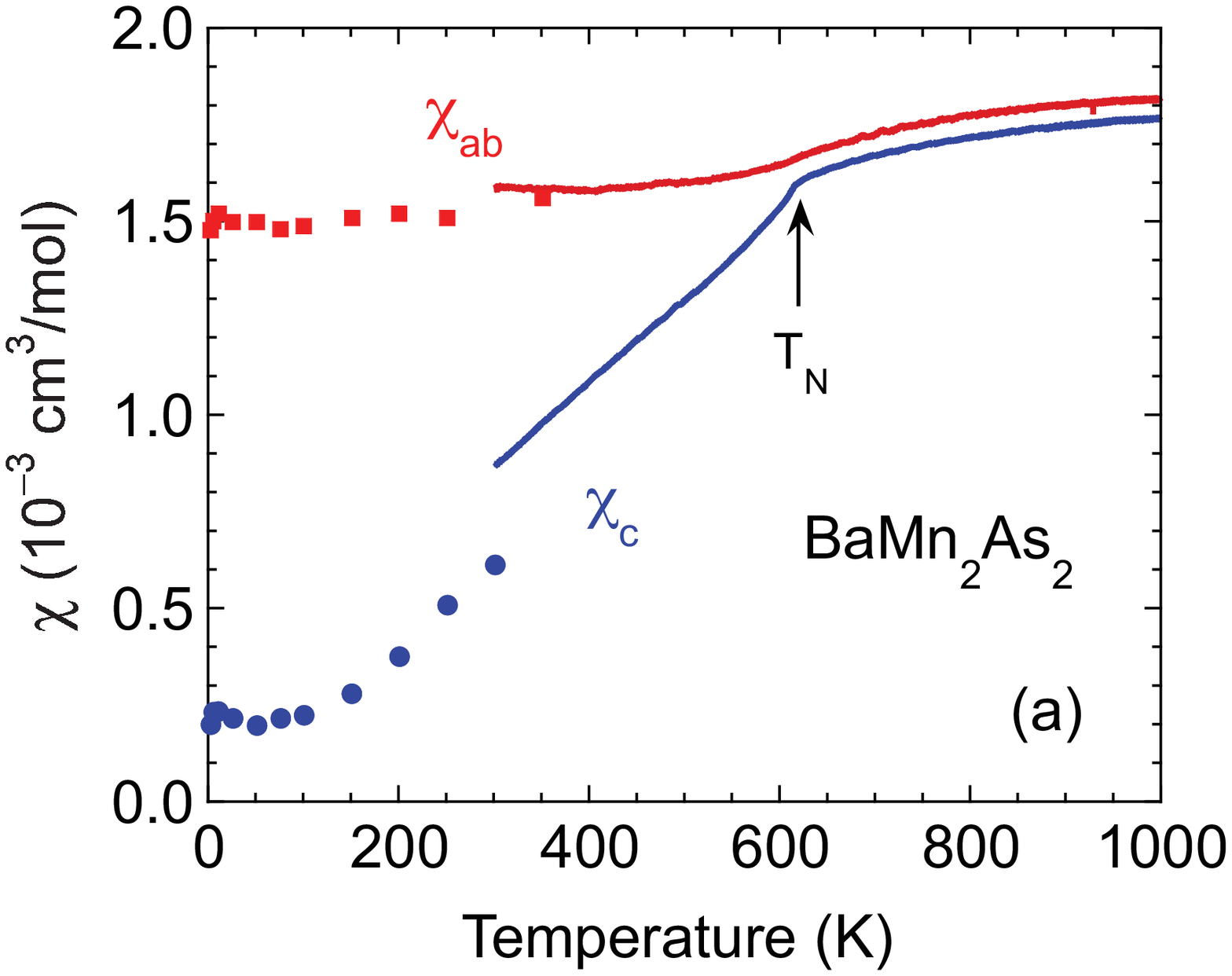}
\includegraphics [width=3.2in]{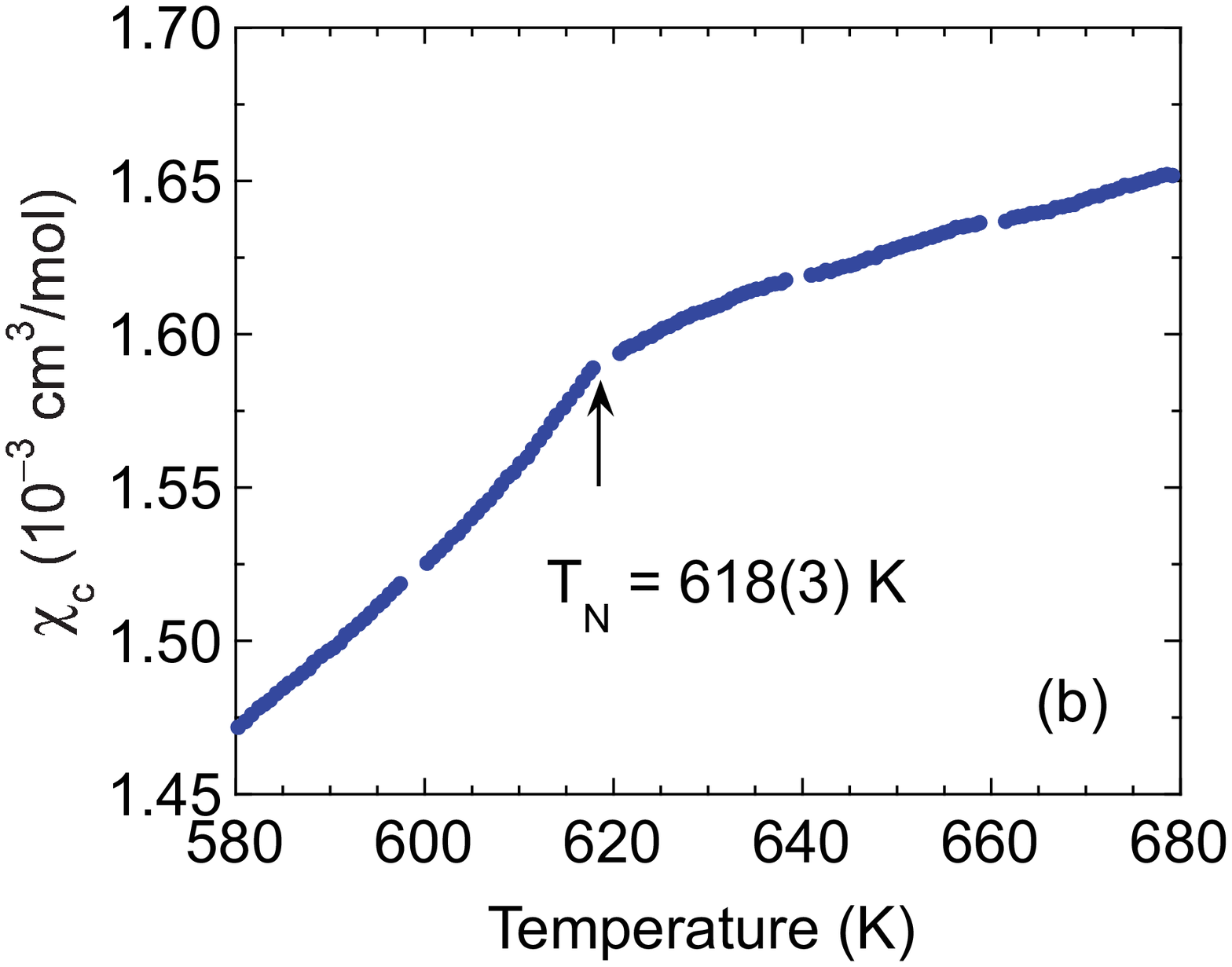}
\caption{(Color online) (a) Magnetic susceptibility $\chi$ versus temperature $T$ of single crystals of ${\rm BaMn_2As_2}$ with the applied magnetic field parallel to the $c$-axis ($\chi_c$) and to the $ab$-plane ($\chi_{ab}$).  The individual symbols are the data previously reported in Ref.~\onlinecite{singh2009}.  The solid curves are the present data obtained in an applied magnetic field $H = 3$~T\@.  The N\'eel temperature $T_{\rm N}$ is indicated.  (b) Expanded plot of $\chi_c(T)$ for temperatures around $T_{\rm N}$.  The temperature of the maximum slope of $\chi_c(T)$ gives $T_{\rm N} = 618$(3)~K\@.}
\label{BaMn2As2_Hi_T_chi}
\end{figure}

\begin{table}
\caption{\label{Tab:ChiAFValues} Parameters describing the magnetic behaviors of ${\rm BaMn_2As_2}$.  Here, $T_{\rm N}$ is the N\'eel temperature, $\chi_{ab}$ is the magnetic susceptibility with the magnetic field aligned in the $ab$-plane, $\chi_{c}$ is the magnetic susceptibility with the magnetic field aligned along the $c$-axis, and $\chi_{\rm ave} = (2\chi_{ab} + \chi_c)/3$ is the powder-averaged value of the susceptibility.  Our $\chi(T)$ results and those of Ref.~\onlinecite{singh2009} are on single crystals.  The $T_{\rm N}$ value listed for Ref.~\onlinecite{YSingh2009} was obtained from magnetic neutron diffraction measurements on a powder sample.}
\begin{ruledtabular}
\begin{tabular}{ccc}
Property & Value & Reference\\
\hline
$\chi_{ab}(10~{\rm K})$ & $1.50(2)\times10^{-3}~{\rm cm^3/mol}$ & \onlinecite{singh2009}\\
$\chi_c(10~{\rm K})$ & $0.20(2)\times10^{-3}~{\rm cm^3/mol}$ & \onlinecite{singh2009}\\
$T_{\rm N}$ & 625(1) K & \onlinecite{YSingh2009}\\
$T_{\rm N}$ & 618(3) K & This work\\
$\chi_{\rm orb}$ & $0.20(2)\times10^{-3}~{\rm cm^3/mol}$ & This work\\
$\chi_{ab}(T_{\rm N})$ & $1.66\times10^{-3}~{\rm cm^3/mol}$ & This work\\
$\chi_c(T_{\rm N})$ & $1.60\times10^{-3}~{\rm cm^3/mol}$ & This work\\
$\chi_{\rm ave}(T_{\rm N})$ & $1.64\times10^{-3}~{\rm cm^3/mol}$ & This work\\
$\chi_{\rm spin}(T_{\rm N})$ & $1.44\times10^{-3}~{\rm cm^3/mol}$ & This work\\
$\chi_{ab}(1000~{\rm K})$ & $1.81\times10^{-3}~{\rm cm^3/mol}$ & This work\\
$\chi_c(1000~{\rm K})$ & $1.76\times10^{-3}~{\rm cm^3/mol}$ & This work\\
$\chi_{\rm ave}(1000~{\rm K})$ & $1.79\times10^{-3}~{\rm cm^3/mol}$ & This work\\
$\chi_{\rm spin}(1000~{\rm K})$ & $1.59\times10^{-3}~{\rm cm^3/mol}$ & This work\\
\end{tabular}
\end{ruledtabular}
\end{table}

The anisotropic magnetic susceptibilites $\chi(T)$ of a single crystal of ${\rm BaMn_2As_2}$ in an applied magnetic field $H = 3$~T are shown in Fig.~\ref{BaMn2As2_Hi_T_chi}(a) for temperatures of 300 to 1000~K, together with our previous data\cite{singh2009} below 350~K\@.  Our $\chi_{ab}(T)$ data are consistent with the previous $\chi_{ab}(T)$ data over the temperature range of overlap (300--400~K),\cite{singh2009} but there is a difference between the $c$-axis data sets over that overlap temperature range for reasons that are not clear to us.  The temperature of the maximum slope of $\chi_c(T)$ from Fig.~\ref{BaMn2As2_Hi_T_chi}(b) gives the N\'eel temperature as $T_{\rm N} = 618$(3)~K, nearly the same as the value of 625(1)~K determined from the previous magnetic neutron diffraction measurements on a powder sample.\cite{YSingh2009}  Above $T_{\rm N}$, the susceptibility is nearly isotropic and exhibits negative curvature.  The susceptibility appears to reach a maximum at a temperature $T^{\rm max} \approx 1000$~K, where the value of the average susceptibility is $\chi_{\rm ave}^{\rm max} = 1.79\times 10^{-3}~{\rm cm^3/mol}$ and a ``mol'' refers to a mole of formula units (f.u.) unless otherwise stated.  The values of the anisotropic susceptibilities at several distinctive temperatures are summarized in Table~\ref{Tab:ChiAFValues}.

One can partition the measured susceptibility $\chi(T)$ of a material into spin $\chi_{\rm spin}$ and orbital $\chi_{\rm orb}$ parts.  Generally the orbital part is independent of $T$ but $\chi_{\rm spin}$ does depend on $T$, so one obtains
\be
\chi(T) = \chi_{\rm orb} + \chi_{\rm spin}(T).
\label{Eq:chiGen}
\ee
The $\chi_{\rm orb}$ generally consists of paramagnetic Van Vleck and diamagnetic core contributions, plus the Landau diamagnetism of conduction electrons which is not significant in semiconducting ${\rm BaMn_2As_2}$.  From Fig.~\ref{BaMn2As2_Hi_T_chi}(a), the measured $\chi(T>T_{\rm N})$ is (nearly) isotropic.  Therefore we infer that $\chi_{\rm orb}$ is isotropic at all $T$.  For a collinear antiferromagnetic insulator (semiconductor) such as ${\rm BaMn_2As_2}$, one expects the spin susceptibility parallel to the ordered moment direction, $\chi_{c\ \rm spin}$ in our case, to be zero at $T\to0$.   From Fig.~\ref{BaMn2As2_Hi_T_chi}(a) we then obtain
\be
\chi_{\rm orb} \approx 0.20(2) \times 10^{-3}\ {\rm cm^3/mol},
\label{Eq:ChiOrb}
\ee
which we have included in Table~\ref{Tab:ChiAFValues}.  Thus the spin susceptibility is given by
\be
\chi_{\rm spin}(T) = \chi(T)-\chi_{\rm orb}.
\label{chiSpin}
\ee

We have listed the values of $\chi_{\rm spin}$ at $T = T_{\rm N}$ and $T = 1000$~K in Table~\ref{Tab:ChiAFValues}.  It appears from Fig.~\ref{BaMn2As2_Hi_T_chi}(a) that $\chi(T)$ reaches a maximum at a temperature $T^{\rm max}\approx 1000$~K\@.  Then one obtains from Table~\ref{Tab:ChiAFValues} the product
\be
\chi_{\rm spin}^{\rm max}T^{\rm max} \approx 0.80~{\rm \frac{cm^3~K}{mol\ Mn}}.
\label{Eq:ChiMaxTMaxExp}
\ee
Note that this value is for a mole of spins, not a mole of formula units.  We will use this value later when comparing theory and experiment.

The temperature dependence of $\chi(T)$ above $T_{\rm N}$ in Fig.~\ref{BaMn2As2_Hi_T_chi}(a) is opposite to that expected for a fully three-dimensional antiferromagnet, where $\chi$ decreases rather than increases above $T_{\rm N}$.\cite{kittel1966}  However, the behavior we observe above $T_{\rm N}$ is common in low-dimensional antiferromagnets such as the tetragonal cuprate compound ${\rm Sr_2CuO_2Cl_2}$ where the intralayer magnetic coupling within the Cu$^{+2}$ spin $S = 1/2$ square lattice is much stronger than the interlayer coupling.\cite{Johnston1997}  Such antiferromagnets exhibit a susceptibility with a broad maximum and the corresponding onset of strong short-range AF ordering at a temperature $T^{\rm max}$ of order the mean-field AF long-range transition temperature [see Eq.~(\ref{Eq:TNfromJ1J2Jc}) below].  However, for the compound ${\rm Sr_2CuO_2Cl_2}$ one estimates $T^{\rm max} \sim 1500$~K but it exhibits long-range AF ordering only at a much lower temperature $T_{\rm N} = 250~{\rm K} \ll T^{\rm max}$.  The interlayer coupling $J_c$ is much smaller than the in-plane coupling $J_{ab}$ in quasi-two-dimensional antiferromagnets.  The suppression of $T_{\rm N}$ with respect to $T^{\rm max}$ is due to fluctuation effects associated with the low dimensionality of the system.

In the following we consider what can be learned about the signs and strengths of the exchange interactions in ${\rm BaMn_2As_2}$ from analysis of our experimental data on this compound in terms of molecular field theory.  Later in Sec.~\ref{SecThy} we develop the theory for fitting the experimental data taking into account the intralayer magnetic correlations that are present above $T_{\rm N}$, which we will then apply to fit the $\chi(T>T_{\rm N})$ data in Fig.~\ref{BaMn2As2_Hi_T_chi}(a) in Sec.~\ref{SecThyExpFit}.

\section{\label{Sec:MFT} Molecular Field Theory (MFT)}

We will be analyzing various experimental data for ${\rm BaMn_2As_2}$ using the Weiss molecular field theory (MFT).  To introduce the MFT, we first consider the known results for a local magnetic moment model on a bipartite spin lattice with equal numbers of spins {\bf S} in the two spin sublattices $i$ and $j$ interacting with the same nearest-neighbor (NN) exchange constant $J$ with the Heisenberg Hamiltonian 
\be
{\cal H} = J{\displaystyle\sum\limits_{\rm NN}}\mathbf{S}_{i}\cdot\mathbf{S}_{j}
 + g\mu_{\rm B}H {\displaystyle\sum\limits_{i}} S^z_{i},
\label{Eq:NNHeisHam}
\ee
where $g$ is the spectroscopic splitting factor ($g$-factor), $\mu_{\rm B}$ is the Bohr magneton and $H$ is the magnitude of the applied magnetic field which is in the $z$-direction.  For such a quantum local moment system of identical spins interacting by NN interactions, if the susceptibility in the absence of $J$ follows the Curie law $\chi_0 = C/T$, then in MFT the $\chi(T)$ above the magnetic ordering temperature follows the Curie-Weiss law\cite{kittel1966}
\be
\chi = \frac{C}{T + \theta},
\label{EqCurieWeiss}
\ee
where the Curie constant $C$ is
\be
C = \frac{Ng^2\mu_{\rm B}^2S(S+1)}{3k_{\rm B}},
\label{CC}
\ee
$N$ is the number of spins and $k_{\rm B}$ is Boltzmann's constant.  Taking $N$ to be Avogadro's number $N_{\rm A}$ and $g = 2$ gives a useful expression for the Curie constant per mole of spins as
\be
C = 0.50020\,S(S+1)\ {\rm \frac{cm^3\,K}{mol\ spins}}.
\label{CC2}
\ee
The Weiss temperature $\theta$ is  
\be
\theta = \frac{zJS(S+1)}{3k_{\rm B}}, 
\label{WT}
\ee
where $z$ is the coordination number of each spin.  Here, positive $\theta$ corresponds to the case when $J$ is positive (AF interactions), whereas a negative $\theta$ corresponds to the case when $J$ is negative (FM interactions).  If $\theta$ is positive, then the magnetic ordering temperature is $T_{\rm N}=\theta$ for AF ordering.  On the other hand, if $\theta$ is negative, then FM ordering occurs at the Curie temperature $T_{\rm C} = |\theta|$.

As discussed in Appendix~\ref{HTSE}, the Curie-Weiss law is not simply a mean-field expression.\cite{Johnston1997, Fisher1962, Rushbrooke1958, Rushbrooke1974}  It arises from the first ($1/T$) term in the exact quantum mechanical high-temperature series expansion of the nearest-neighbor two-spin correlation function and is accurate in the limit that higher order $1/T^n$ terms in the two-spin correlation functions are negligible.  Thus the Curie-Weiss law, and hence our scaling expressions in Eqs.~(\ref{ChivsT}) and~(\ref{ChivsT3}) below, begin to fail when $1/T^2$ and higher order terms in the two-spin correlation functions become significant compared to the $1/T$ term with decreasing $T$.

Another important conclusion from Appendix~\ref{HTSE} is that the Weiss temperature in the Curie-Weiss law results from all the spins that a given spin interacts with, irrespective of the dimensionality of the spin lattice, of whether or not the spin lattice is bipartite (see Sec.~\ref{Sec:MFTJ1J2J3}) or whether all those interactions are the same, but where all spins are equivalent.  Thus if there are different interactions present of a given spin $i$ with other spins $j$, in Eq.~(\ref{WT}) for the Weiss temperature one can make the replacement $zJ\to\sum_{j=1}^zJ_{ij}$, where $z$ is the total number of spins that spin $i$ has interactions with, giving the Weiss temperature as
\be
\theta = \frac{S(S+1)\sum_{j=1}^zJ_{ij}}{3k_{\rm B}}.
\label{WT3}
\ee

\section{\label{Sec:MFTJ1J2J3} The $J_1$-$J_2$-$J_c$ Heisenberg Model Treated in Molecular Field Theory}

The Hamiltonian~(\ref{Eq:HamilJ1J2Jz}) represents a situation where there is coupling both between the two spin sublattices and within each sublattice, where the two sublattices 1 and 2 correspond to the red (up-pointing) and blue (down-pointing) magnetic moments in the top panel of Fig.~\ref{Fig:Magnetic_Structures}, respectively.  Consider a specific spin $i$ in sublattice~1.  This spin $i$ has four in-plane NN in sublattice~2 coupled by $J_1$ and two out-of-plane NN in sublattice~2 coupled by $J_c$.  Within the same sublattice~1, spin $i$ is coupled to four in-plane NNN by $J_2$.  Since there are multiple exchange constants present from a given spin to its NN and NNN spins, we have
\[
\sum_{j=1}^{10} J_{ij} = 2(2J_1+J_c + 2J_2)
\]
and the Weiss temperature~(\ref{WT3}) becomes
\be
\theta = \frac{2(2J_1+J_c + 2J_2)\,S(S+1)}{3k_{\rm B}} .
\label{Eq:ThetaJ1J2Jc}
\ee
We cannot measure $\theta$ for ${\rm BaMn_2As_2}$ because according to Fig.~\ref{BaMn2As2_Hi_T_chi} the temperature range required for the susceptibility measurments to be in the Curie-Weiss regime would be far above 1000~K\@.

In MFT, the magnetic induction ${\bf B}=B\hat{\bf k}$ seen by each sublattice 1 and~2 is the sum of the applied field ${\bf H}=H\hat{\bf k}$ and the respective exchange field ${\bf H}=H_{\rm exch}\hat{\bf k}$, i.e., 
\bea
B_1 &=& H+H_{\rm 1\,exch}\nonumber\\*
B_2 &=& H+H_{\rm 2\,exch}.\label{Eq:B1,2}
\eea
The MFT exchange field $H_{\rm exch}$ seen by each sublattice is respectively
\bea
H_{\rm 1\,exch} &=& \lambda_{\rm s}M_1 + \lambda_{\rm d}M_2 \nonumber\\*
H_{\rm 2\,exch} &=& \lambda_{\rm d}M_1 + \lambda_{\rm s}M_2, \label{Eq:H12exch}
\eea
where $\lambda_{\rm s}$ is the net molecular field coupling parameter for coupling within the {\bf same} sublattice and $\lambda_{\rm d}$ is the net molecular field coupling parameter for coupling between the two {\bf different} sublattices.  We will obtain in Eq.~(\ref{Eq:lambdafromJs}) below expressions for these $\lambda$ values in terms of the $J$ parameters in Hamiltonian~(\ref{Eq:HamilJ1J2Jz}).

We only consider here the limit of low applied fields $H$.  In MFT, the magnetization of each sublattice 1 and~2 is given by the response to the applied field plus the exchange field as
\bea
M_1(T,H) &=& \frac{\chi_0(T)B_1}{2} \nonumber\\*
&=& \frac{\chi_0(T)}{2}(H + \lambda_{\rm s}M_1 + \lambda_{\rm d}M_2)\label{Eq:MFTM1M2}\\*
M_2(T,H) &=& \frac{\chi_0(T)B_2}{2} \nonumber\\*
&=& \frac{\chi_0(T)}{2}(H + \lambda_{\rm s}M_1 + \lambda_{\rm d}M_2),\nonumber
\eea
where $\chi_0(T) \equiv \lim_{H\to0}M/H$ is the temperature-dependent \emph{spin} susceptibility of the whole system in the absence of the explicit exchange fields, the factors of 1/2 are there because each sublattice only has half of the total number of spins, and $M_i$ is the $z$-axis magnetization of the system induced by a magnetic field in the $z$-direction with magnitude $H$.  In the paramagnetic state, $M_2=M_1$ and Eqs.~(\ref{Eq:MFTM1M2}) yield
\[
M_i(H,T)=\frac{\chi_0(T)H/2}{1-(\chi_0/2)(\lambda_{\rm d}+\lambda_{\rm s})},
\] 
where $i=1,2$.  Since $M=2M_i$, one obtains the spin susceptibility $\chi(T) = 2M_i/H$ as
\be
\chi(T) = \frac{\chi_0(T)}{1-(\chi_0/2)(\lambda_{\rm d}+\lambda_{\rm s})}.
\label{Eq:ChiMFT}
\ee

The inverse susceptibility is
\be
\frac{1}{\chi(T)} = \frac{1}{\chi_0(T)}-\frac{\lambda_{\rm d}+\lambda_{\rm s}}{2}.
\label{Eq:chiInvMFT}
\ee
This is typical of molecular field theory, where the molecular exchange field just shifts the inverse susceptibility up or down by a temperature-independent amount that depends on the sign and magnitude of the net molecular field coupling constant.  It is important to note, with respect to fitting experimental data by molecular field theory, that the presence of molecular fields cannot change the temperature of peaks in the susceptibility $\chi_0(T)$ that is assumed in the absence of explicit exchange couplings.  For example, one could take $\chi_0(T)$ to be the susceptibility of the isotropic square lattice Heisenberg antiferromagnet such as in Fig.~\ref{Tsq.J2_all_L80} below, which has a broad peak at $T \sim J/k_{\rm B}$.  If one uses a molecular exchange field to magnetically couple the square lattice layers, this molecular field cannot change the temperature of the broad AF short-range ordering peak.

To determine the magnetic ordering temperature(s) $T_{\rm m}$, we set the applied field $H$ to zero in Eqs.~(\ref{Eq:MFTM1M2}) and solve for nonzero $M_1$ and $M_2$.  For the general case one obtains
\be
\frac{\chi_0(T_{\rm m})}{2}(\lambda_{\rm s}\pm\lambda_{\rm d}) = 1,
\label{Eq:TNChi0}
\ee
so $T_{\rm m}$ depends on the assumed $\chi_0(T)$.  From Eqs.~(\ref{Eq:H12exch}), we see that for G-type AF ordering, we need to have $\lambda_{\rm d}$ to be negative, so we take the minus sign in Eq.~(\ref{Eq:TNChi0}) to get
\be
\frac{\chi_0(T_{\rm N})}{2}(\lambda_{\rm s}-\lambda_{\rm d}) = 1,
\label{Eq:TNChi01}
\ee
where now $T_{\rm m}$ is the antiferromagetic ordering (N\'eel) temperature $T_{\rm N}$.  Now we can use the solution for a $\lambda$ in terms of the related $J$ value(s) from Ref.~\onlinecite{kittel1966} to get
\bea
\lambda_{\rm s} &=& -\left(\frac{2}{Ng^2\mu_{\rm B}^2}\right)(4J_2)\nonumber\\*
\lambda_{\rm d} &=& -\left(\frac{2}{Ng^2\mu_{\rm B}^2}\right)(4J_1+2J_c),
\label{Eq:lambdafromJs}
\eea
which yield
\be
\lambda_{\rm s}-\lambda_{\rm d} = \left(\frac{2}{Ng^2\mu_{\rm B}^2}\right)(4J_1+2J_c-4J_2).
\label{Eq:lmdas-lmdad}
\ee
Inserting this expression into Eq.~(\ref{Eq:TNChi01}) for G-type antiferromagnets gives
\be
\left[\frac{\chi_0(T_{\rm N})}{Ng^2\mu_{\rm B}^2/k_{\rm B}}\right]\frac{4J_1+2J_c-4J_2}{k_{\rm B}} = 1. \label{Eq:JRestrictions2}
\ee
This is a constraint on the exchange parameters in ${\rm BaMn_2As_2}$ in addition to those in Eqs.~(\ref{Eq:JRestrictions}).  If $\chi_0(T)$ is the spin susceptibility per mole of spins, then $N$ is Avogadro's number $N_{\rm A}$.  Taking $g=2$ we have
\be
\frac{N_{\rm A}g^2\mu_{\rm B}^2}{k_{\rm B}} = 1.500~{\rm \frac{cm^3}{mol}}
\ee
and Eq.~(\ref{Eq:JRestrictions2}) becomes
\be
\frac{4J_1+2J_c-4J_2}{k_{\rm B}} = \frac{1.500~{\rm cm^3/mol}}{\chi_0(T_{\rm N})}.
\label{Eq:JsfromChiTm2}
\ee

In the following sections we will assume that the spin susceptibility in the absence of any explicit exchange fields follows a Curie law, $\chi_0(T) = C/T$.  Then Eqs.~(\ref{CC2}) and~(\ref{Eq:JsfromChiTm2}) yield 
\be
\frac{2J_1+J_c-2J_2}{k_{\rm B}} = \frac{3T_{\rm N}}{2S(S+1)},
\label{Eq:JsfromChiTm3}
\ee
or
\be
T_{\rm N} = \frac{2(2J_1+J_c-2J_2)}{3k_{\rm B}}\,S(S+1).
\label{Eq:TNfromJ1J2Jc}
\ee
Substituting Eq.~(\ref{Eq:TNfromJ1J2Jc}) into~(\ref{Eq:lmdas-lmdad}) gives
\be
\lambda_{\rm s}-\lambda_{\rm d} = \frac{6k_{\rm B}T_{\rm N}}{Ng^2S(S+1)\mu_{\rm B}^2}.
\label{Eq:lmdas-lmdad2}
\ee

It is useful to express differently how the NNN intra-sublattice interaction $J_2$ affects $T_{\rm N}$.  From Eq.~(\ref{Eq:TNfromJ1J2Jc}), one obtains
\be
\frac{T_{\rm N}(J_2)}{T_{\rm N}(J_2=0)} = 1-\frac{2J_2}{2J_1+J_c},
\label{Eq:TNfromJ1J2Jc2}
\ee
which is independent of the spin $S$ and only depends on the ratio of the intrasublattice exchange constant $J_2$ to the net intersublattice exchange constant $2J_1+J_c$.  From Fig.~\ref{Fig:Magnetic_Structures} and Eq.~(\ref{Eq:TNfromJ1J2Jc2}), an antiferromagnetic $J_2>0$ is frustrating for G-type AF ordering and hence lowers $T_{\rm N}$, whereas a ferromagnetic $J_2< 0$ is nonfrustrating for G-type AF ordering and instead enhances $T_{\rm N}$. 

\subsection{N\'eel Temperature Reduction Factor $f$}

One can define a N\'eel temperature reduction factor $f$ for antiferromagnets by
\be
f = \frac{\theta}{T_{\rm N}}, 
\label{FrustRatio}
\ee
where $\theta$ is the positive AF Weiss temperature in the Curie-Weiss law in Eq.~(\ref{EqCurieWeiss}).  For molecular-field bipartite antiferromagnets with only nearest-neighbor interactions, $f = 1$.\cite{kittel1966}  However, there are four classes of AF materials in which $f$ can be much different from unity: (1) materials in which fluctuation effects associated with a low-dimensionality (0, 1 or 2) of the spin lattice are strong, (2) three-dimensional materials in which geometric frustration for AF ordering occurs, (3) spin lattices in which the signs of the exchange interactions of a spin with its neighbors frustrate the ordering, and/or (4) spin lattices that are not bipartite; i.e., interactions between spins on the same sublattice occur. In each of these classes of materials, $T_{\rm N}$ can be strongly suppressed, sometimes to $T = 0$, which gives $f \gg 1$.  Alternatively, it can occur that second neighbor interactions can enhance $T_{\rm N}$ but suppress $|\theta|$ as we will see below in Eq.~(\ref{Eq:RatioThetaTN}) if $J_2$ is negative (ferromagnetic).  It can occur that a given compound belongs to more than one class.

One of us has discussed class (1) in the context of low-dimensional copper oxide compounds such as quasi-two-dimensional ${\rm La_2CuO_4}$ containing a ${\rm Cu^{+2}}$ $d^9$ spin-1/2 square lattice and quasi-one-dimensional ${\rm Sr_2CuO_3}$ containing ${\rm Cu^{+2}}$ $d^9$ spin-1/2 chains.\cite{Johnston1997}  In these materials the AF correlation length $\xi$ grows with decreasing $T$.  In ${\rm La_2CuO_4}$, long-range AF ordering occurs at $T_{\rm N} \sim \pi(\xi/a)^2J_c/k_{\rm B}$, where the number of spins within an AF correlated area in the plane is $N_\xi \sim \pi(\xi/a)^2$, $J_c$ is the interplane nearest-neighbor exchange coupling constant and $a$ is the square lattice parameter.  A large number $N_\xi$ of spins within a correlated area amplifies the effect of a small $J_c$.  In ${\rm Sr_2CuO_3}$, $N_\xi$ grows much more slowly with decreasing $T$ than in ${\rm La_2CuO_4}$ because what is relevant here is the number of spins within a correlation length rather than within a correlation area, and the former is much smaller than the latter at the same temperature.  Hence, one expects $f$ for ${\rm Sr_2CuO_3}$ to be much larger than for ${\rm La_2CuO_4}$, as observed.  The $\xi$ and the Weiss temperature $\theta$ are determined by the in-chain or in-plane exchange coupling $J_{ab} \gg J_c$, respectively, and hence $T_{\rm N} \ll \theta$ for both compounds.  

Ramirez has extensively discussed class (2).\cite{Ramirez1994}  In frustrated three-dimensional antiferromagnets, the susceptibility follows a Curie-Weiss-like temperature dependence down to temperatures much less than $\theta$.  One can describe the physics in two equivalent ways.  In one view, the AF correlation length $\xi$ does not grow as fast as one would predict from the Curie-Weiss law where one expects $\xi$ to diverge at the mean-field $T_{\rm N} = \theta$.  An alternate equivalent explanation is that because the Curie-Weiss law holds to low temperatures $T\ll\theta$, which results in $f \gg 1$,  the coefficients of the higher-order $1/T^n$ terms ($n > 1$) in the high temperature series expansions of the two-spin correlation functions in Eqs.~(\ref{GammaExpand}) and~(\ref{ChiFromCorr3}) in Appendix~\ref{HTSE} are strongly suppressed in frustrated antiferromagnets.  ${\rm BaMn_2As_2}$ likely belongs to classes~(1), (3) and~(4). 

Using Eqs.~(\ref{Eq:ThetaJ1J2Jc}) and~(\ref{Eq:TNfromJ1J2Jc}) which assume $\chi_0=C/T$ and $J_1,\ J_c > 0$, the ratio of the Weiss temperature to the N\'eel temperature for G-type antiferromagnets in the $J_1$-$J_2$-$J_c$ model within MFT is
\be
f = \frac{\theta}{T_{\rm N}} = \frac{2J_1+J_c+2J_2}{2J_1+J_c-2J_2} = \frac{1+\frac{2J_2}{2J_1+J_c}}{1-\frac{2J_2}{2J_1+J_c}},
\label{Eq:RatioThetaTN}
\ee
which gives
\be
\frac{2J_2}{2J_1+J_c} =\frac{f-1}{f+1}.
\label{Eq:J2fromf}
\ee
Thus $f$ depends on the sign and magnitude of the NNN in-plane interaction $J_2$.  For an antiferromagnetic $J_2>0$, one gets $f>1$, whereas for a ferromagnetic $J_2<0$, one gets $f<1$.  The constraint on $J_2$ in Eqs.~(\ref{Eq:JRestrictions}) that $J_2<J_1/2$ still applies, giving an upper limit (for which $J_c=0$) of
\be
f < 3. \hspace{0.2in} {\rm (for\ G\ type\ AF\ ordering)}
\label{Eq:fConstraint1}
\ee
Using Eq.~(\ref{Eq:J2fromf}), one can rewrite Eq.~(\ref{Eq:TNfromJ1J2Jc2}) as
\be
\frac{T_{\rm N}(J_2)}{T_{\rm N}(J_2=0)} = \frac{2}{f+1}.
\label{Eq:TNfromJ1J2Jc3}
\ee

In MFT in the paramagnetic state, the spin susceptibility~(\ref{EqCurieWeiss}) follows the Curie-Weiss law $\chi(T) = C/(T+\theta)$, and $\chi(T)$ reaches a maximum  at $T = T_{\rm N}$.  Therefore we obtain the product
\bea
\chi_{\rm spin}^{\rm max}T^{\rm max} &=& \chi_{\rm spin}^{\rm max}(T_{\rm N})T_{\rm N}=\frac{C}{T_{\rm N}+\theta}\,T_{\rm N} \nonumber\\*
&=& \frac{C}{T_{\rm N}(1+f)}\,T_{\rm N} \label{Eq:} = \frac{C}{1+f} \nonumber\\*
&=& \frac{0.5002\,S(S+1)}{1+f}\ {\rm \frac{cm^3\,K}{mol\ spins}},\label{Eq:chimaxtmaxMFTJ1J2Jc}
\eea
where we used Eq.~(\ref{CC2}) in the last equality.  This gives
\be
f = \frac{(0.5002~{\rm cm^3\,K/mol\,spins})S(S+1)}{\chi_{\rm spin}^{\rm max}T^{\rm max}}-1.
\label{Eq:ffromchimaxtmax}
\ee

From Eqs.~(\ref{Eq:JsfromChiTm3}) and~(\ref{Eq:chimaxtmaxMFTJ1J2Jc}) one can solve for $2J_1+J_c$ and $J_2$ to obtain
\bea
\frac{2J_1+J_c}{k_{\rm B}} &=& \frac{(3\,{\rm cm^3\,K/mol})T_{\rm N}}{8\chi_{\rm spin}^{\rm max}T^{\rm max}},\label{Eq:FindJ1J2Jc}\\*
\frac{J_2}{k_{\rm B}} &=& \frac{2J_1+J_c}{2k_{\rm B}}-\frac{3T_{\rm N}}{4S(S+1)}.\nonumber
\eea

Additional useful expressions include the following.  From Eqs.~(\ref{Eq:ThetaJ1J2Jc}), (\ref{Eq:TNfromJ1J2Jc}) and~(\ref{FrustRatio}) one obtains   
\be
2J_1+J_c = \frac{3k_{\rm B}T_{\rm N}(f+1)}{4S(S+1)}.
\label{Eq:j1c}
\ee
Then from Eq.~(\ref{Eq:J2fromf}) one gets
\be
2J_2 = \frac{3k_{\rm B}T_{\rm N}(f-1)}{4S(S+1)}.
\label{Eq:j2}
\ee
Now using Eqs.~(\ref{Eq:j1c}) and~(\ref{Eq:j2}) we can rewrite the molecular field coupling constants in Eqs.~(\ref{Eq:lambdafromJs}) in the simple symmetric forms
\bea
\lambda_{\rm s} &=& -\frac{T_{\rm N}(f-1)}{C}\label{Eq:lambdasdTN}\\*
\lambda_{\rm d} &=& -\frac{T_{\rm N}(f+1)}{C},\nonumber
\eea
where $C$ is the Curie constant in Eq.~(\ref{CC}).

\subsection{Anisotropic $\chi(T)$ below $T_{\rm N}$}

We would like to compare our experimental anisotropic $\chi(T)$ data below $T_{\rm N}$ with the MFT predictions using the $J_1$-$J_2$-$J_c$ model.  We discuss first the perpendicular susceptibility $\chi_\perp$ and then the parallel susceptibility $\chi_\parallel$, where $\chi_\perp$ refers to the susceptibility with the applied magnetic field perpendicular to the easy axis of the collinear antiferromagnetic structure and $\chi_\parallel$ to the susceptibility when the applied magnetic field is parallel to it.  For ${\rm BaMn_2As_2}$, $\chi_\parallel = \chi_c$ and $\chi_\perp = \chi_{ab}$.  In the Heisenberg model, above $T_{\rm N}$ the susceptibility is isotropic and hence $\chi_\parallel = \chi_\perp$.  Below $T_{\rm N}$, $\chi_\perp$ and $\chi_\parallel$ are no longer the same.

Below $T_{\rm N}$ of a collinear antiferromagnet, one always has $\chi_\parallel < \chi_\perp$ (see also Fig.~\ref{Fig:BaMn2As2_Hi_T_chi_All_Fit} below).  Since the magnetic energy of the system at low fields is $-(1/2)\chi H^2$, if the field is aligned along the ordered moment axis the spin system can lower its energy via a ``spin-flop'' transition where the ordered moment axis rotates to be perpendicular to the applied field.  To prevent this from happening, one needs to have an anisotropy energy present that is not included in the Heisenberg Hamiltonian.  Otherwise one could never measure $\chi_\parallel$.  An important example of such an anisotropy energy is the axial single ion anisotropy energy with the form $DS_z^2$ (for $S > 1/2$) and $D<0$, and/or higher order forms, that arise from the spin-orbit interaction of the magnetic moments with the crystalline electric field of the lattice.  Here we assume that an infinitesimal axial anisotropy is present with sufficient magnitude to prevent the ordered moment axis from flopping from the parallel to the perpendicular orientation when we are measuring the parallel magnetization in the limit of an infinitesmal field.  We will not further consider the spin-flop transition in this paper. 

The $\chi_\perp(T)$ and $\chi_\parallel(T)$ are derived for the $J_1$-$J_2$-$J_c$ model at $T  \leq T_{\rm N}$ in Appendix~\ref{Eq:AnisChi}.  For the perpendicular susceptibility, one obtains the constant value
\be
\chi_\perp = \frac{1}{|\lambda_{\rm d}|} = \frac{C}{T_{\rm N}(1+f)}  = \frac{C}{T_{\rm N}+\theta}=\chi(T_{\rm N}), \hspace{0.1in}(T\leq T_{\rm N}),
\label{Eq:ChiPerp}
\ee
using $f \equiv \theta/T_{\rm N}$.  This result is similar to that for a bipartite lattice,\cite{kittel1966} except in that case one has $\theta=T_{\rm N}$ whereas in our case we have $\theta = fT_{\rm N}$ with, in general, $f\neq1$ from Eq.~(\ref{Eq:RatioThetaTN}).  The estimated values of $f$ from Eqs.~(\ref{Eq:fFromchimaxtmaxMFT}) below are $\sim3$--5 in ${\rm BaMn_2As_2}$, i.e., $T_{\rm N}$ is much smaller than $\theta$, but within MFT the susceptibility still follows the Curie-Weiss law $\chi = C/(T+\theta)$ down to $T_{\rm N}$.  This interesting behavior is the result of bond frustration for AF ordering (the antiferromagnetic NNN interaction $J_2$ frustrates the occurrence of G-type AF ordering) and has been noted as a property of geometrically frustrated antiferromagnets.\cite{Ramirez1994}

\begin{figure}
\includegraphics [width=3.in]{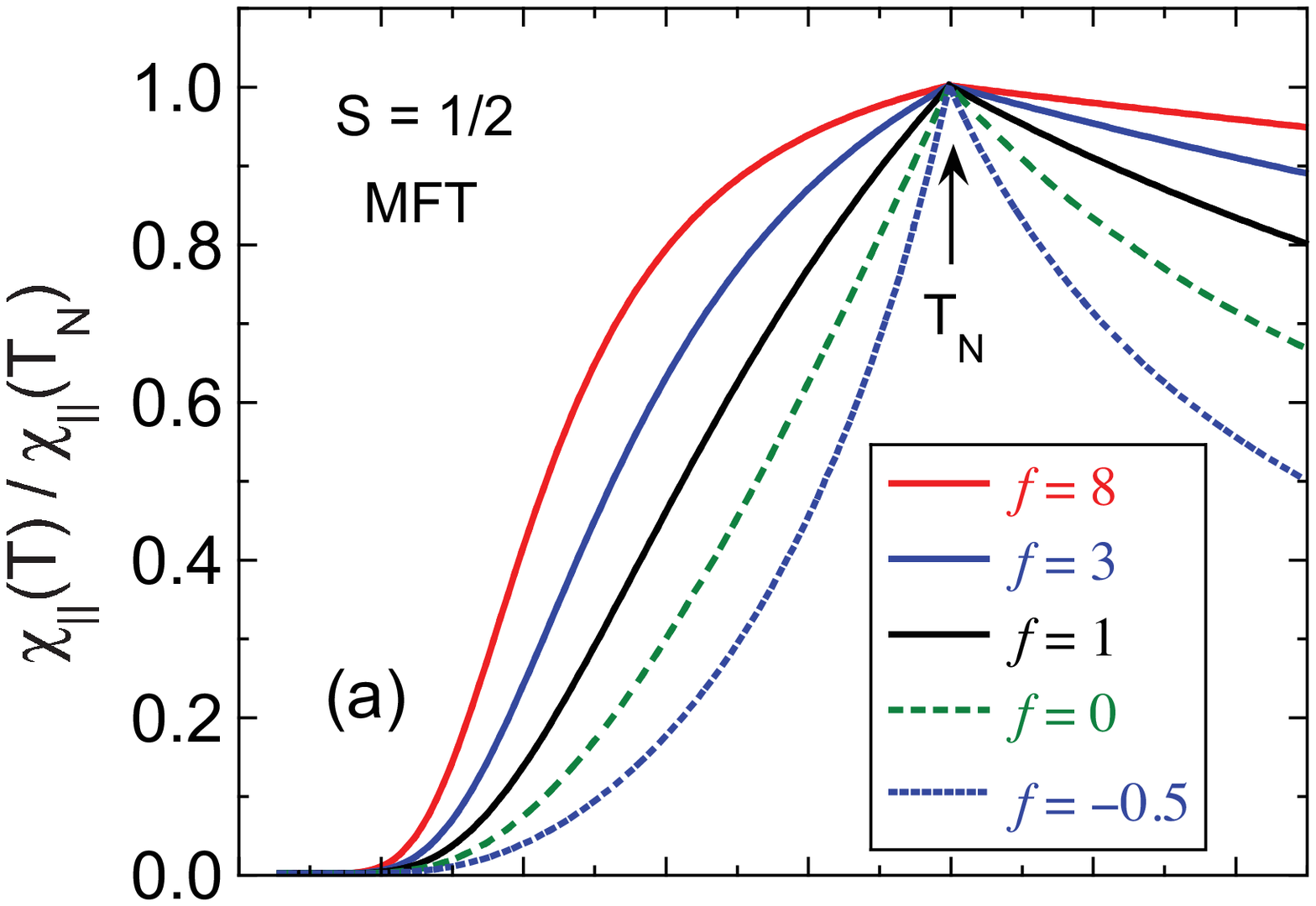}\vspace{-0.23in}
\includegraphics [width=3.in]{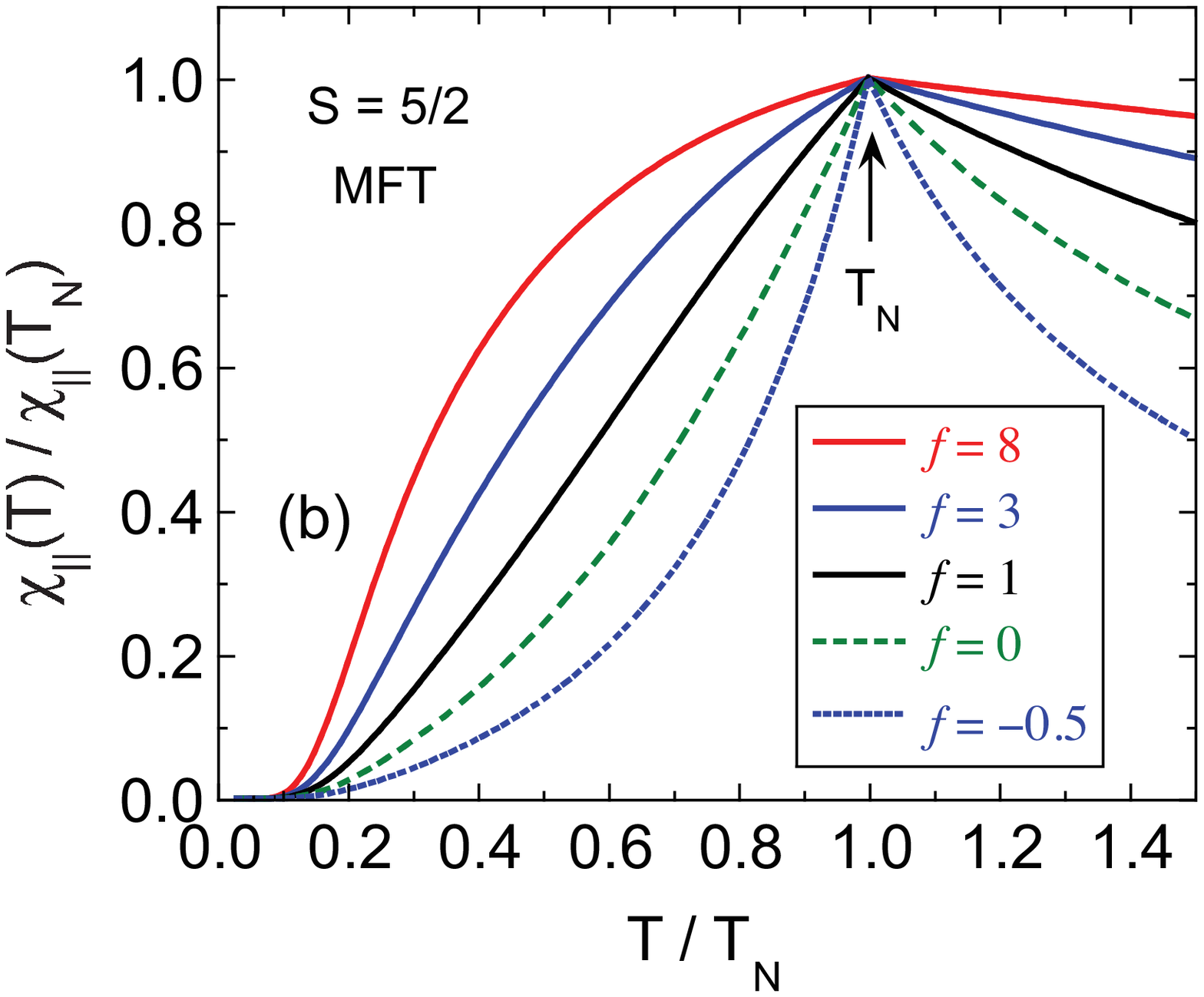}
\caption{(Color online) Parallel susceptibility $\chi_\parallel$ versus temperature $T$ through the N\'eel temperature $T_{\rm N}$ for G-type AF ordering using the $J_1$-$J_2$-$J_c$ model in molecular field theory (MFT) for various values of $f = \theta/T_{\rm N}$, as listed, for spins (a) $S = 1/2$ and (b) $S = 5/2$.  The order of the curves from top to bottom is the same as in the figure legends.  At temperatures  $T > T_{\rm N}$, $\chi$ is isotropic.  For $T < T_{\rm N}$, the perpendicular susceptibility is constant, $\chi_\perp = \chi_\perp(T_{\rm N})$ (not shown).  The G-type AF state is unstable against the stripe AF state for $f>3$.}
\label{Fig:Chi_ParaJ1J2MFT_S1_2}
\end{figure}

\begin{figure}
\includegraphics [width=3in]{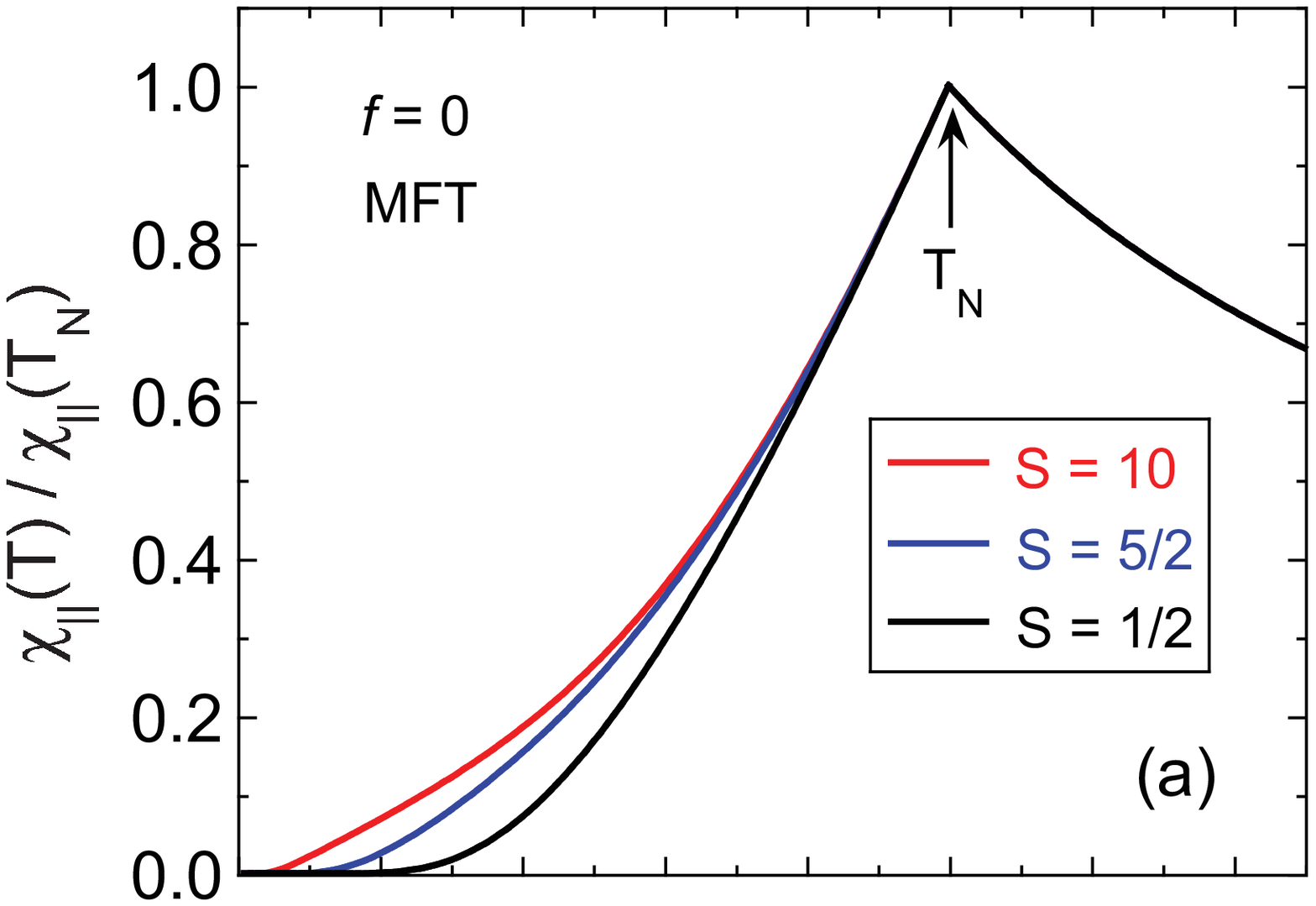}\vspace{-0.24in}
\includegraphics [width=3in]{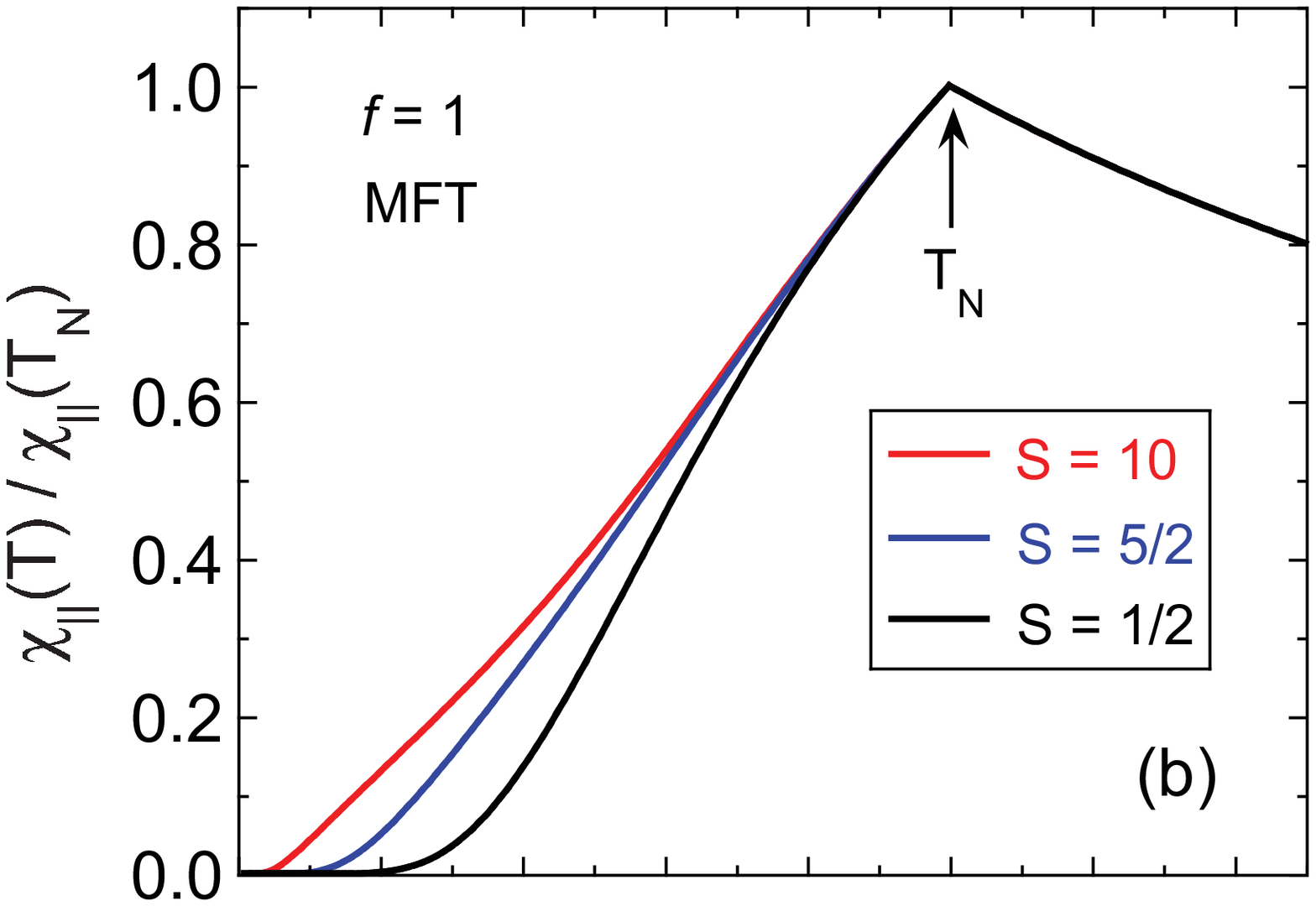}\vspace{-0.24in}
\includegraphics [width=3in]{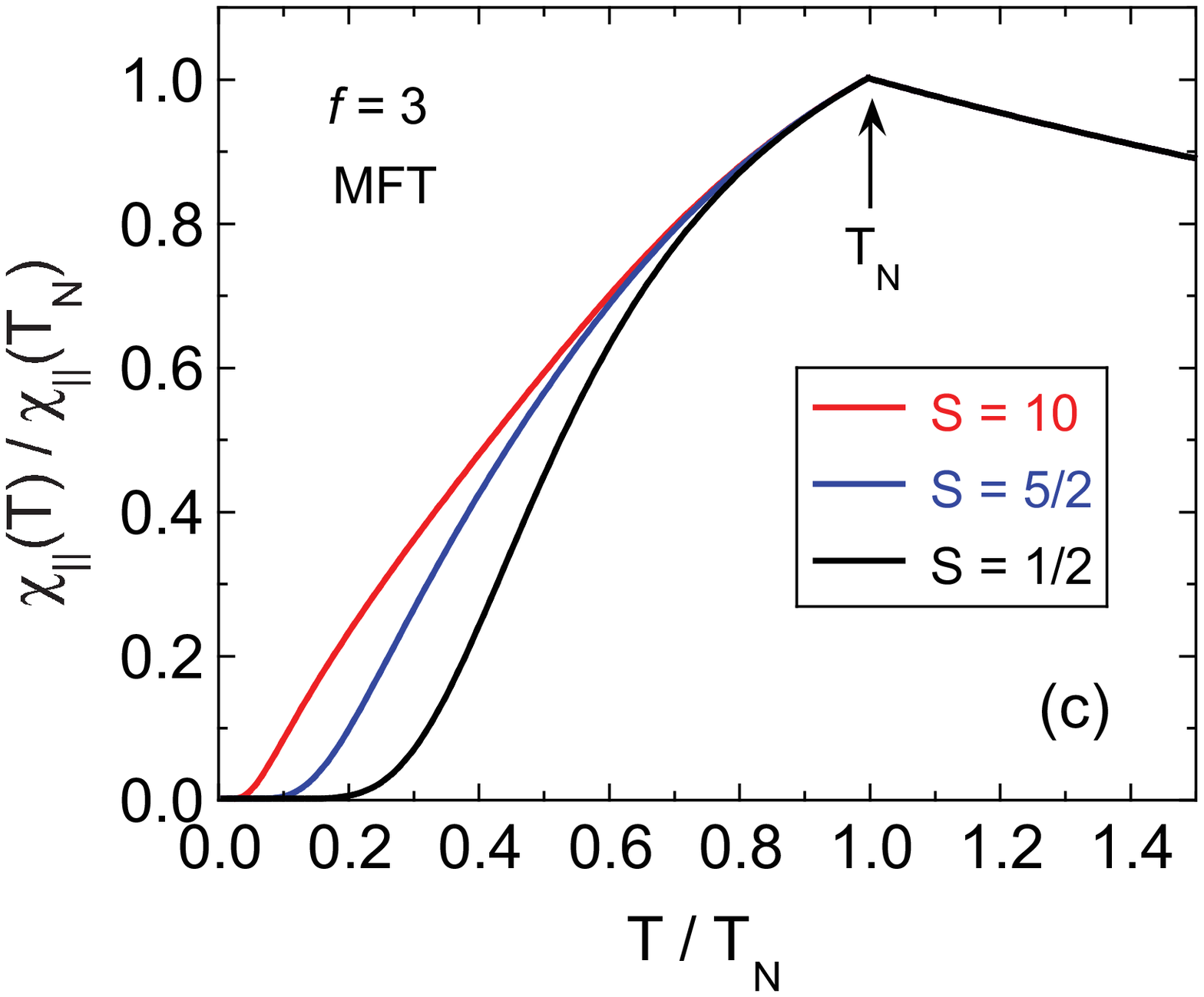}
\caption{(Color online) Parallel susceptibility $\chi_\parallel$ versus temperature $T$ through the N\'eel temperature $T_{\rm N}$ for the $J_1$-$J_2$-$J_c$ model in molecular field theory (MFT) for spins $S = 1/2,$ 5/2 and~10 and $f = \theta/T_{\rm N}$ values of (a) 0, (b) 1 and (c) 3.  The order of the curves in each panel from top to bottom is the same as in the figure legends.  The value $f=1$ corresponds to the conventional bipartite lattice with $J_2=0$.}
\label{Fig:Chi_ParaJ1J2MFT_f0}
\end{figure}

The dependence of $\chi_\parallel(T)/\chi_\parallel(T_{\rm N})$ on $t\equiv T/T_{\rm N}$ determined by solving Eqs.~(\ref{Eq:mu1zmu2zMFT}), (\ref{Eq:barmuparallel}) and~(\ref{Eq:MFTPredictionforChi||}) is shown in Figs.~\ref{Fig:Chi_ParaJ1J2MFT_S1_2}(a) and \ref{Fig:Chi_ParaJ1J2MFT_S1_2}(b) for spins $S = 1/2$ and $S = 5/2$, respectively, for various values of $f = \theta/T_{\rm N}$.  The value $f=1$ corresponds to the conventional nonfrustrated bipartite stacked square spin lattice as in the top panel of Fig.~\ref{Fig:Magnetic_Structures} with $J_2=0$.  Figure~\ref{Fig:Chi_ParaJ1J2MFT_S1_2} shows that the presence of a nonzero diagonal coupling $J_2$ has a strong influence on $\chi_\parallel(T)/\chi_\parallel(T_{\rm N})$.  Complementary plots of $\chi_\parallel(T)/\chi_\parallel(T_{\rm N})$ versus $T/T_{\rm N}$ at fixed $f = 0$, 1 and~3 for $S = 1/2$, 5/2 and~10 are shown in Fig.~\ref{Fig:Chi_ParaJ1J2MFT_f0}.

\subsection{Ordered Moment versus Temperature below $T_{\rm N}$}

\begin{figure}
\includegraphics [width=3.in]{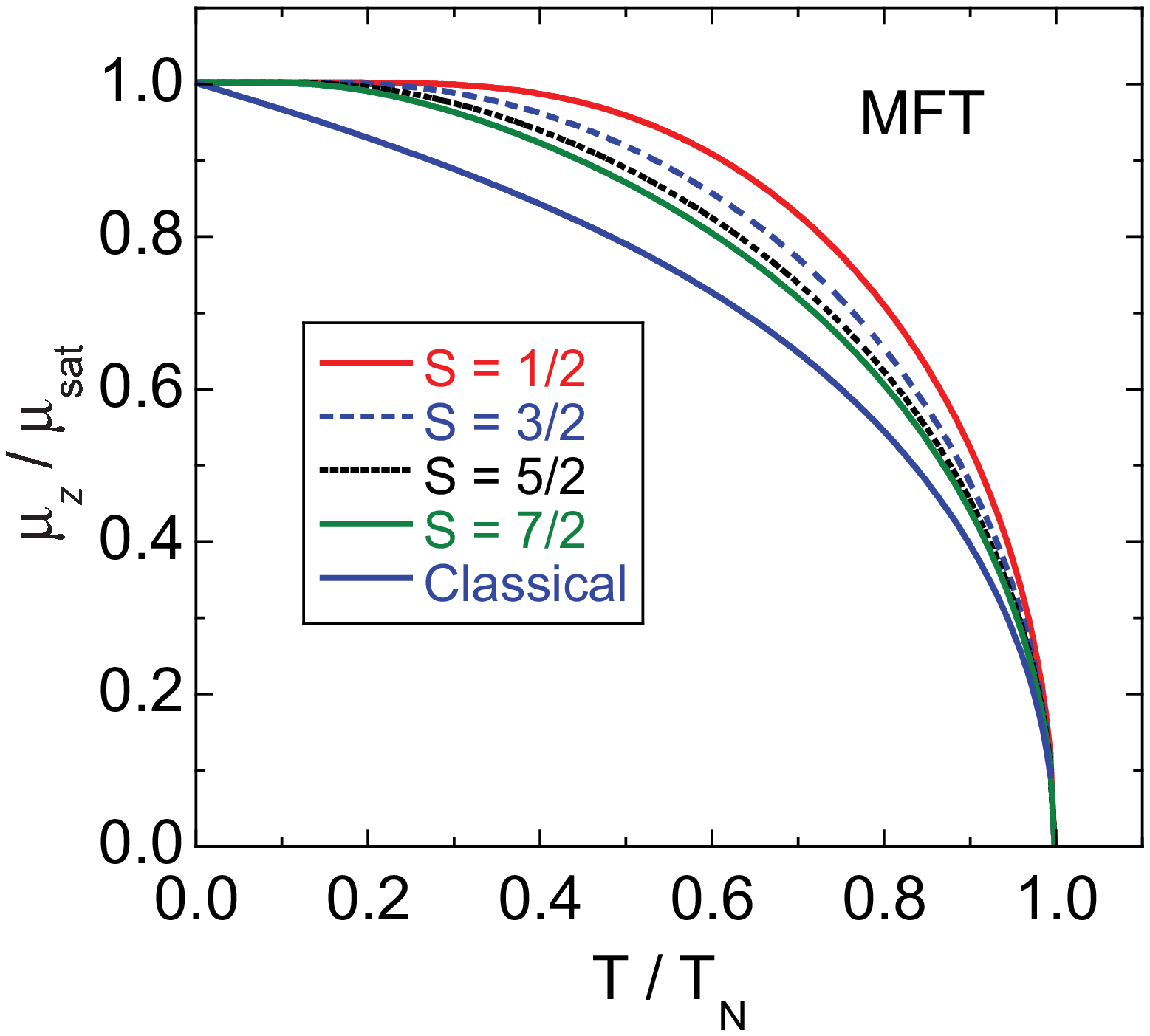}
\caption{(Color online) Ordered moment $\mu_z \equiv \mu_z^\dagger$ versus temperature $T$ from molecular field theory of the $J_1$-$J_2$-$J_c$ model for a collinear antiferromagnet for classical spins and for several quantum spins as listed, where the saturation moment is $\mu_{\rm sat}=gS\mu_{\rm B}$.  The order of the curves from top to bottom is the same as in the figure legend.   Remarkably, the results are independent of $J_2$.}
\label{Fig:OrderedMomentMFT}
\end{figure}

The ordered moment in the antiferromagnetic state of ${\rm BaMn_2As_2}$, which is the staggered moment $\mu_z^\dagger$ in Eq.~(\ref{Eq:muDagger}),  has been previously measured, but not modeled.\cite{YSingh2009}  In Appendix~\ref{Sec:OrdMomentMFT} we determine the MFT predictions on the basis of the $J_1$-$J_2$-$J_c$ model.  In Fig.~\ref{Fig:OrderedMomentMFT} are plotted the solutions of Eq.~(\ref{Eq:OrderedMomentMFT}) for the nonzero ordered moment $\mu_z \equiv \mu_z^\dagger$ versus reduced temperature $T/T_{\rm N}$ for classical spins and for four values of quantum spins.  In contrast to quantum spins for which $\mu_{z}^\dagger$ approaches the respective saturation moment $\mu_{\rm sat} = gS\mu_{\rm B}$ exponentially fast for $T\to0$ due to an energy gap between the ground state and the lowest excited states, the low-temperature classical behavior is linear.  This results in a magnetic heat capacity $C_{\rm mag}\to~{\rm constant}\neq0$ as $T\to0$ for classical spins, which violates the third law of thermodynamics, whereas for  quantum spins $C_{\rm mag}\to0$ as $T\to0$ (see Fig.~\ref{Fig:Mag_Heat_Capacity_MFT} below).

Interestingly, the parameter $f=\theta/T_{\rm N}$ that characterizes the influence of $J_2$ on the magnetism has disappeared from the expression for $\mu_z^\dagger(T)$ in Eq.~(\ref{Eq:muDagger}) when the temperature scale is normalized by $T_{\rm N}$.  Thus Eq.~(\ref{Eq:OrderedMomentMFT}) and the plots in Fig.~\ref{Fig:OrderedMomentMFT} are identical to the corresponding MFT predictions for an AF bipartite spin lattice with $J_2=0$.  However, in our case with $J_2\neq0$, we must keep in mind that $J_2$ has already manifested its influence on the magnetism by changing $T_{\rm N}$.  

\subsection{Zero-Field Magnetic Heat Capacity $C_{\rm mag}$ and Entropy $S_{\rm mag}$ below $T_{\rm N}$}

\begin{figure}
\includegraphics [width=3.in]{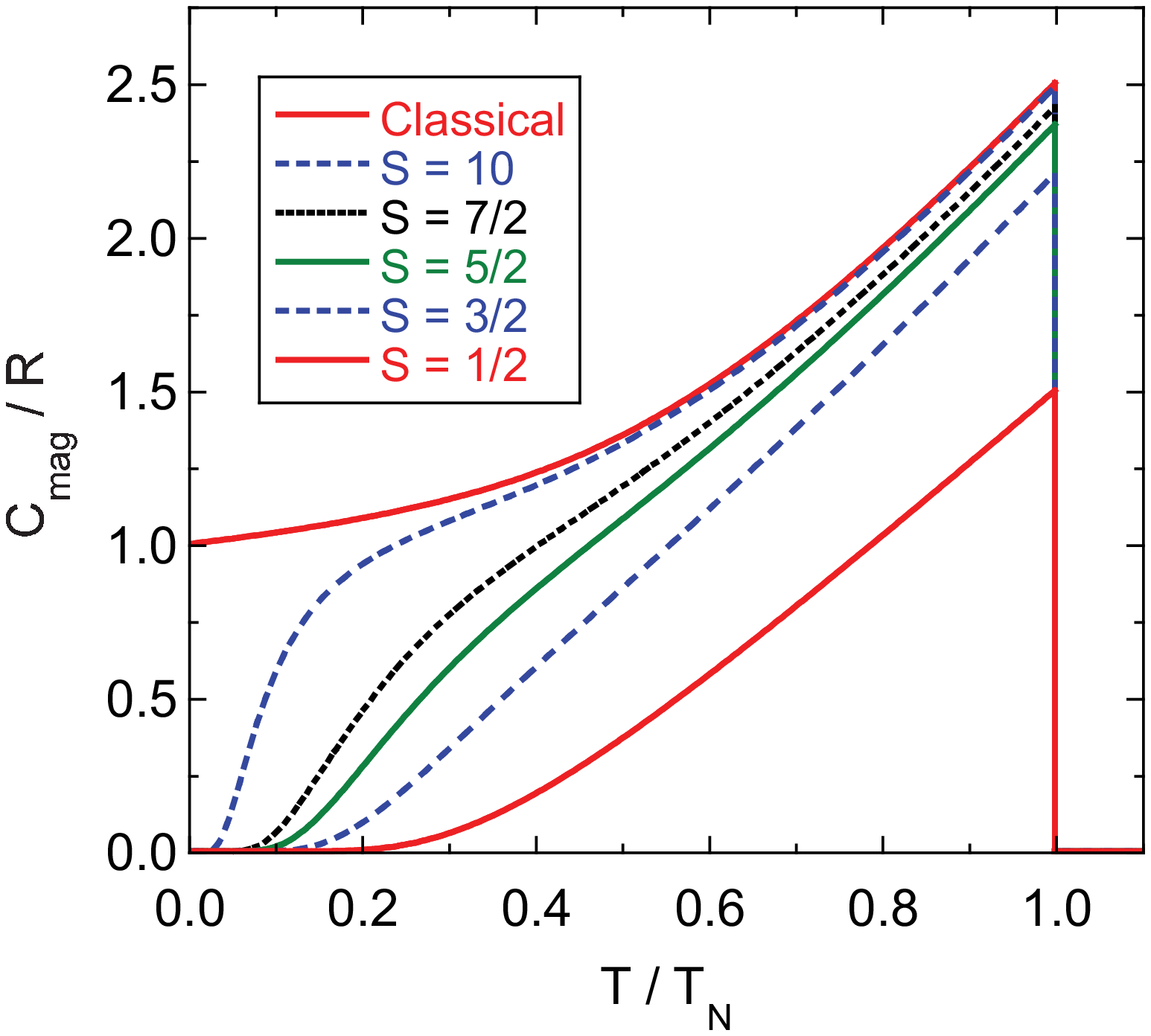}
\caption{(Color online) Magnetic component of the heat capacity $C_{\rm mag}$, divided by the molar gas constant $R$, versus the ratio of the temperature $T$ to the N\'eel temperature $T_{\rm N}$ according to the molecular field theory Eq.~(\ref{Eq:CmagQuantum}) of the $J_1$-$J_2$-$J_c$ model for a collinear antiferromagnet for classical spins and for several quantum spins as listed.  The order of the curves from top to bottom is the same as in the figure legend.  As in Fig.~\ref{Fig:OrderedMomentMFT}, the results are independent of $J_2$.}
\label{Fig:Mag_Heat_Capacity_MFT}
\end{figure}

The zero-field magnetic heat capacity $C_{\rm mag}(T)$ is derived in MFT in Appendix~\ref{Sec:CpMFT} as
\be
\frac{C_{\rm mag}(t)}{R} = -\frac{3S}{S+1}\bar{\mu}_z^\dagger (t)\frac{d{\bar{\mu}_z^\dagger}(t)}{dt},
\label{Eq:CmagQuantum}
\ee
where $t = T/T_{\rm N}$ is the reduced temperature and $\bar{\mu}_z^\dagger = \mu_z^\dagger/\mu_{\rm sat}$ is the reduced ordered (staggered) moment.  Since $\bar{\mu}_z^\dagger$ does not explicitly depend on $J_2$ as discussed above, neither does $C_{\rm ave}(T/T_{\rm N})$, but rather  implicitly via the dependence of $T_{\rm N}$ on $J_2$.
The ${\bar{\mu}_{z}^\dagger}(t)$ is determined by numerically solving Eq.~(\ref{Eq:OrderedMomentMFT}).  Inserting this result into~(\ref{Eq:CmagQuantum}), $C_{\rm mag}(T)$ was calculated for several spin values as plotted in Fig.~\ref{Fig:Mag_Heat_Capacity_MFT}.  One observes a triangular shape for $C_{\rm mag}(t)$ near $T_{\rm N}$ for each $S$, which is characteristic of the mean field solution, with a discontinuous increase (``jump'') in $C_{\rm mag}(T)$ upon decreasing $T$ through $T_{\rm N}$ given by\cite{Smart1966}
\be
\frac{\Delta C_{\rm mag}(T_{\rm N})}{R} = \frac{5}{2}\left[\frac{(2S+1)^2 - 1}{(2S+1)^2 + 1}\right],
\label{Eq:DeltaC}
\ee
where $2S+1$ is the Zeeman degeneracy in zero field for a spin~$S$.  There is not a large range of $\Delta C_{\rm mag}$ possible upon varying the spin $S$\@.  From Eq.~(\ref{Eq:DeltaC}) one obtains
\bea
\frac{\Delta C_{\rm mag}(T_{\rm N})}{R} &=& \frac{3}{2}\hspace{0.2in}\left(S = \frac{1}{2}\right)\nonumber\\*
\frac{\Delta C_{\rm mag}(T_{\rm N})}{R} &=& \frac{5}{2},\hspace{0.2in}\left(S = \infty\right)\nonumber
\eea
consistent with Fig.~\ref{Fig:Mag_Heat_Capacity_MFT}.

The evolution in Fig.~\ref{Fig:Mag_Heat_Capacity_MFT} of the low temperature $C_{\rm mag}(T)$ with increasing spin $S$ is interesting.  It develops a hump at a temperature that decreases with increasing $S$, until in the classical limit $S\to\infty$ the hump merges into the classical finite-value behavior for $T\to0$.  The hump is required in order that the entropy of the disordered spin system increase with increasing $S$ (see below), since $C_{\rm mag}(T)$ is bounded from above by the classical prediction.  For quantum spins, the heat capacity approaches zero exponentially at sufficiently low temperatures irrespective of the (finite) spin value, whereas for classical spins the heat capacity approaches a nonzero finite value for $T\to0$.

\begin{figure}
\includegraphics [width=3.in]{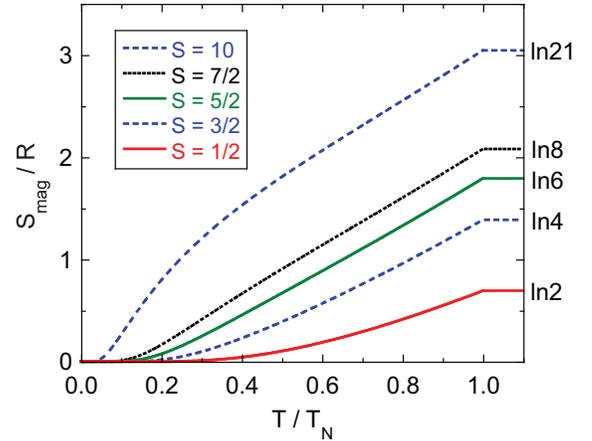}
\caption{(Color online) Magnetic entropy $S_{\rm mag}/R$ for quantum spins versus reduced temperature $T/T_{\rm N}$ according to molecular field theory for the quantum spins $S$ indicated.  The order of the curves from top to bottom is the same as in the figure legend.  In the disordered state at $T > T_{\rm N}$, the magnetic entropy is the constant value $S_{\rm mag} = R\ln(2S+1)$ for each $S$, as indicated along the right-hand ordinate.}
\label{Fig:Smag_vs_T_MFT}
\end{figure}

The magnetic entropy $S_{\rm mag}$ is determined from the magnetic heat capacity via
\be
\frac{S_{\rm mag}(t)}{R} = \int_0^t \frac{C_{\rm mag}(t)/R}{t}\,dt.
\label{Eq:Smag}
\ee
The magnetic entropy obtained from Eq.~(\ref{Eq:Smag}) and from the data in Fig.~\ref{Fig:Mag_Heat_Capacity_MFT} is plotted versus temperature for quantum spins in Fig.~\ref{Fig:Smag_vs_T_MFT}.  The constant values for $T\geq T_{\rm N}$ as indicated by the notations on the right-hand ordinate agree with the values expected for disordered spins given by the molar magnetic entropy $S_{\rm mag}=R\ln(2S+1)$.  For classical spins the calculated entropy for $T\to0$ is $S_{\rm mag}(T\to0^+)/R = \lim_{T\to0}\ln[T/(0~{\rm K})] = \infty$, which violates the third law of thermodynamics.

\section{\label{Sec:MFT vs Expt} Comparison of Theoretical Predictions with Experimental Data for ${\rm\bf BaMn_2As_2}$}

\subsection{Comparisons with Molecular Field Theory}

We expect the Mn$^{+2}$ $d^5$ ion in ${\rm BaMn_2As_2}$ to have the high-spin configuration with spin $S = 5/2$.  On the other hand, the observed ordered moment is $\mu = 3.9(1)~\mu_{\rm B}$/Mn,\cite{YSingh2009} suggesting from the relation $\mu = gS\mu_{\rm B}$ with $g=2$ that $S = 2$.  Therefore in the following we will consider both of these possibilities.  

\subsubsection{N\'eel Temperature}

Using Eq.~(\ref{Eq:JsfromChiTm3}) and $T_{\rm N} = 625$~K for ${\rm BaMn_2As_2}$, one obtains
\bea
\frac{2J_1+J_c-2J_2}{k_{\rm B}} &=& {\rm 156\ K} \hspace{0.25in}(S=2)\nonumber\\*
 &=& {\rm 107\ K}. \hspace{0.15in}(S=5/2)
\label{Eq:JaveMFT}
\eea

\begin{table}
\caption{\label{Tab:J1J2JcMFT}  Parameters determined from a fit of magnetic susceptibility data by molecular field theory of G-type antiferromagnetic ${\rm BaMn_2As_2}$ with NN ($J_{1}$), NNN ($J_{2}$) and interlayer ($J_{c}$) exchange interactions.}
\begin{ruledtabular}
\begin{tabular}{lcc}
Quantity & $S=2$ & $S=5/2$ \\
\hline
$f=\theta/T_{\rm N}$ & 2.75 & 4.47 \\
$(2J_1+J_c-2J_2)/k_{\rm B}$ (K) & 156 & 107 \\
$(2J_1+J_c)/k_{\rm B}$ (K) & 293 & 293 \\
$2J_1+J_c$ (meV) & 25.2 & 25.2 \\
$S(2J_1+J_c)/k_{\rm B}$ (K) & 586 & 733 \\
$S(2J_1+J_c)$ (meV) & 50.5 & 63.2 \\
$J_2/k_{\rm B}$ (K) & 68 & 93\\
$J_2$ (meV) & 5.9 & 8.0\\
$SJ_2/k_{\rm B}$ (K) & 136 & 233\\
$SJ_2$ (meV) & 11.7 & 20.1\\
\end{tabular}
\end{ruledtabular}
\end{table}

\subsubsection{Magnetic Susceptibility}

Inserting the experimental $\chi_{\rm spin}^{\rm max}T^{\rm max}$ value from Eq.~(\ref{Eq:ChiMaxTMaxExp}) into~(\ref{Eq:ffromchimaxtmax}) gives
\bea
f &=& 2.75\hspace{0.28in} (S = 2) \label{Eq:fFromchimaxtmaxMFT}\\*
&=& 4.47.\hspace{0.2in} (S = 5/2) \nonumber
\eea
According to Eq.~(\ref{Eq:fConstraint1}), the value of $f$ for $S=5/2$ is not possible for G-type AF ordering and hence $S=5/2$ is ruled out by this criterion.  The $f$ value for $S = 2$ suggests that interlayer coupling might have a significant effect on the observed magnetic susceptibility above $T_{\rm N}$.  On the other hand, for the layered cuprate ${\rm La_2CuO_4}$ one has $z = 4$, $J/k_{\rm B} = 1600$~K, $S = 1/2$, and $T_{\rm N} = 325$~K,\cite{Johnston1997} which yields a Weiss temperature $\theta = 1600$~K and $f = 4.9$, and the magnetism of this compound is known to be described very well by two-dimensional physics in the temperature range above $T_{\rm N}$.\cite{Johnston1997}  As a further comparison, the quasi-one-dimensional spin-1/2 chain compound ${\rm Sr_2CuO_3}$ has $z = 2$, $J/k_{\rm B} = 2200$~K, $S = 1/2$, and $T_{\rm N} = 5.4$~K,\cite{Johnston1997} which yields $\theta = 1100$~K and $f = 200$.  This large $f$ value is the reason that ${\rm Sr_2CuO_3}$ is often considered to be a model quasi-one-dimensional Heisenberg antiferromagnet.\cite{Johnston1997}

Again using the experimental $\chi_{\rm spin}^{\rm max}T^{\rm max}$ value from Eq.~(\ref{Eq:ChiMaxTMaxExp}), Eqs.~(\ref{Eq:FindJ1J2Jc}) yield
\bea
\frac{2J_1+J_c}{k_{\rm B}} &=& {\rm 293\ K} \hspace{0.2in}(S=2)\label{Eq:J1JcJ2MFT}\\*
 &=& {\rm 293\ K} \hspace{0.15in}(S=5/2)\nonumber\\*
\frac{J_2}{k_{\rm B}} &=& {\rm 68\ K} \hspace{0.25in}(S=2)\nonumber\\*
 &=& {\rm 93\ K}. \hspace{0.15in}(S=5/2)\nonumber
\eea
The above results, summarized in Table~\ref{Tab:J1J2JcMFT}, are only approximate qualitative constraints on the exchange parameters in ${\rm BaMn_2As_2}$, because they assume that the susceptibility follows the Curie-Weiss law above $T_{\rm N}$, which Fig.~\ref{BaMn2As2_Hi_T_chi} shows is not accurate.  In particular, if $J_c/J_1\approx 0.1$ as determined from the neutron scattering results and the theoretical predictions in Sec.~\ref{Sec:BandTheory}, one obtains unrealistically large $J_2/J_1=0.50$ and~0.67 for $S=2$ and $S=5/2$, respectively.  The problem stems from the fact that $T^{\rm max}$ and $T_{\rm N}$ do not coincide, which is an inconsistency in the analysis.

\begin{figure}
\includegraphics [width=3.3in]{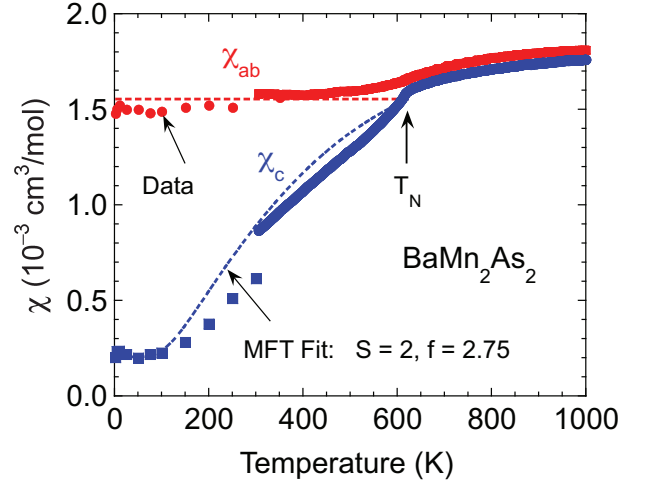}
\caption{(Color online) Comparison of the MFT prediction below $T_{\rm N}$ of the anisotropic susceptibilities in Eqs.~(\ref{Eq:ChiPerp}) and~(\ref{Eq:MFTPredictionforChi||}) versus temperature with the experimental data from Fig.~\ref{BaMn2As2_Hi_T_chi}(a).}
\label{Fig:BaMn2As2_Hi_T_chi_All_Fit}
\end{figure}

A comparison of the MFT predictions below $T_{\rm N}$ of the anisotropic susceptibilities in Eqs.~(\ref{Eq:ChiPerp}) and~(\ref{Eq:MFTPredictionforChi||}) with the experimental data from Fig.~\ref{BaMn2As2_Hi_T_chi}(a) is shown in Fig.~\ref{Fig:BaMn2As2_Hi_T_chi_All_Fit}.  For the MFT dashed-line predictions we used Eq.~(\ref{Eq:chiGen}) with $\chi_{\rm orb} = 0.2\times10^{-3} {\rm \ cm^3/mol}$ and $\chi_{\rm spin}(T_{\rm N}) = 1.35\times10^{-3} {\rm \ cm^3/mol}$.  We used the value $T_{\rm N} = 625$~K and the MFT parameter $f=2.75$ listed in Table~\ref{Tab:J1J2JcMFT} for $S = 2$.  The temperature dependences of the MFT predictions for the anisotropic susceptibilities are seen to be in semiquantitative agreement with the experimental data.  We do not consider the case $S = 5/2$ because the large $f=4.47 > 3$ in Eqs.~(\ref{Eq:JaveMFT}) and Table~\ref{Tab:J1J2JcMFT} for $S = 5/2$ makes the G-type AF structure unstable with respect to the stripe AF structure in Fig.~\ref{Fig:Magnetic_Structures}. 

\subsubsection{Ordered Moment}

\begin{figure}
\includegraphics [width=3.in]{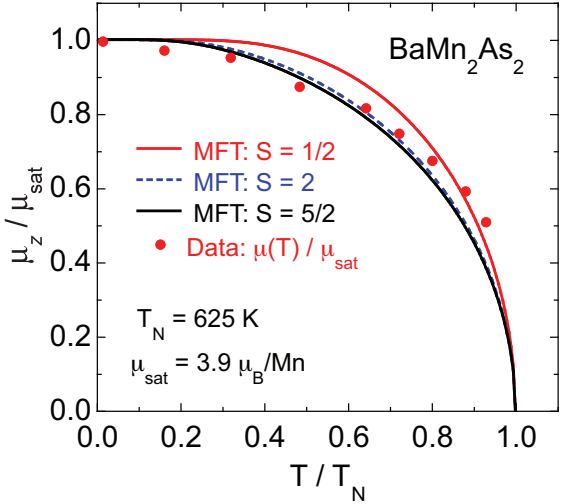}
\caption{(Color online) Ordered moment $\mu_z$ versus temperature $T$ measured for ${\rm BaMn_2As_2}$ from Ref.~\onlinecite{YSingh2009} (filled red circles).  The N\'eel temperature $T_{\rm N}$ and saturation moment $\mu_{\rm sat}$ are given in the figure.  Also shown are the molecular field theory (MFT) predictions for quantum spins $S=1/2$, 2 and 5/2 from Eq.~(\ref{Eq:OrderedMomentMFT}) (solid and dashed curves), where the order of the curves from top to bottom is the same as in the figure legend.  }
\label{Fig:BaMn2As2_ord_mom_Fit}
\end{figure}

The theoretical MFT results for the ordered moment versus temperature in Fig.~\ref{Fig:OrderedMomentMFT} for $S=3/2$ to $S=5/2$ are nearly the same, so we do not expect to be able to differentiate between the two possibilities of $S = 2$ and $S = 5/2$ for the Mn spins in ${\rm BaMn_2As_2}$ on the basis of the observed temperature dependence of the ordered moment. This expectation is confirmed in Fig.~\ref{Fig:BaMn2As2_ord_mom_Fit} where we compare the MFT predictions for $S=1/2$, 2 and~5/2 from Eq.~(\ref{Eq:OrderedMomentMFT}) with the experimental data from magnetic neutron diffraction measurements in Ref.~\onlinecite{YSingh2009}.  Although the overall temperature dependence of the data agrees with MFT, the data are not quantitatively fitted by the prediction for any particular fixed $S$ value.

\subsubsection{High-Temperature Magnetic Heat Capacity}

Here we will compare our experimental heat capacity $C_{\rm p}$ data for ${\rm BaMn_2As_2}$ single crystals at temperatures up to 350~K with the prediction of MFT for the magnetic heat capacity $C_{\rm mag}$ at high temperatures, i.e., near room temperature.  To do this we will need to estimate the lattice heat capacity contribution using the Debye model.  

\begin{figure}
\includegraphics [width=3.in]{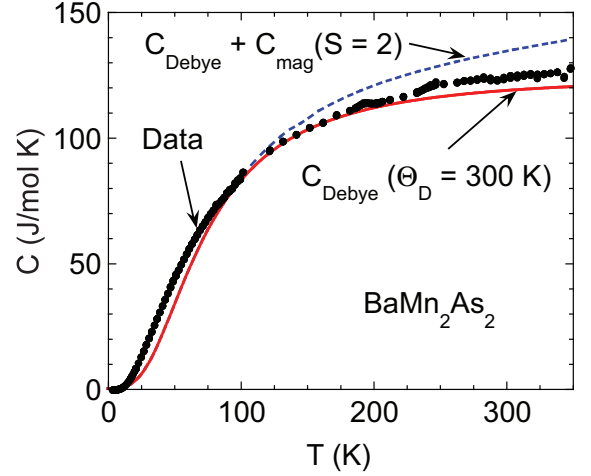}
\caption{(Color online) Heat capacity $C$ versus temperature $T$.  The experimental $C_{\rm p}(T)$ data obtained for a single crystal of ${\rm BaMn_2As_2}$ (Ref.~\onlinecite{singh2009}) are shown as the filled black circles.  The lattice heat capacity for a Debye temperature $\Theta_{\rm D} = 300$~K is plotted versus $T$ as the solid red curve.  The sum of the lattice heat capacity and the magnetic heat capacity for spin $S = 2$ is shown as the dashed blue curve.}
\label{Fig:C(T)-BaMn2As2-Xtal}
\end{figure}

The heat capacity at constant pressure $C_{\rm p}(T)$ for a single crystal of ${\rm BaMn_2As_2}$, previously reported by Singh et al.,\cite{singh2009} is shown in Fig.~\ref{Fig:C(T)-BaMn2As2-Xtal} for the measured temperature range 2--350~K\@.  We fitted the data by the Debye function for the molar lattice heat capacity of acoustic phonons at constant volume, given by\cite{kittel1966}
\be
C_{\rm Debye} = 9nR\left(\frac{T}{\Theta_{\rm D}}\right)^3\int_0^{\Theta_{\rm D}/T}\frac{x^4e^x}{(e^x -1 )^2}\,dx,
\label{Eq:DebyeCV}
\ee
where $n$ is the number of atoms per formula unit ($n=5$ here) for various values of the Debye temperature $\Theta_{\rm D}$.  In order that $C_{\rm Debye}(T)$ does not lie above the experimental data over any temperature range, the minimum value of $\Theta_{\rm D}$ is about 300~K, for which the Debye function is plotted as the solid red curve in Fig.~\ref{Fig:C(T)-BaMn2As2-Xtal}.  The $\Theta_{\rm D}$ is evidently temperature-dependent because the deviation of the curve from the experimental data varies nonmonotonically with temperature.   From the same set of experimental $C_{\rm p}(T)$ data,\cite{singh2009} at low temperatures $T\leq5$~K a value $\Theta_{\rm D} = 246(4)$~K was deduced using the Debye $T^3$ law [the low-temperature limit of Eq.~(\ref{Eq:DebyeCV})] given by\cite{kittel1966}
\bea
C_{\rm Debye} &=& \beta_{\rm D} T^3\nonumber\\*
\beta_{\rm D} &=& 0.65(3)\ {\rm \frac{mJ}{mol\,K^4}}\label{Eq:DebyeT3}\\*
\Theta_{\rm D} &=& \left(\frac{12\pi^4Rn}{5\beta_{\rm D}}\right)^{1/3}.\nonumber
\eea
The experimental data at the highest temperatures lie above the lattice heat capacity curve for $\Theta_{\rm D} = 300$~K in Fig.~\ref{Fig:C(T)-BaMn2As2-Xtal}, suggesting the presence of one or more heat capacity contributions in addition to that due to acoustic phonons.

We calculated the difference  $C_{\rm p}-C_{\rm V}$ for the lattice heat capacity for the compound ${\rm BaFe_2As_2}$, where $C_{\rm V}$ is the lattice heat capacity at constant volume, according to the thermodynamic relation $C_{\rm p}-C_{\rm V} = V_{\rm M}\beta_{\rm V}^2(T)B(T)T$,
where $V_{\rm M}$ is the molar volume, $\beta_{\rm V}$ is the volume thermal expansion coefficient, and $B$ is the bulk modulus.  For the 200--300~K temperature range, using the values $\beta_{\rm V} \approx 4.8\times10^{-5}$~K$^{-1}$,\cite{Budko2010} $B = 6.6\times10^{12}$~dyne/cm$^2$,\cite{Mittal2011} and $V_{\rm M}= 61.5$~cm$^3$/mol,\cite{Johnston2010} we obtained $C_{\rm p}-C_{\rm V} = ({\rm 9.3\,mJ/mol\,K^2})T.$  This gives $(C_{\rm p}-C_{\rm V})$(300\,K) = 2.8~J/mol\,K, which is about a factor of two too small to account for the difference between the data and the Debye curve.  It was not possible to calculate a value of $C_{\rm p}-C_{\rm V}$ specific to ${\rm BaMn_2As_2}$ because $\beta_{\rm V}(T)$ and $B(T)$ have not been measured for this compound.

The magnetic contribution $C_{\rm mag}(T)$ to the heat capacity at high temperatures was calculated using the MFT prediction in Eq.~(\ref{Eq:CmagQuantum}).  We chose to calculate it for spin $S=2$ because the $S=5/2$ possibility was ruled out by the large value of $f>3$ for spin $S=5/2$ in Eq.~(\ref{Eq:fFromchimaxtmaxMFT}).  Using $T_{\rm N}=625$~K, the $C_{\rm mag}(T)$ from two moles of spins~$S = 2$ per mole of ${\rm BaMn_2As_2}$ was added to the Debye heat capacity and is plotted as the dashed blue curve in Fig.~\ref{Fig:C(T)-BaMn2As2-Xtal}.  Now the calculated curve lies above the experimental data around room temperature, indicating that the magnetic heat capacity is smaller than predicted by MFT\@.  The $\chi(T)$ data in Fig.~\ref{BaMn2As2_Hi_T_chi}(a) appear to be approaching a maximum at a temperature $T^{\rm max}\approx 1000$~K that is far above $T_{\rm N}=625$~K, indicating the occurrence of strong short-range AF ordering above $T_{\rm N}$ (see also Sec.~\ref{SecThy} below).  This removes spin entropy and decreases $C_{\rm mag}$ below the value expected from MFT at temperatures below $T_{\rm N}$.  This may be the reason for the suppression of $C_{\rm mag}(T)$ in our measurements around room temperature.

\subsection{\label{Sec:SWTCmag} Comparison of Experiment with Spin Wave Heat Capacity Theory at Low Temperatures in the $J_1$-$J_2$-$J_c$ Model}

\subsubsection{\label{Sec:SWTCmagTheory} Theory}

The lack of significant susceptibility anisotropy above $T_{\rm N}$ in Fig.~\ref{BaMn2As2_Hi_T_chi} indicates that single-ion anisotropy is small.  This anisotropy, if present, gives rise to an energy gap in the spin wave excitation spectrum.  Here we assume that the anisotropy gap is infinitesmally small and calculate the low-temperature magnetic heat capacity of AF spin waves in the $J_1$-$J_2$-$J_c$ model.  This is an extension of the standard treatment for simple cubic spin lattices with isotropic NN exchange interactions.

The original 1952 papers by Anderson\cite{Anderson1952} and by Kubo\cite{Kubo1952} give a clear prescription of how to do this using a spin wave model with two AF sublattices~1 and~2 containing a total of $N$ spins $S$.  Their starting Heisenberg Hamiltonian is
\be
{\cal H} = J\sum_{\langle ij\rangle}{\bf S}_i\cdot{\bf S}_j
\ee
where there is only a single $J$ and the sum is over distinct nearest-neighbor spin pairs.  In zero field and in the absence of significant anisotropy the diagonalized spin wave Hamiltonian contains the following term involving the excitation energies of spin waves
\be
E = \sum_{\bf q}\ (n_{1\bf q}\hbar\omega_{1 \bf q}+n_{2\bf q}\hbar\omega_{2 \bf q}),
\label{ESWHAM}
\ee
where {\bf q} is the wave vector of a spin wave excitation, $n_{i\bf q}$ is the occupation number of the mode for sublattice $i$, and the two terms correspond to excitations on the two degenerate spin wave branches $\omega_{1 \bf q}$ and $\omega_{2 \bf q}$ associated with the two spin sublattices, respectively.  Since $n_{1\bf q}\hbar\omega_{1 \bf q} = n_{2\bf q}\hbar\omega_{2 \bf q} \equiv n_{\bf q}\hbar\omega_{\bf q}$ are degenerate, the excitation energy of the system can be written
\be
E = 2 \sum_{\bf q}n_{\bf q}\hbar\omega_{\bf q}. 
\label{Eq:Eqnq}
\ee
The thermal-average energy of the spin waves is then
\be
E_{\rm ave} = 2 \sum_{\bf q}\frac{\hbar\omega_{\bf q}}{e^{\hbar\omega_{\bf q}/k_{\rm B}T}-1}.
\label{Eq:EaveSW}
\ee
where $\langle n_{\bf q}\rangle=1/(e^{\hbar\omega_{\bf q}/k_{\rm B}T}-1)$ is the Planck distribution function for the thermal-average number of quanta in an oscillator at energy $\hbar\omega_{\bf q}$.  One converts the sum into an integral over {\bf q} for a three-dimensional spin lattice via
\be
\sum_{\bf q} \to \frac{N}{2}\frac{V_{\rm spin}}{(2\pi)^3}\int_{-\pi/a}^{\pi/a}\int_{-\pi/b}^{\pi/b}\int_{-\pi/c}^{\pi/c}d{\bf q},
\ee
where $V_{\rm spin}$ is the volume per spin.  The factor of $N/2$ arises because each spin sublattice has $N/2$ spins.  Then Eq.~(\ref{Eq:EaveSW}) becomes
\bea
E_{\rm ave} = \frac{NV_{\rm spin}}{(2\pi)^3}\int_{-\pi/a}^{\pi/a}dq_x\int_{-\pi/b}^{\pi/b}dq_y\nonumber\\*
\int_{-\pi/c}^{\pi/c}dq_z \frac{\hbar\omega_{\bf q}}{e^{\hbar\omega_{\bf q}/k_{\rm B}T}-1}.
\label{Eq:EaveSWB}
\eea

Note that the integration in Eq.~(\ref{Eq:EaveSWB}) is over the entire \emph{Brillouin zone of the primitive direct lattice} (containing a single spin), not over the Brillouin zone of the magnetic lattice.  The reason for this important fact is that integrating over the Brillouin zone of a primitive space lattice with one spin in the basis sums up the response of a single spin, whereas if one were to integrate over an antiferromagnetic Brillouin zone, this zone would include the response of more than one spin.  Indeed, the average energy per spin calculated this way does not depend on the type of magnetic ordering at all, even if the magnetic ordering is ferromagnetic or incommensurate.  The only relevant difference between the thermal average energy per spin of different magnetic ordering configurations is the difference between the specific $\omega_{\bf q}$ functions and their degeneracies over the Brillouin zone of the primitive space lattice.

\begin{figure}
\includegraphics [width=\columnwidth]{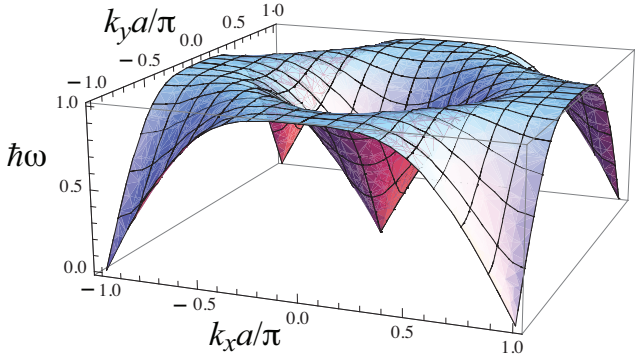}
\caption{(Color online) Spin wave dispersion relation $\hbar\omega/4J_1S$ of the isotropic two-dimensional square lattice over the Brillouin zone of the primitive tetragonal \emph{space} lattice.  The dispersion relation is doubly degenerate everywhere.  At low temperatures, there are two distinct doubly degenerate spin wave branches that are relevant, one at the $\Gamma$ point at (0,0) and the other at $(\frac{\pi}{a},\frac{\pi}{a})$ (and equivalent points).}
\label{Fig:SquareLattDispReln}
\end{figure}

The dispersion relation for a general spin lattice is 
\be
\hbar\omega_{\bf q} = zJS\sqrt{1-\gamma_{\bf q}^2}
\label{Eq:DispRelnSW}
\ee
where
\be
\gamma_{\bf q} = \frac{1}{z}\sum_{i=1}^z e^{i{\bf q}\cdot {\bf r}_i},
\label{Eq:gammaDef}
\ee
$z$ is the coordination number of a spin on one sublattice by spins on the other sublattice, and {\bf r}$_i$ is a vector from a spin to one of its $z$ neighbors.  We now need to make a point that will be illustrated using the spin wave spectrum of an isotropic two-dimensional square spin-$S$ lattice ($z=4$).  In this case Eq.~(\ref{Eq:gammaDef}) yields 
\[
\gamma_{\bf q}=\frac{1}{2}[\cos(k_xa) + \cos(k_ya)]
\]
and Eq.~(\ref{Eq:DispRelnSW}) gives the doubly degenerate dispersion relation as
\be
\hbar\omega_{\bf q} = 4JS\sqrt{1-[\cos(k_xa) + \cos(k_ya)]^2/4}.
\label{Eq:2DDispReln}
\ee
This dispersion relation is plotted in Fig.~\ref{Fig:SquareLattDispReln}.  One sees that $\omega_{\bf q}$ has doubly degenerate branches arising from zero energy at the $\Gamma$ point (0,0), as expected, but also at the corners of the Brillouin zone at $(\frac{\pi}{a},\frac{\pi}{a})$ and equivalent points.  In a three-dimensional spin lattice with $J_c \neq 0$, using the dispersion relation in Eq.~(\ref{Eq:SpnWaveDispersionsMbased}), one sees that the low-energy points of the dispersion relation move from the $(\frac{\pi}{a},\frac{\pi}{a},0)$ points in the corners of the two-dimensional Brillouin zone to the $(\frac{\pi}{a},\frac{\pi}{a},\frac{\pi}{c})$ and equivalent points at the other four corners of the three-dimensional Brillouin zone.  Thus in either case there is another multiplicative factor of two to include in Eq.~(\ref{Eq:EaveSWB}) if we only integrate over the two degenerate $\Gamma$ point branches for $T\to0$.  

Equation~(\ref{Eq:EaveSWB}) is evaluated in Appendix~\ref{App:CmagSW} to yield the magnetic heat capacity per mole of spins at low temperatures due to the spin waves as
\bea
C_{\rm mag} &=& \left(\frac{4\pi^2Rk_{\rm B}^3V_{\rm spin}}{15\hbar^3v_xv_yv_z}\right)\,T^3, \hspace{0.2in} (T \ll T_{\rm N})\label{Eq:CmagSW}
\eea
where $R$ is the molar gas constant, $V_{\rm spin}$ is the volume per spin, and $v_x,\ v_y,\ v_z$ are the spin wave velocities along the $a$-, $b$- and $c$-axes, respectively.  This expression includes the contribution of the low energy spin waves at the Brillouin zone corners, and can be written in a form analogous to Eq.~(\ref{Eq:DebyeT3}) for phonons as
\bea
C_{\rm mag} &=& \beta_{\rm SW}T^3,\nonumber\\*
\beta_{\rm SW} &=& 2\left(\frac{2\pi^2Rk_{\rm B}^3V_{\rm spin}}{15\hbar^3v_xv_yv_z}\right).\label{Eq:betaSW}
\eea
By writing the Debye temperature in Eqs.~(\ref{Eq:DebyeT3}) in terms of its constituent quantities,\cite{kittel1966} one obtains the lattice heat capacity coefficient $\beta_{\rm D}$ per mole of atoms as
\be
\beta_{\rm D} = 3\left(\frac{2\pi^2Rk_{\rm B}^3V_{\rm atom}}{15\hbar^3v^3}\right), \label{Eq:CmagSWD}
\ee
where $v$ is the sound wave speed, assumed isotropic, and $V_{\rm atom}$ is the volume per atom.  This expression is similar to Eq.~(\ref{Eq:betaSW}) except that the prefactor is three instead of two, due to the three sound wave polarization directions for each sound wave mode (two mutually perpendicular transverse polarizations and one longitudinal polarization) which are assumed to have the same wave speed $v$ in the Debye model.

\subsubsection{\label{Sec:SWTCmagBaMn2As2} Application of the Spin Wave Theory for the Magnetic Heat Capacity to the $J_1$-$J_2$-$J_c$ Heisenberg Model and ${\rm BaMn_2As_2}$}

From the expressions for the spin wave velocities in the $J_1$-$J_2$-$J_c$ model in Eq.~(\ref{Eq:SWvels}), one has
\bea
v_xv_yv_z=v_{ab}^2v_c &=& \frac{4\sqrt{2}(J_1S)^3a^2c}{\hbar^3}\left(1+\frac{J_c}{2J_1}\right)^{3/2}\nonumber\\*
&&\hspace{0.2in}\times\ \left(1 - \frac{2J_2}{J_1}\right)\sqrt{\frac{J_c}{J_1}}.
\label{Eq:vxvyvz}
\eea
For ${\rm BaMn_2As_2}$, there are two formula units, or four Mn atoms, per unit cell with volume $a^2c$.  The volume per spin is thus
\be
V_{\rm spin} = \frac{a^2c}{4}.
\label{Eq:VspinBaMn2As2}
\ee
Dividing Eq.~(\ref{Eq:vxvyvz}) by Eq.~(\ref{Eq:VspinBaMn2As2}) gives
\be
\frac{v_xv_yv_z}{V_{\rm spin}} = \frac{16\sqrt{2}(J_1S)^3}{\hbar^3}\left(1+\frac{J_c}{2J_1}\right)^{3/2} \left(1 - \frac{2J_2}{J_1}\right)\sqrt{\frac{J_c}{J_1}}.
\label{Eq:vvvV}
\ee
Inserting Eq.~(\ref{Eq:vvvV}) into~(\ref{Eq:betaSW}) gives 
\bea
\beta_{\rm SW} &=& \frac{\pi^2R}{60\sqrt{2}(J_1S/k_{\rm B})^3}\label{Eq:ThetaSW4}\\*
&& \times\ \left[\left(1+\frac{J_c}{2J_1}\right)^{3/2} \left(1 - \frac{2J_2}{J_1}\right)\sqrt{\frac{J_c}{J_1}}\right]^{-1}.\nonumber
\eea

From Table~\ref{tbl2}, the exchange constants from the neutron data are $J_1S/k_{\rm B}=380$~K, $J_2/J_1=0.29$ and $J_c/J_1=0.09$.  Inserting these values into Eq.~(\ref{Eq:ThetaSW4}) gives the calculated value
\be
\beta_{\rm SW} = 0.13~{\rm mJ/mol\,spins\,K^4\hspace{0.2in} for\ BaMn_2As_2}.
\label{Eq:betaBaMn2As2}
\ee
From Eq.~(\ref{Eq:DebyeT3}), the observed $\beta$ value per mole of Mn spins is $0.325$~mJ/mol~spins~K$^4$.  The calculated $\beta_{\rm SW}$ value is thus 40\% of the measured value, so the observed $\beta$ value contains a significant magnetic contribution if the anisotropy gap in the spin wave spectrum is negligible.  However, an anisotropy gap would reduce the spin wave contribution to the heat capacity to exponentially small values at low temperatures.

\section{\label{SecThy} Monte Carlo Simulations of the Magnetic Susceptibility and Magnetic Heat Capacity in the $J_1$-$J_2$-$J_c$ Model}

Both our classical and quantum Monte Carlo simulations were carried out within the framework of the $J_1$-$J_2$-$J_c$ Heisenberg model introduced above in Sec.~\ref{Sec:J1J2J3Model}.  We have calculated the magnetic heat capacity and magnetic spin susceptibility versus temperature for various size lattices of quantum spins $S = 1/2$, 1, 3/2, 2, and~5/2, and for the classical model.  We first motivate the scaling of the axes of our theoretical plots of $\chi(T)$, remark on the temperature regime over which this scaling is expected to hold, and then present our Monte Carlo simulation results.  Then we will compare our predictions for the magnetic susceptibility with the experimental susceptibility data for ${\rm BaMn_2As_2}$ above $T_{\rm N}$ in Fig.~\ref{BaMn2As2_Hi_T_chi} to obtain additional estimates of the exchange constants in this compound.

\subsection{Scaling of the Theoretical $\chi(T)$ Axes}

Using Eqs.~(\ref{CC}) and~(\ref{WT}) in the Heisenberg ``$J$ model'' for a bipartite spin lattice with equal NN exchange, the Curie-Weiss law~(\ref{EqCurieWeiss}) can be rewritten as
\be
\frac{\chi J}{Ng^2\mu_{\rm B}^2} = \frac{1}{\frac{3k_{\rm B}T}{JS(S+1)} + z}.
\label{CWuniversal}
\ee
The quantity on the left-hand side of Eq.~(\ref{CWuniversal}) is the theorist's definition of ``$\chi$'', which is the susceptibility per spin, in units of $1/J$, with $g\mu_{\rm B}$ set equal to 1.  On the right-hand side, we see that if we use a temperature scale defined by $k_{\rm B}T/[JS(S+1)]$, then all spin lattices with the same coordination number $z$ but with different $J$ and/or $S$ will all follow the same universal curve at high temperatures.  Therefore in this paper we scale the calculated susceptibilities when $J_2,J_c = 0$ as
\be
\frac{\chi J_1}{N g^2 \mu_{\rm B}^2} \ \ \ {\rm versus}\ \ \ \frac{k_{\rm B}T}{J_1S(S+1)}.
\label{ChivsT}
\ee 
This is the same scaling of the temperature axis as for the magnetic heat capacity in Eq.~(\ref{Eq:CmagScaling}).

In the $J_1$-$J_2$-$J_c$ model, according to Fig.~\ref{Fig:Magnetic_Structures} there are four in-plane next-nearest-neighbor interactions 
\be
J_2= \alpha J_1,
\label{DefineAlpha}
\ee 
two NN interactions along the $c$-axis
\be
J_c= \gamma J_1,
\label{DefineGamma}
\ee 
in addition to the $z_1 = 4\equiv z$ nearest-neighbor interactions $J_1$.  When these additional interactions are present, according to Eq.~(\ref{WT3}) the Weiss temperature becomes
\be
\theta = \frac{zJ_1(1+\alpha+\gamma/2)S(S+1)}{3k_{\rm B}}, 
\label{WT2}
\ee
and the form of the new Curie Weiss law corresponding to Eq.~(\ref{CWuniversal}) is
\be
\frac{\chi J_1(1 + \alpha+\gamma/2)}{Ng^2\mu_{\rm B}^2} = \frac{1}{\frac{3k_{\rm B}T}{J_1(1 + \alpha+\gamma/2)S(S+1)} + z}.
\label{CWuniversal2}
\ee
A more accurate high-temperature scaling is obtained in this case by replacing $J_1$ in Eq.~(\ref{ChivsT}) by $J_1+J_2 +J_c/2 = J_1(1+\alpha+\gamma/2)$ and scaling the data according to
\be
\frac{\chi J_1(1+\alpha + \gamma/2)}{N g^2 \mu_{\rm B}^2} \ \ \ {\rm versus}\ \ \ \frac{k_{\rm B}T}{J_1(1+\alpha + \gamma/2)S(S+1)}.
\label{ChivsT3}
\ee

The scalings in Eqs.~(\ref{ChivsT}) and~(\ref{ChivsT3}) are expected to be universal with respect to the spin and the  exchange constants only at ``high'' temperatures.  Appendix~\ref{HTSE} shows that the calculations begin to deviate from the Curie-Weiss behavior when $1/T^2$ and higher order terms in the two-spin correlation functions become significant compared to the $1/T$ terms with decreasing $T$.

\subsection{\label{SecCMC} Classical Monte Carlo Simulations}

The classical Monte Carlo (CMC) simulations were performed on periodic $L\times L$ clusters for $J_c=0$ and for $L\times L \times L_c$ clusters for $J_c \neq 0$ using a hybrid algorithm that combines Metropolis and over-relaxation sweeps.\cite{Creutz}  In order to obtain statistically reliable data we have generated $\sim 10^5$ configurations at each temperature and then averaged the results over 50 independent annealing runs.

The spin Hamiltonian for our classical Monte Carlo simulations is the classical analogue of the quantum spin Hamiltonian~(\ref{Eq:HamilJ1J2Jz}), given by
\bea
{\cal H}_{\rm classical} &=& J_1S^2 \sum_{\langle ij \rangle}\hat{{\bf S}}_i \cdot \hat{{\bf S}}_j + J_2S^2 \sum_{\langle ik \rangle}\hat{{\bf S}}_i \cdot \hat{{\bf S}}_k \nonumber\\*
&& + J_cS^2 \sum_{\langle il\rangle}\hat{{\bf S}}_i \cdot \hat{{\bf S}}_k +\ g\mu_{\rm B}H \sum_{i}S^z_i,\nonumber\\ 
\label{EqJ1J2}
\eea
where $S$ is the magnitude of the spin, $\hat{{\bf S}}$ is a classical spin unit vector, and $\hat{{\bf S}}_i \cdot \hat{{\bf S}}_k = \cos\theta_{ij}$. According to Eq.~(\ref{EqJ1J2}), the exchange parameters $J_\alpha$ are always combined with the classical spin magnitude $S$ in the combination  $J_\alpha S^2$.  

In the following, we first consider our simulations for $J_c = 0$ and then for $J_c \neq 0$.

\begin{figure}
\includegraphics [width=3.2in]{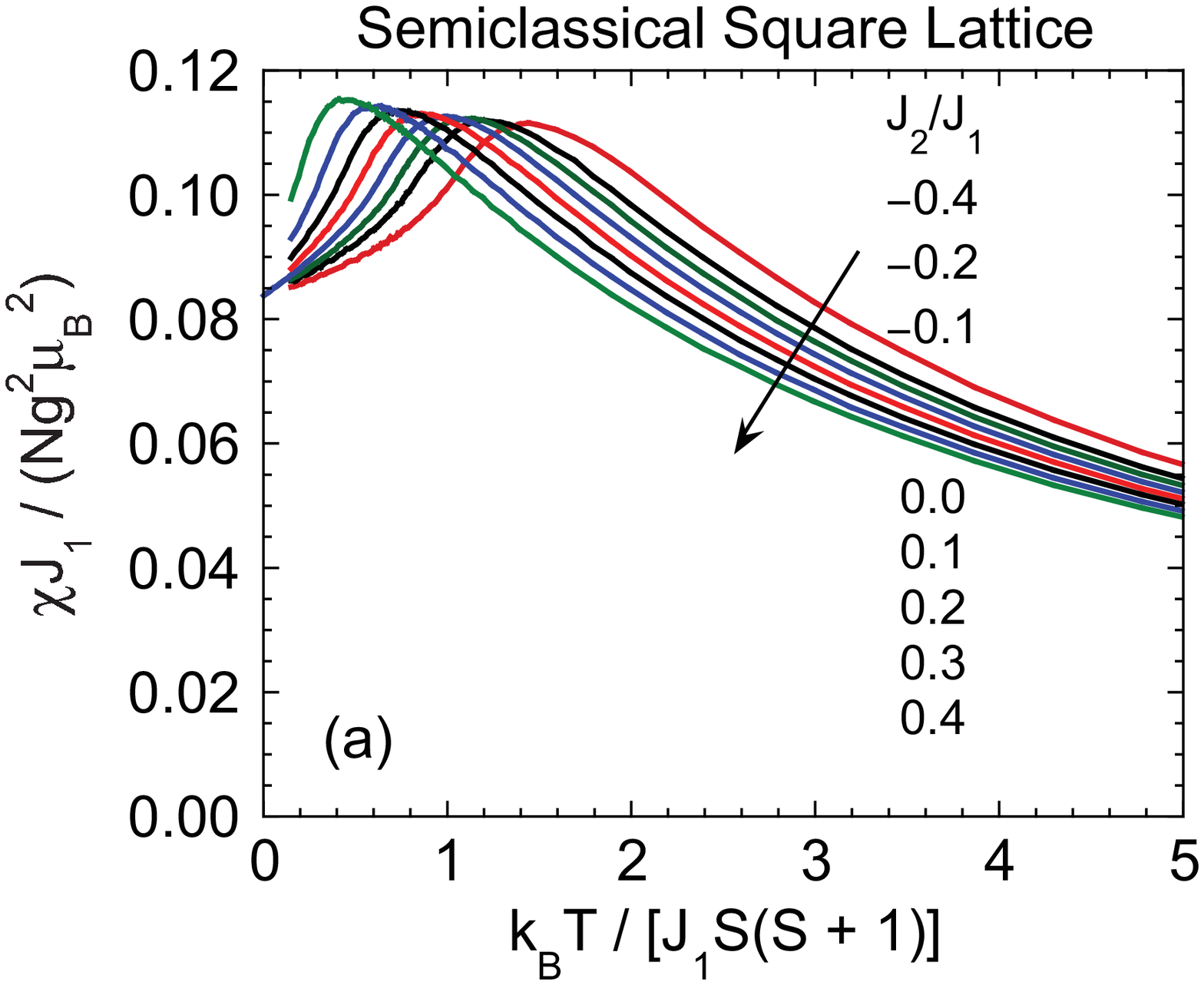}
\includegraphics [width=3.3in]{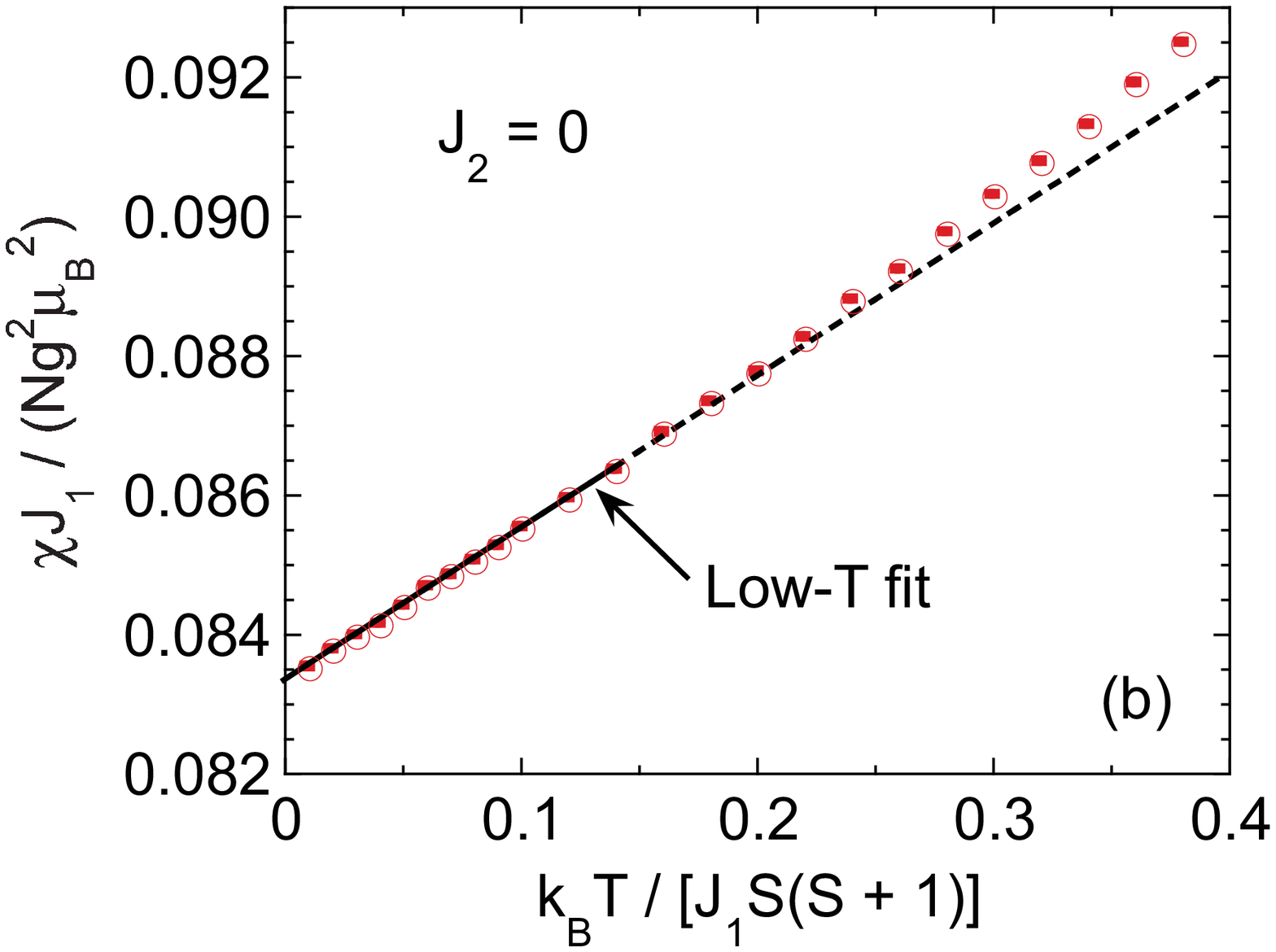}
\caption{(Color online) (a) Normalized magnetic spin susceptibility $\chi J_1/(N g^2 \mu_{\rm B}^2)$ versus normalized temperature $k_{\rm B}T/[J_1 S(S + 1)]$ for the classical spin $S$ Heisenberg square lattice with $J_c=0$ and $J_2/J_1 = -0.4$ to 0.4, with $S(S+1)$ replacing $S^2$.  The lattice size is $80 \times 80$ in each case.  (b) Expanded plot at low temperatures (open circles) of the data for $J_2 = 0$ in (a).  The error bars are shown and are inside the open circles.  A linear fit to the data up to a reduced temperature of 0.14 is shown as the solid straight line, and the dashed line is an extrapolation of the fit.  The coefficients of the fit are listed in Eq.~(\ref{J20Fitpars}).}
\label{Tsq.J2_all_L80}
\end{figure}

\subsubsection{$J_c=0$}

The semiclassical magnetic spin susceptibilities $\chi$ versus $T$ for the square lattice calculated using CMC simulations on $80\times80$ spin lattices are shown in Fig.~\ref{Tsq.J2_all_L80}(a) for $J_c=0$ and $J_2/J_1 = -0.4$ to~0.4.  Here, the term ``semiclassical'' means that $S^2$ in the final result of the classical simulations is replaced by the quantum mechanical expectation value $\langle S^2 \rangle = S(S+1)$.  This replacement allows the classical simulations to merge smoothly with the quantum Monte Carlo simulations (see Fig.~\ref{square_chi_J2} below).  We carried out simulations of various other $L\times L$ lattice sizes with $L = 10$--100 for $J_2/J_1 = 0$ and 0.2 and found that finite-size corrections to both the calculated magnetic susceptibility and magnetic heat capacity are negligible for $L\geq50$.

The $\chi(T)$ data in Fig.~\ref{Tsq.J2_all_L80}(a) show two interesting trends.  First, at high temperatures the Curie-Weiss law $C/(T + \theta)$ is obtained, in which the (positive) Weiss temperature $\theta$ is proportional to the sum of all interactions of a given spin with its neighbors according to Eq.~(\ref{WT3}).  Thus for a negative (ferromagnetic) $J_2$ that partially cancels the positive $J_1$, the susceptibility increases at a fixed $T$, and for a positive $J_2$ it decreases. Second, at low temperatures this trend is reversed.  A negative ferromagnetic $J_2$ is nonfrustrating with respect to $J_1$, and reinforces the short-range ordering that causes the peak in $\chi(T)$.  This moves the peak up in temperature and suppresses the susceptibility in the short-range ordered state at low temperatures below the peak temperature.  The opposite behavior is found for a positive AF $J_2$ which is frustrating with respect to $J_1$.  This $J_2$ suppresses the short-range AF ordering, which decreases the peak temperature and increases the susceptibility below the peak temperature compared to the case when $J_2 = 0$.  

\begin{figure}
\includegraphics [width=3.3in]{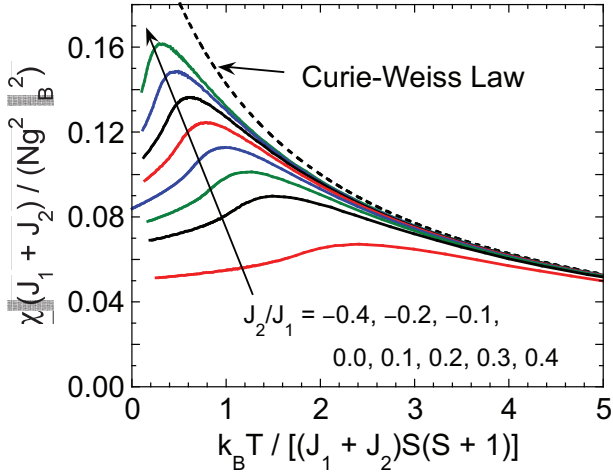}
\caption{(Color online) Normalized magnetic spin susceptibility $\chi (J_1+J_1)/(N g^2 \mu_{\rm B}^2)$ versus normalized temperature $k_{\rm B}T/[(J_1 + J_2) S(S + 1)]$ determined from classical Monte Carlo simulations for the spin $S$ Heisenberg square lattice with $J_c=0$ and $J_2/J_1 = -0.4$ to 0.4, as indicated, with $S(S+1)$ replacing the classical $S^2$.  The lattice size is $80 \times 80$ in each case.  The axis scaling is superior at high temperatures to that in Fig.~\ref{Tsq.J2_all_L80}(a).  A plot of the Curie-Weiss law in Eq.~(\ref{CWuniversal2}) is shown by the black dashed curve.}
\label{Tsq.J2_all_L80_scaled}
\end{figure}

These trends are illustrated in a different way if the best high-temperature scaling for these plots, given in Eq.~(\ref{ChivsT3}), is used, as shown in Fig.~\ref{Tsq.J2_all_L80_scaled}.  In addition, the Curie-Weiss law from Eq.~(\ref{CWuniversal2}) is plotted in Fig.~\ref{Tsq.J2_all_L80_scaled} as the blue dashed line.  From a comparison of the simulation data with the Curie-Weiss prediction, one sees that the two-spin correlations higher order than present in the Curie-Weiss regime ($\sim 1/T$) begin to become observable on the scale of the figure for $T \lesssim 5(J_1/k_{\rm B})S(S+1)$.   According to Eq.~(\ref{WT2}), this latter value is about four times the Weiss temperature $\theta$, which has the value 4/3 on the horizontal scale in Fig.~\ref{Tsq.J2_all_L80_scaled}.

The data in Fig.~\ref{Tsq.J2_all_L80}(a) for $J_2 = 0$ were obtained down to a reduced temperature of 0.01 as shown in the expanded plot in Fig.~\ref{Tsq.J2_all_L80}(b).  The lowest temperature data are linear in $T$.  A linear fit yielded
\be
\frac{\chi J_1}{Ng^2\mu_{\rm B}^2} = 0.08333(2) + 0.0218(3) \frac{k_{\rm B}T}{J_1S(S+1)},
\label{J20Fitpars}
\ee
as shown by the solid line in Fig.~\ref{Tsq.J2_all_L80}(b).  According to Takahashi's modified spin wave theory for the AF square lattice, the classical limit~(A9) in Ref.~\onlinecite{Takahashi1989} reads
\be
\frac{\chi J_1}{Ng^2\mu_{\rm B}^2} = \frac{1}{12} + \frac{1}{24\pi} \frac{k_{\rm B}T}{J_1S^2} + {\cal O}(T^3).
\label{Takahashi}
\ee
The zero-temperature reduced susceptibility $1/12 \approx 0.08333$ in Eq.~(\ref{Takahashi}) is the same as our value in Eq.~(\ref{J20Fitpars}) to within the errors of our Monte Carlo data, but the theoretical initial slope $1/(24\pi) \approx 0.01326$ is too small compared to our Monte Carlo value in Eq.~(\ref{J20Fitpars}).  On the other hand, in a $1/D$ expansion where $D$ is the dimensionality of the spins ($D = 3$ here), for the classical square spin lattice at low $T$ Hinzke et al.\cite{Hinzke2000} obtained 
\be
\frac{\chi J_1}{Ng^2\mu_{\rm B}^2} = \frac{1}{12} + \frac{1}{32} \frac{k_{\rm B}T}{J_1S^2}.
\label{Hinzke}
\ee
The zero temperature susceptibility is the same as our and Takahashi's value but Hinzke et al.'s initial slope is $1/32 = 0.03125$, which this time is larger than our Monte Carlo value in Eq.~(\ref{J20Fitpars}).  Thus our value of the initial slope is bracketed by the predictions of the modified spin wave theory and the $1/D$ expansion.

\begin{figure}
\includegraphics [width=3.3in]{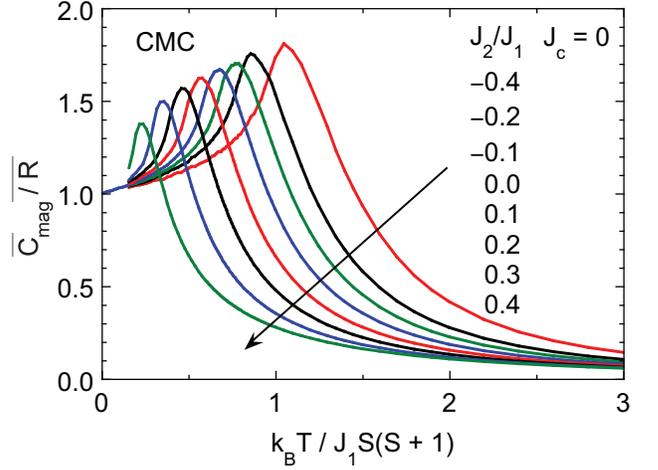}
\caption{(Color online) Classical Monte Carlo (CMC) simulations on $80\times80$ spin lattices of the magnetic heat capacity $C_{\rm mag}$ divided by the molar gas constant $R$ versus the scaled temperature $k_{\rm B}T/J_1S(S+1)$ for $J_c = 0$ and $J_2/J_1 = -0.4$ to~0.4 as shown.}
\label{Fig:C80x80CMC}
\end{figure}

\begin{table}
\caption{\label{CpCalcData} Parameters obtained from semiclassical (SC) Monte Carlo and quantum Monte Carlo simulations of the magnetic heat capacity $C_{\rm mag}(T)$ for the square lattice with no interlayer coupling ($J_c = 0$).  Here, $C_{\rm mag}^{\rm max}$ is the value of $C_{\rm mag}$ at a magnetic ordering peak, $T^{\rm max}$ is the temperature at which the maximum occurs, and SC means we have replaced $S^2$ by $S(S+1)$ in the temperature scaling of the classical Monte Carlo data. Note that the value of $T^{\rm max}$ is different than the temperature of the maximum in the magnetic susceptibility.}
\begin{ruledtabular}
\begin{tabular}{lcccc}
$S$ & lattice size & $J_2/J_1$  &  $\frac{C_{\rm mag}^{\rm max}}{R}$ & $\frac{k_{\rm B}T^{\rm max}}{J_1S(S+1)}$ \\
\hline
1/2 &  $32\times 32$  & 0  & 0.4606(7) & 0.801(2) \\
1 &  $32\times 32$  & 0 & 0.885(2) & 0.690(4)  \\
  &  $64\times 64$  & 0 & 0.879(2) & 0.700(3)   \\
3/2 &  $32\times 32$  & 0 & 1.159(2) & 0.674(1) \\
2 &  $32\times 32$  & 0 & 1.325(2) & 0.673(1)   \\
  &  $64\times 64$  & 0 & 1.295(2) & 0.684(2)  \\
5/2 &  $32\times 32$  & 0 & 1.428(2) & 0.673(2) \\
\hline
SC & $80\times 80$  & $-0.4$ & 1.801(5)  & 1.055(3) \\
SC & $80\times 80$  & $-0.2$ & 1.752(3)  & 0.861(3) \\
SC & $80\times 80$  & $-0.1$ & 1.699(1)  & 0.777(2) \\
SC & $80\times 80$  & 0 & 1.666(1)  & 0.678(1) \\
SC & $80\times 80$  & 0.1 & 1.621(2)  & 0.575(1) \\
SC & $80\times 80$  & 0.2 & 1.567(2)  & 0.471(2) \\
SC & $80\times 80$  & 0.3 & 1.498(3)  & 0.352(3) \\
SC & $80\times 80$  & 0.4 & 1.378(3)  & 0.232(2) \\
\end{tabular}
\end{ruledtabular}
\end{table}

The magnetic heat capacity is plotted in Fig.~\ref{Fig:C80x80CMC} according to Eq.~(\ref{Eq:CmagScaling}) versus the scaled temperature for exchange constant ratios $J_2/J_1 = -0.4$ to 0.4 on $80\times80$ spin lattices.  The broad peaks in the curves decrease in temperature with increasing $J_2$.  This is understandable in terms of the enhancement of short-range AF order for ferromagnetic (negative) $J_2$, which increases the temperature of the peak, and the frustration effect for antiferromagnetic (positive) $J_2$, which decreases the temperature of the peak.  The peak values and the temperatures at which they occur are listed in Table~\ref{CpCalcData}.  It is interesting that the variation in $C_{\rm mag}(T)$ with $J_2$ depends on the sign of $J_2$, in contrast with expectation from the first term in the HTSE in Eq.~(\ref{Eq:HTSECmag}) in which the uniform $J$ appears as the square and is hence independent of the sign.  Thus one cannot replace $zJ^2$ in Eq.~(\ref{Eq:HTSECmag}) by $\sum_{j}J_{ij}^2$.  This constraint is not present when calculating the Weiss temperature in the Curie-Weiss law from Eq.~(\ref{WT3}), in which one includes the interactions of a given spin with all of its neighbors algebraically and on the same footing.

\subsubsection{$J_c \neq 0$}

The classical Monte Carlo simulations do not produce the same results as the molecular field theory does because the interaction between a spin and its neighbors is not approximated by the interaction of the spin with the \emph{average} spin of its neighbors as in the molecular field theory.  In particular, according to the Mermin-Wagner theorem,\cite{Mermin1966} a Heisenberg spin system in one or two dimensions, as in the $J_1$-$J_2$ model with only intraplanar exchanges, should not show long-range magnetic ordering at finite temperature.  This theorem is respected in our classical simulations, but not in molecular field theory.  On the other hand, when the simulations are carried out with $J_c \neq 0$, we find that long-range AF ordering does occur, as expected.  Because a uniform magnetic field does not directly couple to the AF order parameter [the staggered moment, see Eq.~(\ref{Eq:muDagger})], these AF phase transitions have rather subtle effects on the calculated uniform susceptibility.  They are much more clearly manifested in the magnetic heat capacity which we will also present, and would also be clearly delineated in calculations of the staggered susceptibility in which the applied magnetic field has opposite directions for the two sublattices.

\begin{figure}
\includegraphics [width=3.3in]{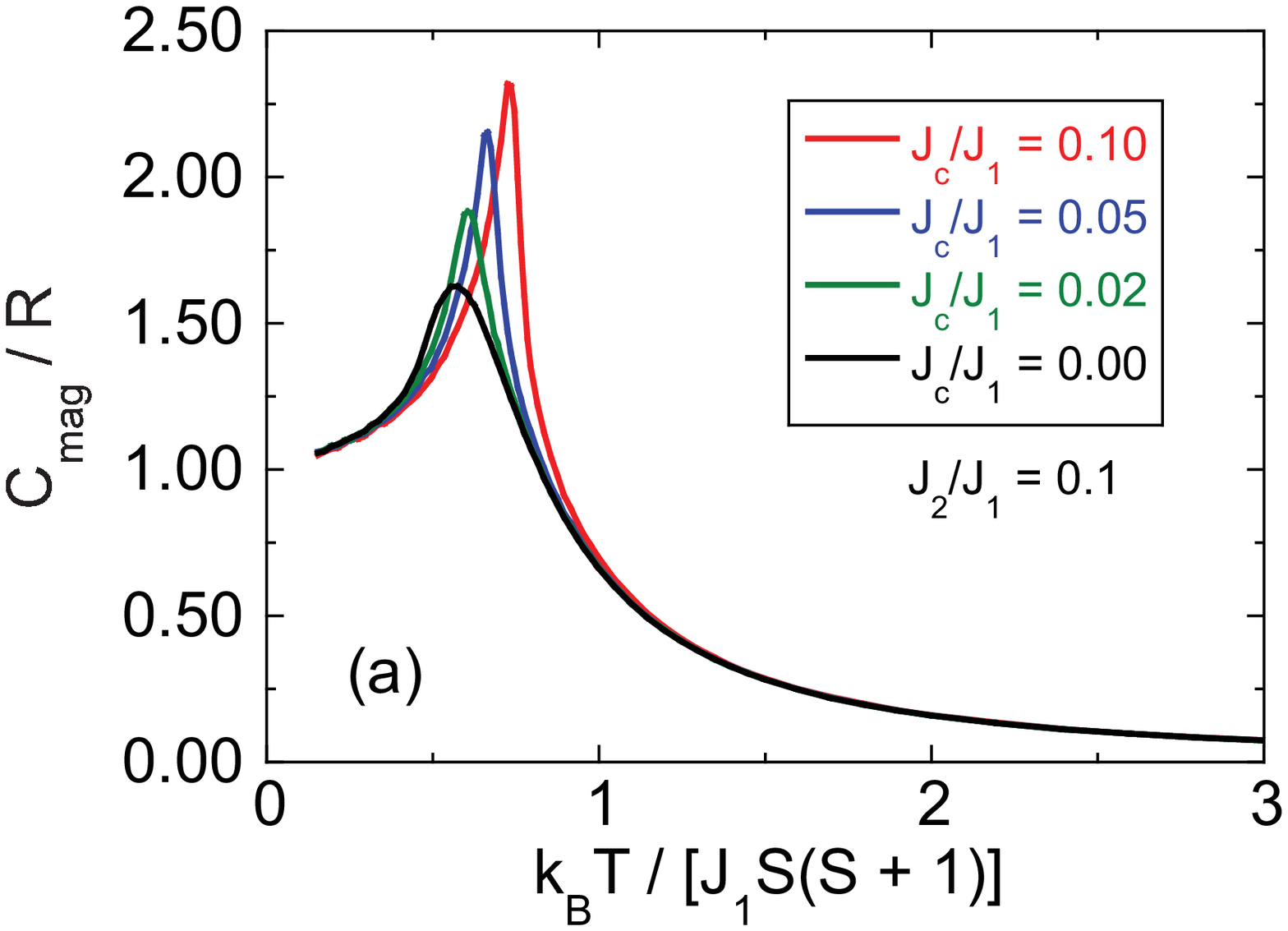}
\includegraphics [width=3.3in]{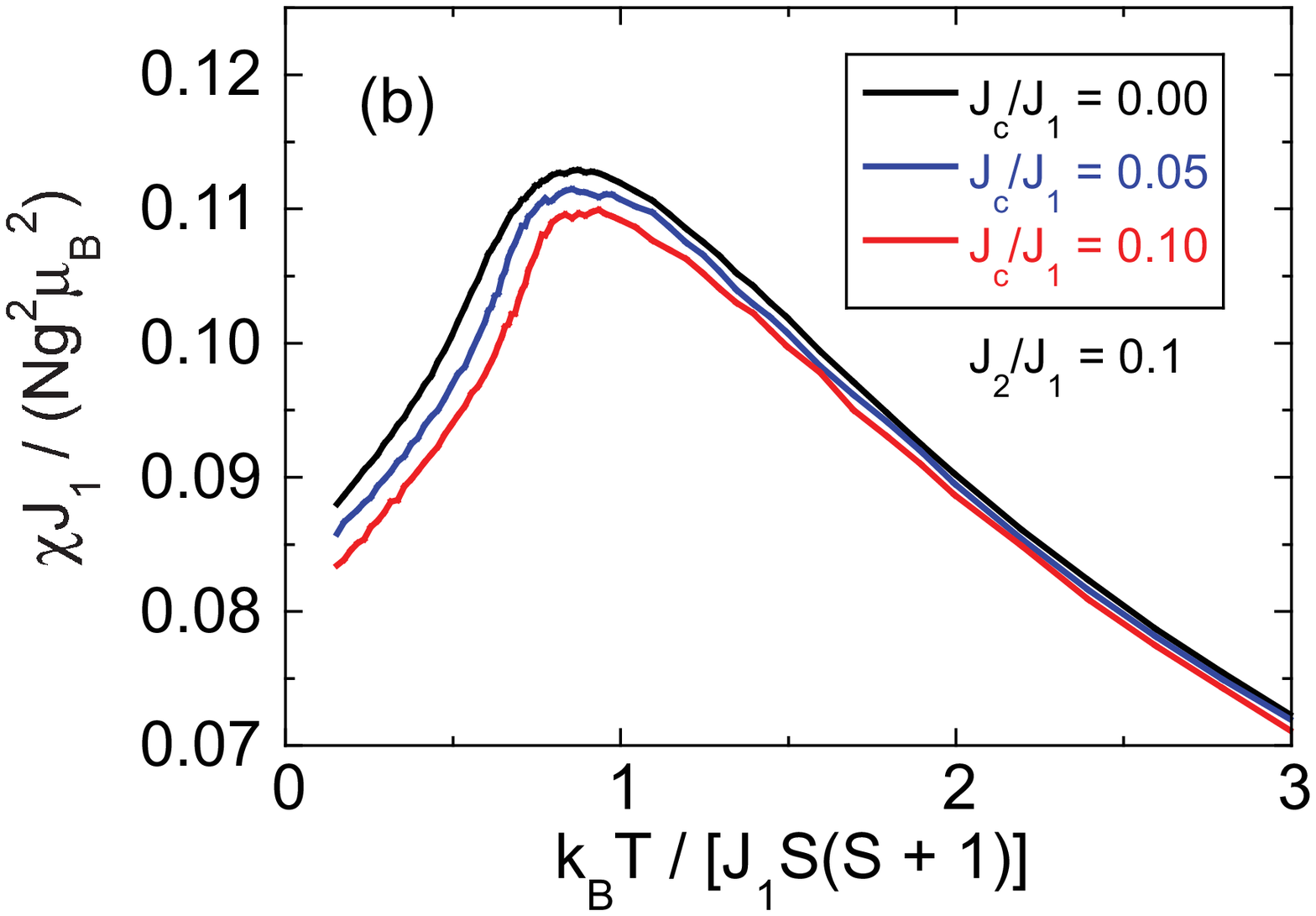}
\caption{(Color online) (a) Magnetic heat capacity and (b) magnetic susceptibility $\chi$ versus temperature $T$ from classical Monte Carlo simulations for the spin $S$ Heisenberg square lattice with $J_2/J_1 = 0.1$ and with $J_c/J_1 = 0$, 0.02, 0.05 and 0.1, as indicated.  The order of the curves from top to bottom is the same as in the respective figure legends.  The lattice size in each case is $20\times20\times10$.}
\label{Fig:CMCCmagJ200.1Jc00.1}
\end{figure}

\begin{figure}
\includegraphics [width=3.3in]{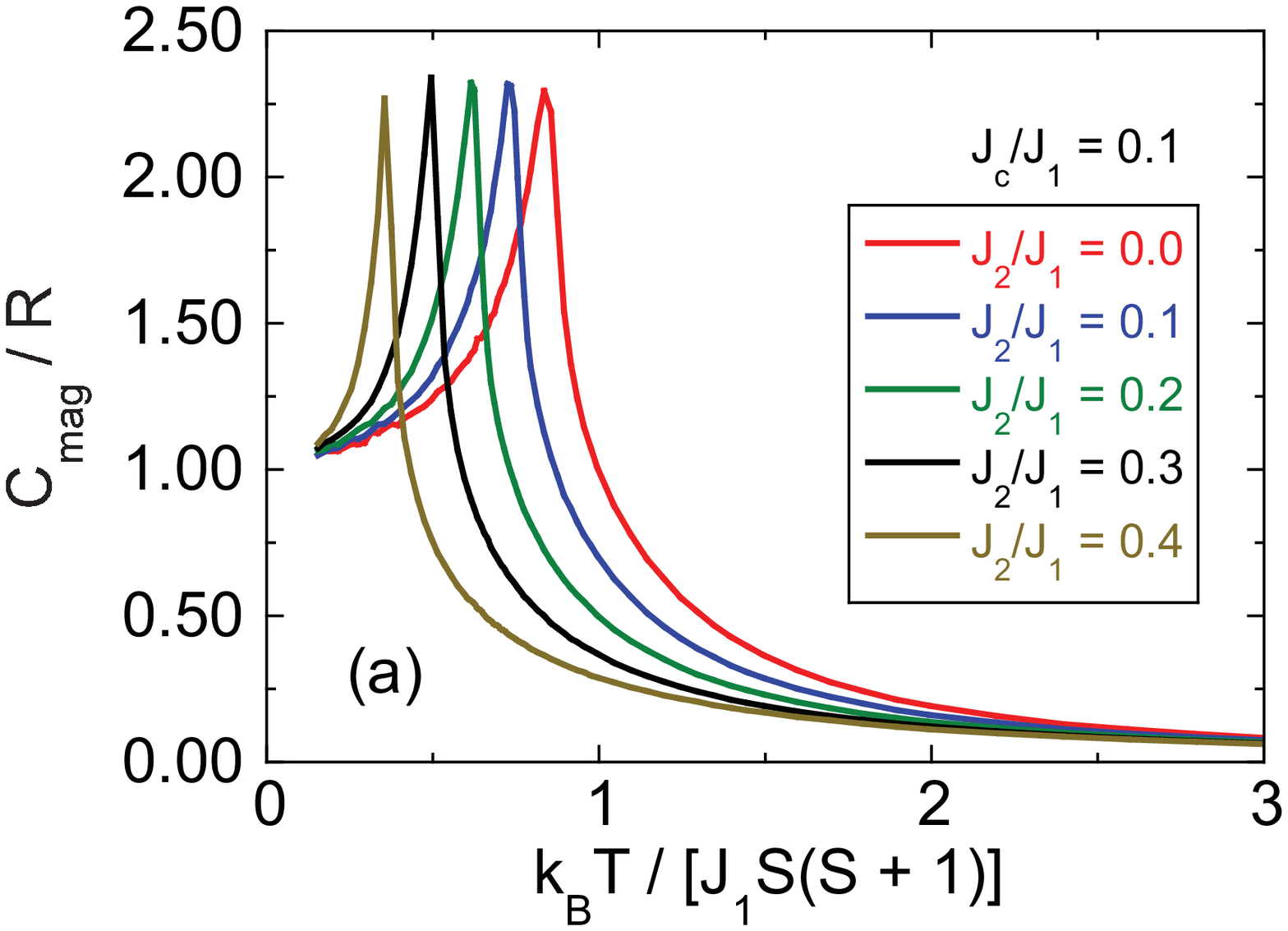}
\includegraphics [width=3.3in]{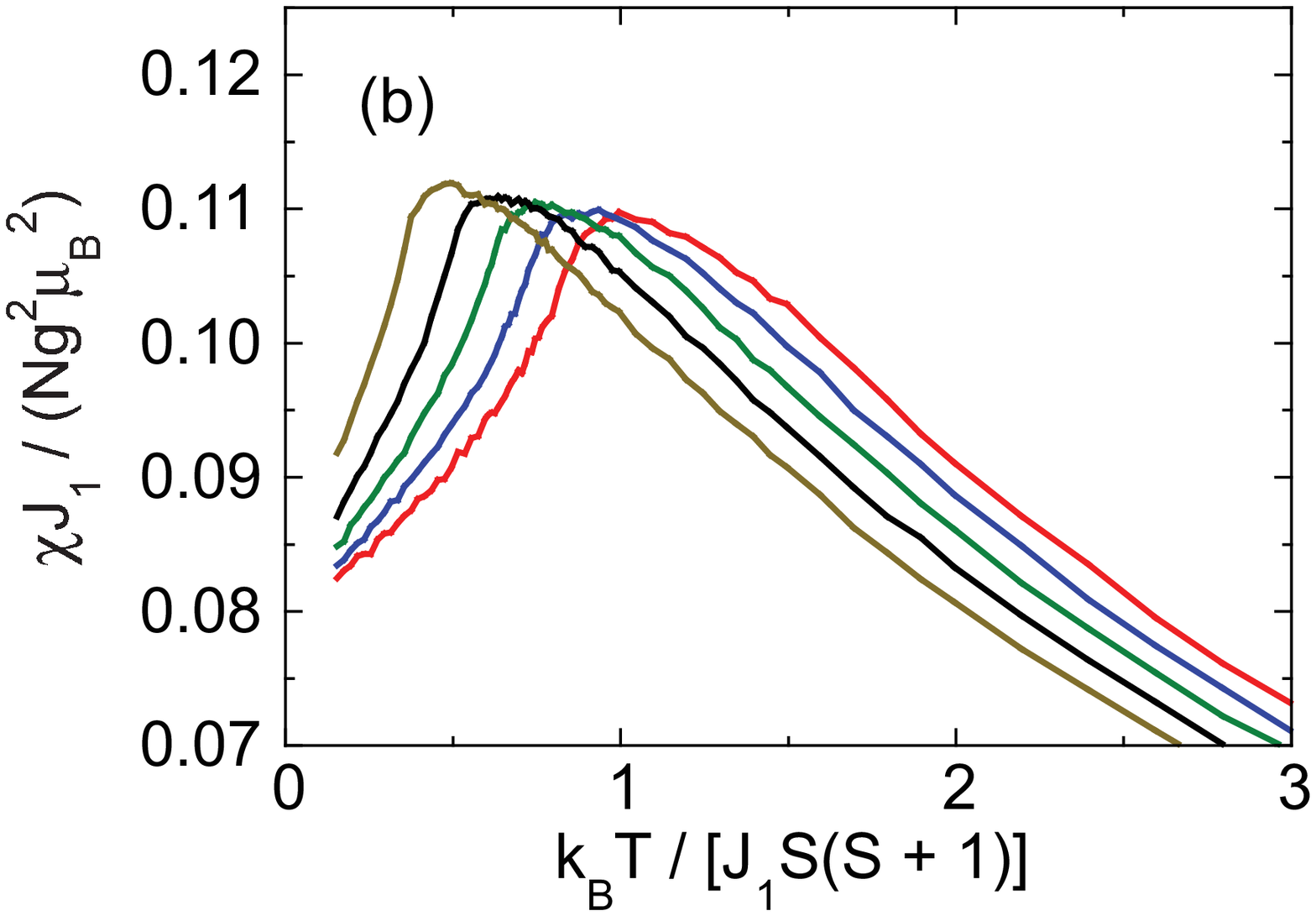}
\caption{(Color online) (a) Magnetic heat capacity $C_{\rm mag}$ and (b) spherically averaged magnetic susceptibility $\chi$ versus temperature $T$ from classical Monte Carlo simulations for the spin~$S$ Heisenberg square lattice with $J_2/J_1 = 0$ to~0.4, as indicated, and with fixed $J_c/J_1 = 0.1$.  The order of the curves from right to left is given in the figure~(a) legend.  The lattice size in each case is $20\times20\times10$.}
\label{Fig:CMCCmagJ204Jc0.1}
\end{figure}

\begin{table}
\caption{\label{Tab:CMCTN}  Temperatures $T_{\rm N}$ of the magnetic ordering peaks in the molar magnetic heat capacity $C_{\rm mag}$ and the values of $C_{\rm mag}(T_{\rm N})$ at the peak, versus the exchange constant ratios $J_2/J_1$ and $J_c/J_1$, obtained from classical Monte Carlo simulations.  The $T_{\rm N}$ and $C_{\rm mag}$ values may be affected by finite size effects.  Here $R$ is the molar gas constant.}
\begin{ruledtabular}
\begin{tabular}{cccl}
$J_2/J_1$ & $J_c/J_1$ & $k_{\rm B}T_{\rm N}/[J_1S(S+1)]$ & \hspace{-0.18in} $C_{\rm mag}(T_{\rm N})/R$ \\
\hline
0 & 0.02 & 0.717(2) & 1.890(3) \\
& 0.05 & 0.773(1) & 2.139(5) \\
& 0.10 & 0.842(1) & 2.29(2) \\
0.1 & 0.02 & 0.611(1) & 1.879(4) \\
& 0.05 & 0.667(1) & 2.151(6) \\
& 0.10 & 0.735(1) & 2.318(5) \\
0.12 & 0.06 & 0.661(1) & 2.190(6§) \\
0.2 & 0.02 & 0.502(1) & 1.880(8) \\
& 0.05 & 0.557(1) & 2.14(1) \\
& 0.10 & 0.619(1) & 2.32(1) \\
0.3 & 0.02 & 0.390(1) & 1.839(2) \\
& 0.05 & 0.439(1) & 2.125(9) \\
& 0.10 & 0.497(1) & 2.343(7) \\
0.4 & 0.02 & 0.263(1) & 1.784(8) \\
& 0.05 & 0.313(1) & 2.06(1) \\
& 0.10 & 0.360(1) & 2.26(1) \\
\end{tabular}
\end{ruledtabular}
\end{table}

Throughout this section, we replace the classical variable $S^2$ by its quantum-mechanical counterpart $S(S+1)$.  We show our results in two formats.  First, in Fig.~\ref{Fig:CMCCmagJ200.1Jc00.1} are shown the magnetic heat capacity $C_{\rm mag}$ and the spherically-averaged magnetic susceptibility $\chi$ versus temperature $T$ for fixed $J_2/J_1=0.1$ and variable $J_c=0$, 0.02, 0.05 and 0.1, where $J_1$, $J_2$ and $J_c$ are all antiferromagnetic.  From \ref{Fig:CMCCmagJ200.1Jc00.1}(a), one sees that $C_{\rm mag}(T)$ for $J_c=0$ just shows a broad peak characteristic of short-range AF order.  However, the $C_{\rm mag}$ quickly and clearly shows a cusp-like behavior with increasing $J_c$ at temperatures $T_{\rm N}$ corresponding to long-range AF order.  Second, in Fig.~\ref{Fig:CMCCmagJ204Jc0.1} are shown $C_{\rm mag}(T)$ and $\chi(T)$ for fixed $J_c/J_1=0.1$ and variable $J_2/J_1=0$--0.4, where $J_1$, $J_2$ and $J_c$ are again all antiferromagnetic. For all combinations of $J_2/J_1$ and $J_c/J_1$ we have studied, to within the errors the peak in $C_{\rm mag}$ at $T_{\rm N}$ coincides in temperature with the peak in $d(\chi T)/dT$ on the low-$T$ side of the broad peak in $\chi(T)$, in agreement with the Fisher relation.\cite{Fisher1962}  We note that the $T_{\rm N}$ and shape/magnitude of $C_{\rm mag}$ at $T_{\rm N}$ determined in our simulations may be affected by finite size effects.  

\begin{figure}
\includegraphics [width=3.3in]{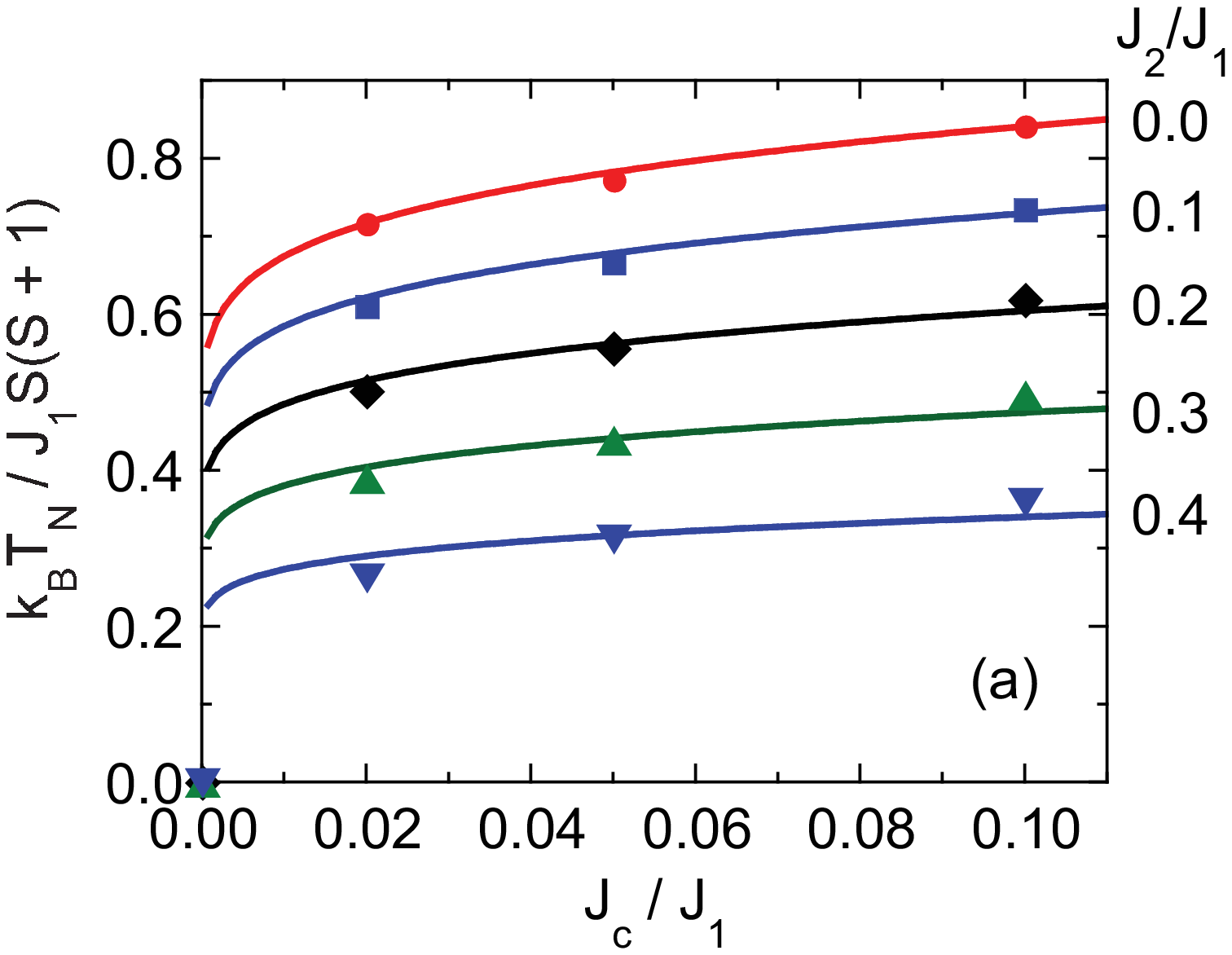}
\includegraphics [width=3.in]{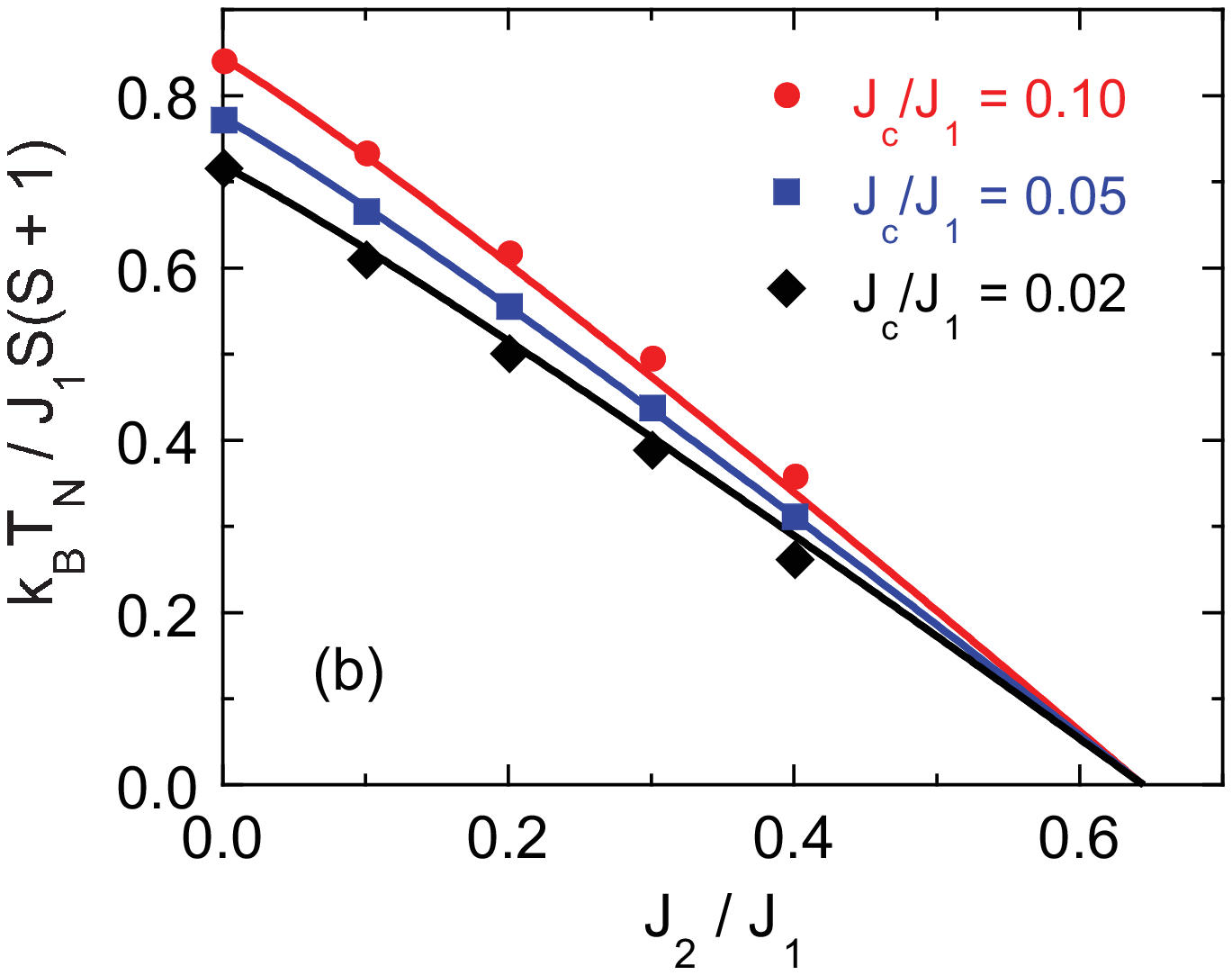}
\caption{(Color online) Antiferromagnetic ordering temperature $T_{\rm N}$ versus (a) the interlayer coupling $J_c$ for fixed values of $J_2/J_1$ and (b) $J_2/J_1$ at fixed $J_c/J_1$.  In (b), the order of the curves from top to bottom is the same as in the figure legend.  The solid curves are a global fit to all the data by Eq.~(\ref{Eq:FitTN(JcJ2)}), using the parameters in Eqs.~(\ref{Eq:ABvals}) and~(\ref{Eq:Rvalues}).}
\label{Fig:CMC_TN_vs_JcFit}
\end{figure}

The temperatures $T_{\rm N}$ of the peaks in $C_{\rm mag}(T)$ and the values of $C_{\rm mag}(T_{\rm N})$ at the peak, versus the exchange constant ratios $J_2/J_1$ and $J_c/J_1$, are listed in Table~\ref{Tab:CMCTN} and plotted in Figs.~\ref{Fig:CMC_TN_vs_JcFit}(a) and~(b).  For $J_2=0$, our $T_{\rm N}$ values are lower by $\lesssim 1$\% than the values obtained by Yasuda et al.\ for $J_c/J_1 = 0.02$, 0.05 and 0.1.\cite{Yasuda2005}  For $J_2 = 0$, a good fit to $T_{\rm N}$ versus $J_c$ for various spin values was obtained in Ref.~\onlinecite{Junger2009} using the expression\cite{Yasuda2005}
\be
\frac{k_{\rm B}T_{\rm N}}{J_1S(S+1)}= \frac{A}{B-\ln(J_c/J_1)},
\label{Eq:FitTN(Jc)}
\ee 
where different values of the constants $A$ and $B$ were required for different spin values.  We fitted the classical Monte Carlo data for $J_2=0$ in Fig.~\ref{Fig:CMC_TN_vs_JcFit}(a) using Eq.~(\ref{Eq:FitTN(Jc)}) and obtained a good fit with the values
\bea
A&=& 7.70\label{Eq:ABvals}\\*
B&=& 6.87,\nonumber
\eea
which are both about a factor of two larger than the respective values $A=3.96$ and $B=3.01$ obtained in Ref.~\onlinecite{Junger2009} for the classical limit $S=\infty$.  The fit is shown as the solid red curve for $J_2/J_1=0$ in Fig.~\ref{Fig:CMC_TN_vs_JcFit}(a).

From Fig.~\ref{Fig:CMC_TN_vs_JcFit}(a), a positive antiferromagnetic $J_2$ frustrates the G-type AF ordering and depresses $T_{\rm N}$ approximately linearly with $J_2$, and the $T_{\rm N}$ for each $J_c$ value extrapolates to zero at $J_2/J_1 \approx 0.6$, which is close to the value of 0.5 from Eq.~(\ref{Eq:JRestrictions}) at which one classically expects the G-type AF order to become unstable with respect to the stripe AF order.  Therefore, we fitted the dependence of $T_{\rm N}$ on $J_2/J_1$ of all the data for $J_c/J_1 = 0.02$, 0.05 and~0.1 together using the expression
\be
\frac{T_{\rm N}(J_c,\ J_2)}{T_{\rm N}(J_c,\ J_2=0)} = 1 - \left(\frac{J_2/J_1}{R_1}\right)^{R_2}
\label{Eq:FitTNvsJ2}
\ee
(not shown) and obtained the values 
\bea
R_1 &=& 0.644\label{Eq:Rvalues}\\*
R_2 &=& 1.082.\nonumber
\eea
The fits are shown as the solid curves in Fig.~\ref{Fig:CMC_TN_vs_JcFit}(b).  Thus the global function to fit all of our $T_{\rm N}(J_c/J_1,\ J_2/J_1)$ data is
\be
\frac{k_{\rm B}T_{\rm N}}{J_1S(S+1)}= \frac{A}{B-\ln(J_c/J_1)}\left[1 - \left(\frac{J_2/J_1}{R_1}\right)^{R_2}\right].
\label{Eq:FitTN(JcJ2)}
\ee 
The fits for $T_{\rm N}$ versus $J_c$ at fixed $J_2/J_1=0.1$--0.4 are shown as the solid curves in Fig.~\ref{Fig:CMC_TN_vs_JcFit}(a).  We see that Eq.~(\ref{Eq:FitTN(JcJ2)}), together with the four parameters in Eqs.~(\ref{Eq:ABvals}) and~(\ref{Eq:Rvalues}), provide a good global fit to all fifteen $T_{\rm N}(J_c/J_1,\ J_2/J_1)$ data points in Fig.~\ref{Fig:CMC_TN_vs_JcFit}(a) from our CMC simulations.

\subsection{\label{SecQMC} Quantum Monte Carlo Simulations}

\subsubsection{Magnetic Susceptibility}

\begin{figure}
\includegraphics [width=3.3in]{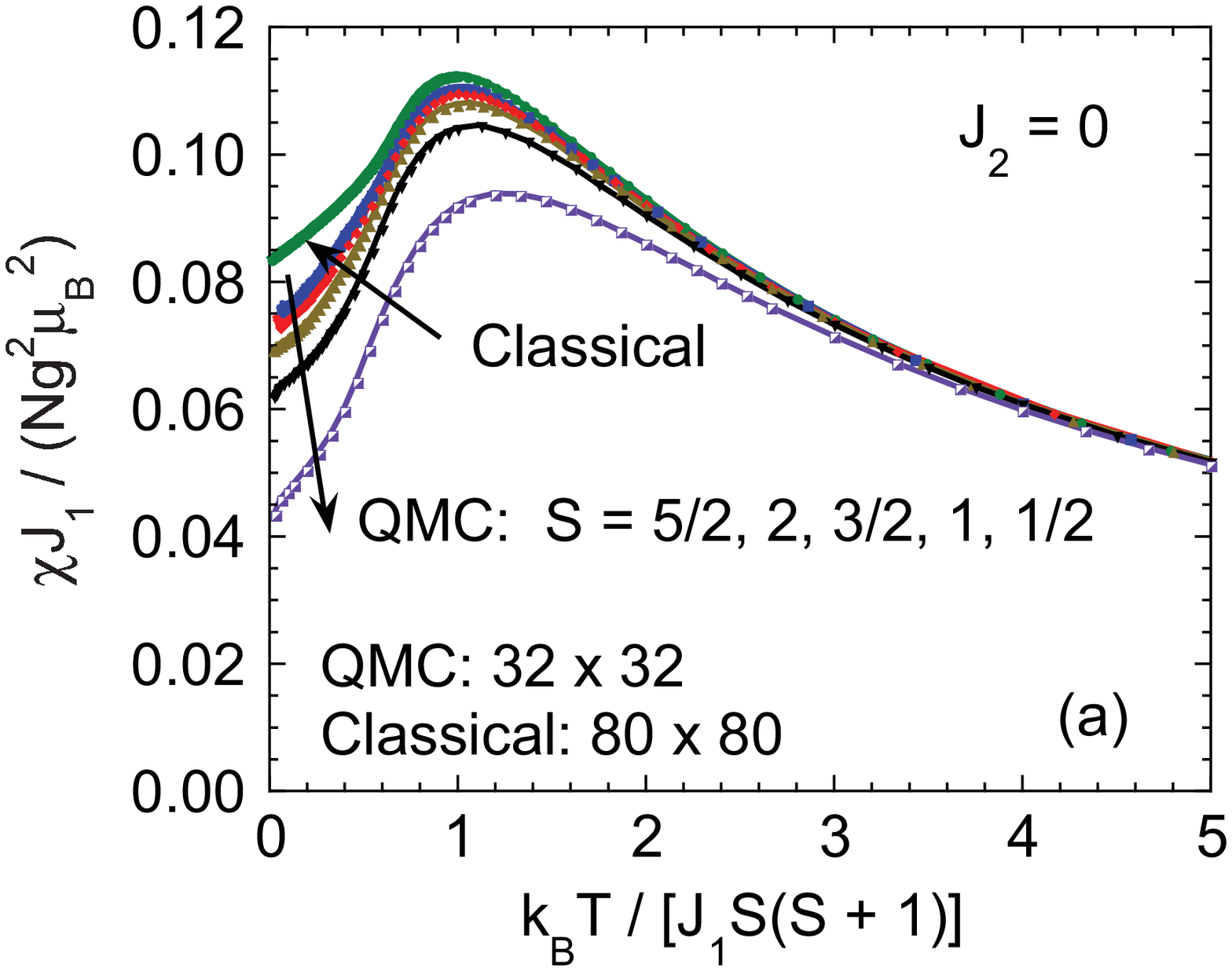}
\includegraphics [width=3.3in]{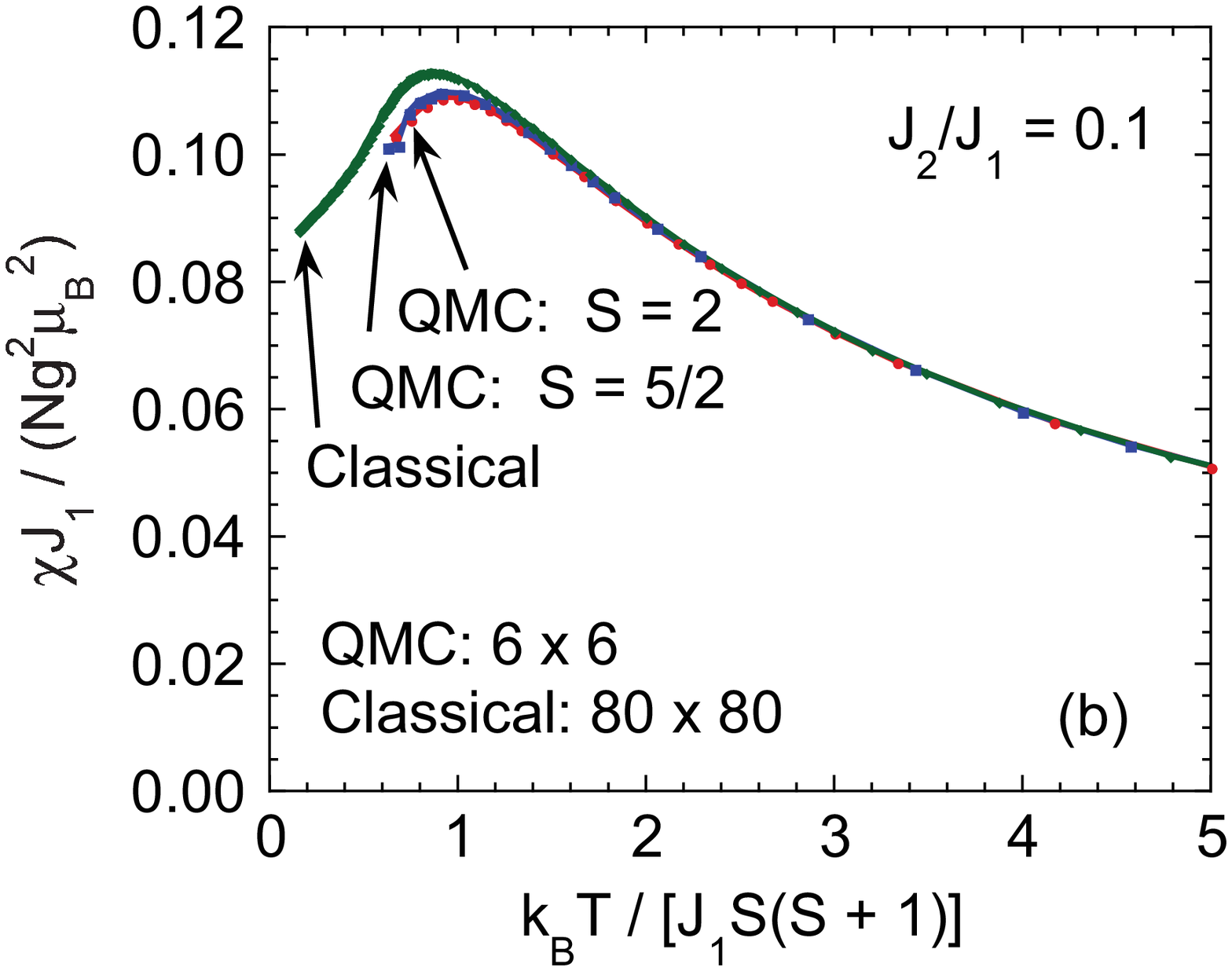}
\caption{(Color online) Normalized magnetic spin susceptibility $\chi J_1/(N g^2 \mu_{\rm B}^2)$ versus normalized temperature $k_{\rm B}T/[J_1 S(S + 1)]$ determined from quantum Monte Carlo (QMC) simulations for quantum Heisenberg square lattices with (a) $J_2/J_1 = 0$ and spins $S = 5/2,$ 2, 3/2, 1, and 1/2, and (b) $J_2/J_1 = 0.1$ and spins $S = 5/2$ and~2.  The data for the semiclassical model (green) with $J_2/J_1 = 0$ and 0.1 are included in the two panels, respectively.    The lattice size for each simulation is indicated in the respective figure.}
\label{square_chi_J2}
\end{figure}

\begin{figure}
\includegraphics [width=3.4in]{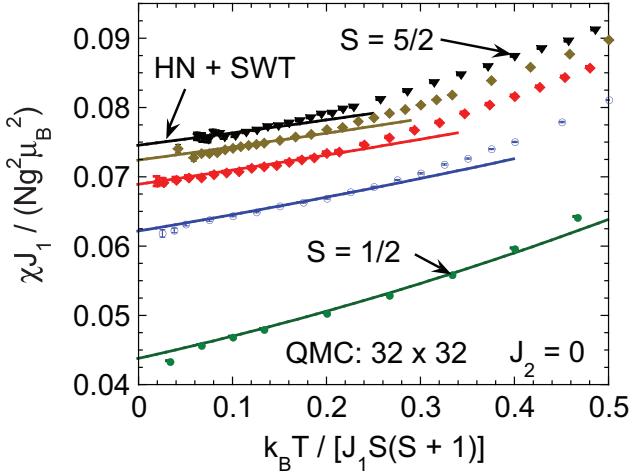}
\caption{(Color online) Normalized QMC magnetic spin susceptibility $\chi J_1/(N g^2 \mu_{\rm B}^2)$ versus normalized temperature $k_{\rm B}T/[J_1 S(S + 1)]$ data (symbols) at low temperatures from Fig.~\ref{square_chi_J2}(a) for spins $S = 1/2$ (bottom), 1, 3/2, 2, and 5/2~(top).  The error bars for the QMC data are also plotted.  The corresponding Hasenfratz-Niedermayer+spin wave theory predictions (HN + SWT) for the low-temperature behaviors\cite{Hasenfratz1993, Hamer1994} are also shown for these spin values as solid curves.}
\label{FitQMC_HF+SWT}
\end{figure}

Our quantum Monte Carlo (QMC) simulations were carried out with the ALPS\cite{alps1,alps2} directed loop application\cite{alps-sse} in the stochastic series expansion framework\cite{Sandvik} using version ALPS~1.3. Up to about $1\times 10^9$ sweeps were carried out for the $32\times 32$ lattice and the sign-free situation $J_2=0$. In order to compensate for the sign problem introduced by $J_2>0$ this was increased to about $2\times 10^{11}$ sweeps on the $6\times6$ lattice for $J_2=0.1$.  QMC simulations have been previously reported for $S=1$ over the temperature range $0.5 \leq k_{\rm B}T/J_1S(S+1)\leq 5$ by Harada et al.\cite{Harada1998}  We have extended these simulations to much lower temperatures (see Fig.~\ref{FitQMC_HF+SWT}).

Our QMC simulations of the magnetic spin susceptibilities versus temperature for the square lattice with quantum spins~1/2 to~5/2 with $J_2 = 0$ are shown in Fig.~\ref{square_chi_J2}(a), and for $J_2/J_1 = 0.1$ and $S = 2$ and~5/2 in Fig.~\ref{square_chi_J2}(b).  Various parameters obtained from these and the above semiclassical data are listed in Table~\ref{ChiCalcData} as described in the caption.  We checked by comparison of the data for the $32\times32$ lattice with $64\times64$ lattice data for $S = 1$ and $S = 2$ (see Table~\ref{ChiCalcData}) that the $32\times32$ lattice data in Figs.~\ref{square_chi_J2} and~\ref{FitQMC_HF+SWT} (below) are close to the thermodynamic limit; i.e., finite size effects are smaller than the size of the symbols (except probably the lowest temperature datum for $S = 1/2$ in Fig.~\ref{FitQMC_HF+SWT}). 

Hasenfratz and Niedermayer obtained the low temperature limit of the spin susceptibility of the Heisenberg antiferromagnet on a square lattice from chiral perturbation theory, given by\cite{Hasenfratz1993}
\be
\chi(T) = \frac{2\chi_\perp(0)}{3}\left[1 +\left(\frac{k_{\rm B}T}{2\pi\rho_S}\right) +\left(\frac{k_{\rm B}T}{2\pi\rho_S}\right)^2 + \cdots\right],
\label{HN}
\ee
where $\rho_S$ is the spin wave stiffness, $\chi_\perp(0)$ is the zero-temperature perpendicular susceptibility given by
\be
\frac{\chi_\perp(0) J_1}{Ng^2\mu_{\rm B}^2} = \frac{\rho_SJ_1a^2}{(\hbar c)^2}, 
\ee
$c$ is the spin wave velocity, and $a$ is the square lattice parameter.  The $\chi_\perp(0)$ and $\rho_S$ depend on the spin $S$ and were calculated using spin wave theory (SWT) by Hamer et al.\ as\cite{Hamer1994}
\bea
\frac{\chi_\perp(0) J_1}{Ng^2\mu_{\rm B}^2} &=& \frac{1}{8} - \frac{0.034\,446\,959\,42}{S} \nonumber\\*
&& +\ \frac{0.002\,040\,06(7)}{S^2} + {\cal O}(S^{-3})\\*
\nonumber \\*
{\rm and}\ \ \ \ \ \frac{\rho_S}{J_1} &=& S^2 - 0.117\,628\,254\,4\, S - 0.010\,207\,987\,3 \nonumber\\*
&& -\ \frac{0.003\,16(2)}{S} + {\cal O}(S^{-2}).
\eea

The low-temperature QMC data from Fig.~\ref{square_chi_J2}(a) are shown in Fig.~\ref{FitQMC_HF+SWT} together with the above predictions of Hasenfratz and Niedermayer (HN) combined with the SWT results of Hamer et al.  The lowest-temperature QMC data for spins~1/2 to~5/2 in Fig.~\ref{FitQMC_HF+SWT} are all seen to be in good agreement with the HN + SWT predictions.  High resolution calculations of $\chi(T)$ from the literature for the $S = 1/2$ square lattice Heisenberg antiferromagnet also confirm the form of Eq.~(\ref{HN}).\cite{Kim1998, Sandvik2010}  For $S=1$ our value of $\chi(0)$ from Fig.~\ref{FitQMC_HF+SWT} disagrees with the value 0.07197 given in Ref.~\onlinecite{Junger2009}.  It was claimed in Ref.~\onlinecite{Takahashi1989} that on the basis of spin wave theory, the next-order term above the $T^1$ term in Eq.~(\ref{HN}) is ${\cal O}(T^3)$, as in Eq.~(\ref{Takahashi}) above, in disagreement with Eq.~(\ref{HN}).  However, the next higher order term is indeed the $T^2$ term.\cite{Hasenfratz2011}

\begin{table*}
\caption{\label{ChiCalcData} Calculated values of (i) the maximum spin susceptibilities $\chi^{\rm max}$ and (ii) the temperatures $T^{\rm max}$ at which they occur for different lattice sizes, $J_2/J_1$  values and quantum and semiclassical $S$ values, (iii) the product $\chi^{\rm max}T^{\rm max}/3C$ where $C$ is the  Curie constant per mole of spins in Eq.~(\ref{CC}),  (iv) the value of the product $\chi^{\rm max}T^{\rm max}$ for the listed $S$ assuming $g=2$, and (v) the value of the product for an alternate semiclassical value of $S=5/2$ assuming $g=2$.  By ``semiclassical'' (SC) is meant that $S^2$ in the final result of the classical calculation is replaced by the quantum mechanical expectation value $\langle S^2 \rangle = S(S+1)$.  In the last column is listed the molecular field N\'eel temperature $T_{\rm N}$, normalized by $J_1S(S+1)/k_{\rm B}$, according to Eq.~(\ref{Eq:TNfromJ1J2Jc}).}
\begin{ruledtabular}
\begin{tabular}{lccccccccc}
$S$ & lattice size & $J_2/J_1$  & $J_c/J_1$  &  $\frac{\chi^{\rm max}J_1}{Ng^2\mu_{\rm B}^2}$& $\frac{k_{\rm B}T^{\rm max}}{J_1S(S+1)}$ &$\frac{\chi^{\rm max}T^{\rm max}}{3C}$ & $\chi^{\rm max}T^{\rm max}$ &   $\chi^{\rm max}T^{\rm max}$ & $\frac{k_{\rm B}T_{\rm N}}{J_1S(S+1)}$ \\
 & && &&&  &${\rm (cm^3\,K/mol)}$ &${\rm (cm^3\,K/mol)}$ \\ 
\hline
1/2 &  $32\times 32$  & 0 & 0 & 0.09370(3) & 1.248(3) & 0.1169(3) & 0.1316(4) & & 1.33 \\
1 &  $32\times 32$  & 0 & 0 & 0.10438(5) & 1.090(3)  & 0.1138(3) &  0.3415(11) & & 1.33\\
  &  $64\times 64$  & 0 & 0 & 0.10424(3) & 1.085(3)  & 0.1131(4) & 0.3394(11) \\
3/2 &  $32\times 32$  & 0 & 0 & 0.10790(3) & 1.050(4) & 0.1133(5) & 0.638(3)  && 1.33\\
2 &  $32\times 32$  & 0 & 0 & 0.10952(4) & 1.030(4)  & 0.1128(5) & 1.016(4) && 1.33 \\
  &  $64\times 64$  & 0 & 0 & 0.10957(6) & 1.038(5)  & 0.1137(6) & 1.024(6)\\
5/2 &  $32\times 32$  & 0 & 0 & 0.11040(3) & 1.018(3) & 0.1144(8) & 1.50(1)  && 1.33\\
\hline
2 & $6\times 6$ & 0.1 & 0 & 0.10882(6) & 0.966(5) & 0.1051(6) & 0.946(6)  && 1.20\\
5/2 & $6\times 6$ & 0.1 & 0 & 0.10966(6) & 0.959(4) & 0.1052(5) & 1.381(6) && 1.20 \\
\hline
&&&&&&& \underline{$S=2$} & \underline{$S= 5/2$}\\
SC & $80\times 80$  &$-0.4$ & 0 & 0.11115(5)  & 1.453(4)  & 0.1615(5) &  1.454(5) & 2.121(8) & 1.87  \\
SC & $80\times 80$  &$-0.2$ & 0 & 0.11173(5)  & 1.225(4)  & 0.1369(5) &  1.232(5) & 1.797(8) & 1.60  \\
SC & $80\times 80$  &$-0.1$ & 0 & 0.11210(4)  & 1.120(3)  & 0.1256(4) &  1.130(4) & 1.649(5) & 1.47  \\
SC & $80\times 80$  &$0$ & 0 & 0.11235(3)  & 0.999(2)  & 0.1122(3) & 1.011(3) & 1.470(4) & 1.33 \\
SC & $80\times 80$  &$0.1$ & 0 & 0.11274(3)  & 0.876(3)  & 0.0988(4) & 0.889(3) & 1.297(5) & 1.20 \\
SC & $20\times 20\times 10$  & $0.1$ & 0.02 & 0.11216(6)  & 0.874(5)  & 0.0980(6) & 0.883(6) & 1.287(8) & 1.21 \\
SC & $20\times 20\times 10$   & $0.1$ & 0.05 & 0.1113(1)  & 0.878(7)  & 0.0977(9) & 0.880(8) & 1.28(1) & 1.23 \\
SC & $20\times 20\times 10$ & $0.1$ & 0.1 & 0.1098(1)  & 0.896(5)  & 0.0984(6) & 0.886(6) & 1.29(1) & 1.27 \\
SC & $20\times 20\times 10$ & $0.12$ & 0.06 & 0.11126(8)  & 0.853(7)  & 0.0949(9) & 0.854(8) & 1.25(1) & 1.21 \\
SC & $80\times 80$  & $0.2$ & 0 & 0.11325(2)  & 0.750(2)  & 0.0849(3) & 0.765(2) & 1.115(3) & 1.07 \\
SC & $20\times 20\times 10$ & $0.2$ & 0.02 & 0.11275(5)  & 0.749(10)  & 0.0844(12) & 0.76(1) & 1.11(1) & 1.08\\
SC & $20\times 20\times 10$ & $0.2$ & 0.05 & 0.11187(10)  & 0.755(10)  & 0.0845(12) & 0.76(1) & 1.11(1) & 1.10\\
SC & $20\times 20\times 10$ & $0.2$ & 0.1 & 0.11029(5)  & 0.767(7)  & 0.0846(8) & 0.762(7) & 1.11(1) & 1.13 \\
SC & $80\times 80$  & $0.3$ & 0 & 0.11391(3)  & 0.616(2)  & 0.0702(3) & 0.632(3) & 0.921(3) & 0.93 \\
SC & $20\times 20\times 10$ & $0.3$ & 0.02 & 0.1134(10)  & 0.614(10)  & 0.070(1) & 0.63(2) & 0.91(3) & 0.95 \\
SC & $20\times 20\times 10$ & $0.3$ & 0.05 & 0.11257(7)  & 0.627(7)  & 0.0706(8) & 0.635(8) & 0.93(1) & 0.97\\
SC & $20\times 20\times 10$ & $0.3$ & 0.1 & 0.1108(10)  & 0.652(7)  & 0.072(2) & 0.65(1) & 0.95(2) & 1.00 \\
SC & $80\times 80$  & $0.4$ & 0 & 0.11509(4)  & 0.468(2)  & 0.0539(3) & 0.485(2) & 0.707(3) & 0.80 \\
SC & $20\times 20\times 10$ & $0.4$ & 0.02 & 0.1147(10)  & 0.468(7)  & 0.054(2) & 0.48(1) & 0.70(2) & 0.81 \\
SC & $20\times 20\times 10$ & $0.4$ & 0.05 & 0.1136(10)  & 0.467(10)  & 0.053(2) & 0.48(1) & 0.70(2) & 0.83\\
SC & $20\times 20\times 10$ & $0.4$ & 0.1 & 0.11175(5)  & 0.497(3)  & 0.0555(4) & 0.500(3) & 0.729(5) & 0.87\\
\end{tabular}
\end{ruledtabular}
\end{table*}

\subsubsection{Magnetic Heat Capacity}

\begin{figure}
\includegraphics [width=3.3in]{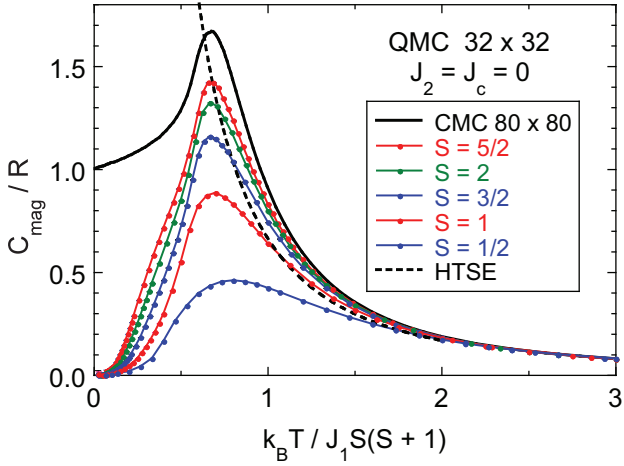}
\caption{(Color online) Magnetic heat capacity $C_{\rm mag}$ divided by the molar gas constant $R$ versus normalized temperature $k_{\rm B}T/[J_1 S(S + 1)]$ for the square lattice with only nearest-neighbor couplings $J_1$ (exchange constants $J_2 = J_c = 0$).  Quantum Monte Carlo (QMC) data (filled circles) for spins $S = 1/2$ (bottom), 1, 3/2, 2, and 5/2~(top) are shown, together with classical Monte Carlo (CMC) data from Fig.~\ref{Fig:CMCCmagJ200.1Jc00.1}(a)  (solid black curve at the top) and the first term in the high-temperature series expansion HTSE for the magnetic heat capacity from Eq.~(\ref{Eq:HTSECmag}) using $z=4$ (dashed black curve).  The error bars for the QMC data are also plotted, and the lines connecting the data points are guides to the eye.  The order of the Monte Carlo data from top to bottom is the same as in the figure legend.}
\label{Fig:C32x32QMC}
\end{figure}

\begin{figure}
\includegraphics [width=3.3in]{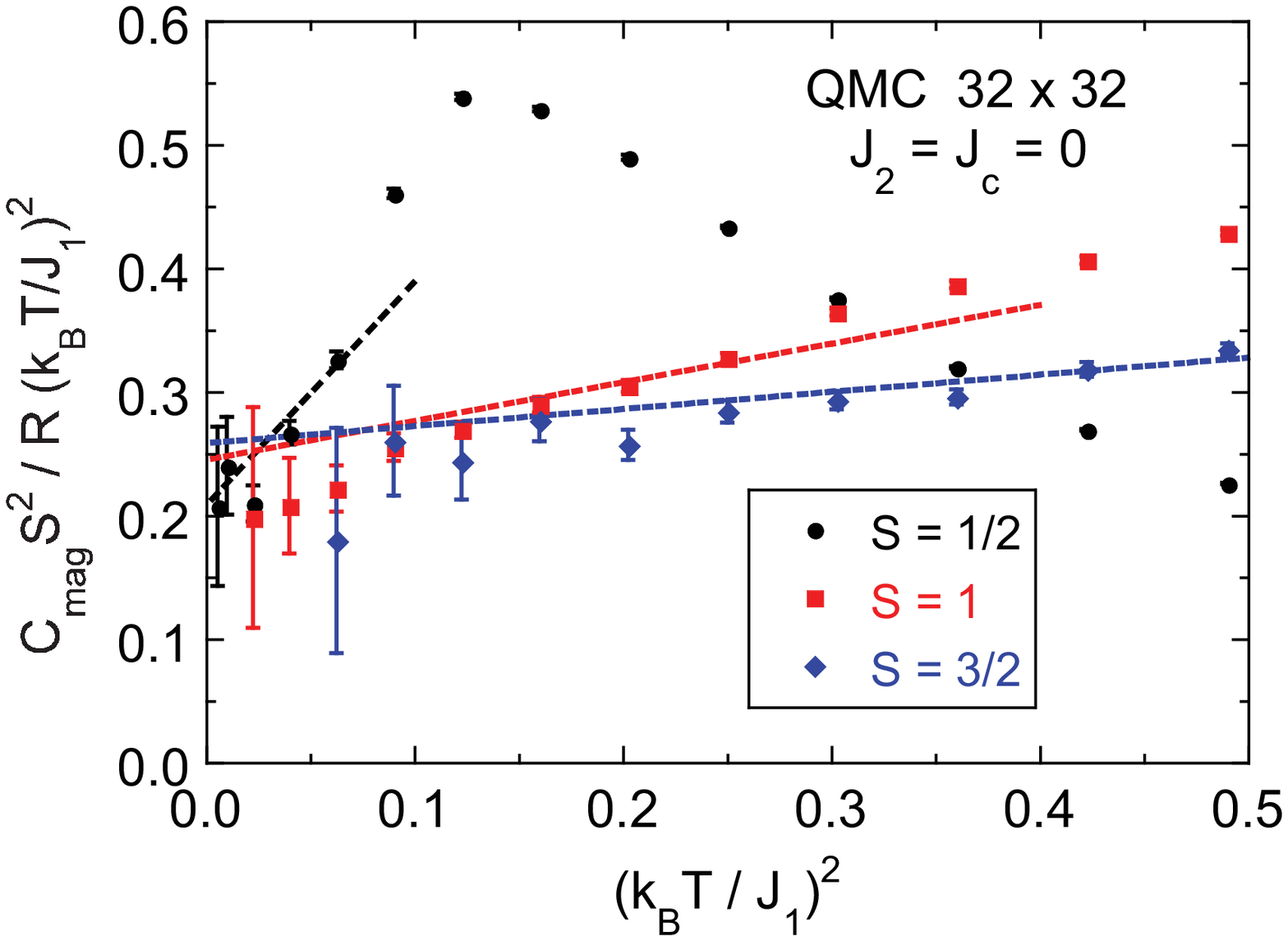}
\caption{(Color online) Low-temperature magnetic heat capacity $C_{\rm mag}$ times $S^2$ divided by the molar gas constant $R$ and $T^2$ versus normalized temperature squared, $(k_{\rm B}T/J_1)^2$, for the square lattice with only nearest-neighbor couplings $J_1$ (exchange constants $J_2 = J_c = 0$).  Quantum Monte Carlo (QMC) data (filled symbols) for spins $S = 1/2$, 1, and 3/2 are shown, together with error bars.  The dotted lines are fits of the data by Eq.~(\ref{Eq:CmagSeries}) assuming the specific $A(S)$ $y$-intercepts given in the text below Eq.~(\ref{Eq:CmagSeries}).  Only the slopes $D(S)$ were fitted.  The fitting ranges of $T^2$ were 0--0.06 for $S = 1/2$, 0--0.25 for $S = 1$ and 0--0.5 for $S = 3/2$.}
\label{Fig:CoT2T232x32S1o2B}
\end{figure}

The magnetic heat capacity $C_{\rm mag}$ versus temperature data from our QMC simulations for the square lattice with only NN interactions ($J_2=J_c=0$) are shown in Fig.~\ref{Fig:C32x32QMC} for spins $S = 1/2$, 1, 3/2, 2 and~5/2.  Also shown for comparison are our CMC heat capacity data for $J_2=J_c=0$ from Fig.~\ref{Fig:CMCCmagJ200.1Jc00.1}(a) and the first term in the HTSE $C_{\rm mag}\propto 1/T^2$ for the magnetic heat capacity from Eq.~(\ref{Eq:HTSECmag}) using the nearest-neighbor coordination number $z=4$.  The CMC and QMC data exhibit this HTSE behavior for temperatures $T \gtrsim 2J_1S(S+1)/k_{\rm B}$.  The values of the heat capacities of the peaks in the simulation data for the spin values $S = 1$ to $S = 5/2$ and the temperatures at which they occur are listed in Table~\ref{CpCalcData} above.

We checked finite-size effects associated with the QMC data by simulating $C_{\rm mag}$ for $64\times64$ $S = 1$ and $S = 2$ lattices for comparison with the $32\times32$ lattices in Fig.~\ref{Fig:C32x32QMC}.  On the scale of the figure, the $64\times64$ data (not shown) were close to the $32\times32$ lattice size data.  For example, the peak heights differ by less than 2 percent between the simulations for the different size lattices (see Table~\ref{CpCalcData}).

According to Eq.~(\ref{Eq:TNfromJ1J2Jc}), the N\'eel temperature in MFT occurs in Fig.~\ref{Fig:C32x32QMC} at a value of 4/3 on the horizontal scale and with a heat capacity jump on cooling below $T_{\rm N}$ given by Eq.~(\ref{Eq:DeltaC}) as $\Delta C_{\rm mag}/R = 3/2$ for $S = 1/2$ and = 5/2 for $S = \infty$.  The data in Fig.~\ref{Fig:C32x32QMC} are very different from these MFT predictions due to the presence of short-range magnetic ordering and the lack of long-range magnetic ordering\cite{Mermin1966} in these two-dimensional spin lattices at finite temperatures.

The expression of Hasenfratz and Niedermayer for the low temperature magnetic heat capacity of the Heisenberg antiferromagnet on a square lattice from chiral perturbation theory, per mole of spins, is\cite{Hasenfratz1993}
\be
C_{\rm mag} = \frac{6\,\zeta(3)R}{\pi (\hbar v/a)^2}(k_{\rm B}T)^2 + O(T^4),
\label{Eq:CmagHN}
\ee
where $\zeta(x)$ is the Riemann zeta function with $\zeta(3)\approx 1.20206$ and $a$ is the length of an edge of the square lattice unit cell.  The spin wave velocity $v$ for the AF Heisenberg square lattice is\cite{Hamer1994, Weihong1993}
\be
\frac{\hbar v}{a} = 2\sqrt{2}SJ_1\left[ 1 + \frac{0.157\,947\,421}{2S}+\frac{0.021\,52}{(2S)^2}+O(S^{-3})\right].
\label{Eq:hbarvHamer}
\ee
From Eqs.~(\ref{Eq:CmagHN}) and~(\ref{Eq:hbarvHamer}) one obtains
\bea
\frac{C_{\rm mag}S^2}{R(k_{\rm B}T/J_1)^2} &=& \frac{3\,\zeta(3)}{4\pi}\left[ 1 + \frac{0.157\,947\,421}{2S}+\frac{0.021\,52}{(2S)^2}\right]^{-2}\nonumber\\*
&& +\ D(S)\left(\frac{k_{\rm B}T}{J_1}\right)^2\nonumber\\*
&\equiv& A(S) + D(S)\left(\frac{k_{\rm B}T}{J_1}\right)^2.
\label{Eq:CmagSeries}
\eea
The factor $A(S)$ is 0.2063, 0.2441, 0.2578, 0.2649 and 0.2692 for $S = 1/2,\ 1, \ldots,\ 5/2$, respectively. Figure~\ref{Fig:CoT2T232x32S1o2B} shows our low-temperature $C_{\rm mag}(T)$ data for $S = 1/2$, 1 and~3/2 plotted according to Eq.~(\ref{Eq:CmagSeries}).  The approximate extrapolated zero-temperature values are in accord with the above $A(S)$ values to within the data error bars.  After setting $A(S)$ to the above respective fixed values, the initial slopes were estimated by fitting the data by Eq.~(\ref{Eq:CmagSeries}), yielding $D(S) = 1.8$, 0.31 and 0.14 for $S=1/2$, 1 and~3/2, respectively, as shown by the respective dotted lines in Fig.~\ref{Fig:CoT2T232x32S1o2B}.  The slope $D$ decreases significantly with increasing $S$ but remains positive from $S = 1/2$ up to $S=3/2$.  The sign of $D(S)$ was indeed predicted by Hofmann to be positive using the effective Lagrangian method.\cite{Hofmann2010} 

\section{\label{SecThyExpFit} Comparison of Monte Carlo Simulations of the Magnetic Properties with Experiment}

\subsection{N\'eel Temperature}

Using Eq.~(\ref{Eq:FitTN(JcJ2)}), we can predict the N\'eel temperature from the values $J_1S/k_{\rm B}=380$~K, $J_2/J_1=0.29$, and $J_c/J_1 = 0.09$ in Table~\ref{tbl2} obtained from the fit of the inelastic neutron scattering data by spin wave theory.  Note that the neutron fit only provides products of the $J$ values with $S$.  Using the above-given parameters, Eq.~(\ref{Eq:FitTN(JcJ2)}) predicts
\begin{enumerate}
\item {$T_{\rm N} = 550$~K for $S = 2$}
\item {$T_{\rm N} = 640$~K for $S = 5/2$.}
\end{enumerate}
A comparison of these values with the experimental value $T_{\rm N}=620$--625~K clearly favors spin~5/2 over spin~2 for the Mn ions.  Indeed, the $T_{\rm N}$ calculated for $S = 5/2$ is remarkably close to the observed value.

\subsection{Magnetic Susceptibility}

\begin{figure}
\includegraphics [width=3.3in]{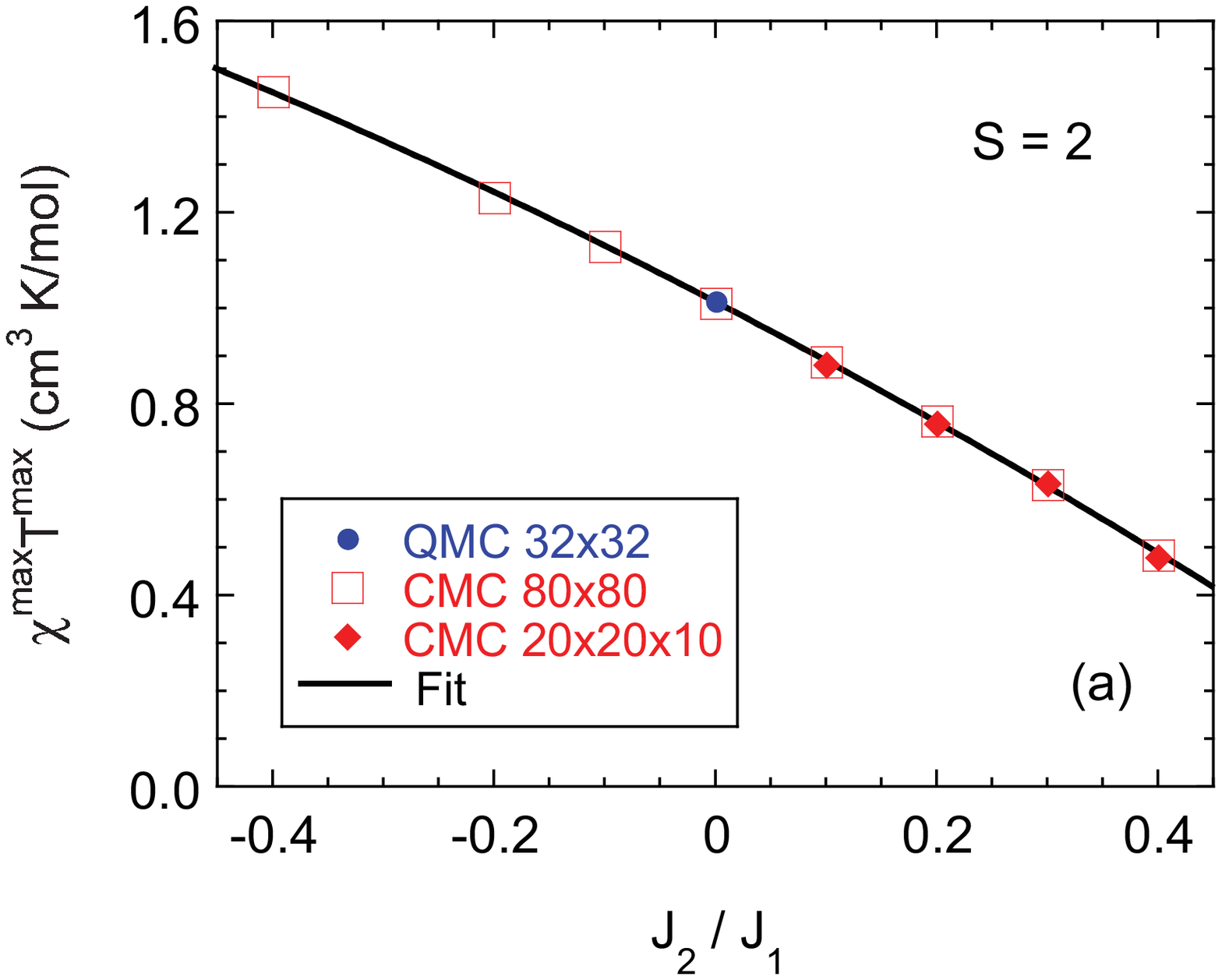}
\includegraphics [width=3.3in]{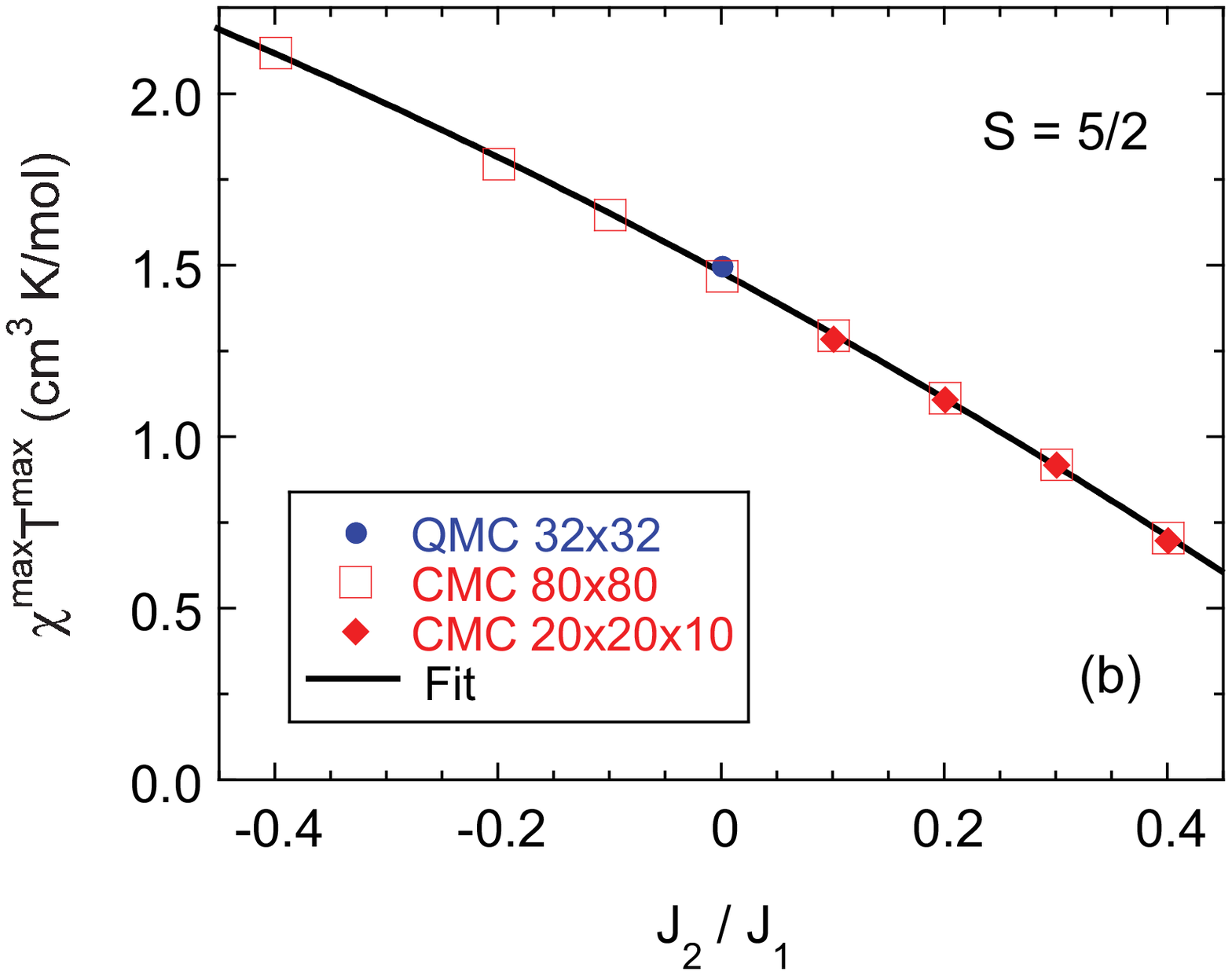}
\caption{(Color online) Product of the maximum susceptibility $\chi^{\rm max}$ per mole of spins and the temperature $T^{\rm max}$ at which it occurs for the $J_1$-$J_2$ model on a Heisenberg square spin lattice for spins (a) $S=2$ and (b) $S=5/2$, versus the ratio $J_2/J_1$ of the diagonal next-nearest-neighbor coupling to the nearest neighbor coupling.  Here a $g$-factor $g=2$ was assumed.  The data were obtained using quantum Monte Carlo (blue, QMC) and classical Monte Carlo (red, CMC) simulations on the lattice sizes indicated.  The $\chi^{\rm max}T^{\rm max}$ values obtained from CMC for the three-dimensional lattices are very insensitive to the coupling $J_c$ between layers up to at least $J_c/J_1 = 0.1$ as shown in Table~\ref{ChiCalcData}.  The solid black curves in (a) and (b) are least square fits to the respective data by the second order polynomials in Eqs.~(\ref{Eq:ChimaxTmaxFcns}).}
\label{ChimaxTmax}
\end{figure}

Our tables of calculated susceptibities are in the form of Eq.~(\ref{ChivsT}).  A very useful quantity for comparison with experimental susceptibility data is the product of the scaled maximum susceptibility $\chi^{\rm max}J_1/Ng^2\mu_{\rm B}^2$ and the scaled temperature at which the maximum occurs $k_{\rm B}T^{\rm max}/J_1S(S+1)$.  Setting $N$ equal to Avogadro's number $N_{\rm A}$ so that $\chi$ is the susceptibility per mole of spins, the product of these two variables is
\be
\left(\frac{\chi^{\rm max}J_1}{N_{\rm A}g^2\mu_{\rm B}^2}\right)\left[\frac{k_{\rm B}T^{\rm max}}{J_1S(S+1)}\right] =  \frac{\chi^{\rm max}T^{\rm max}}{3C},
\ee
where $C$ is the Curie constant in Eq.~(\ref{CC}).  This product does not contain any exchange constants and hence is a potential diagnostic for the value of the spin $S$ from experimental data.  One cannot hope to obtain a good fit to an experimental $\chi(T)$ data set by the theoretical predictions unless one can at least fit the experimental $\chi^{\rm max}T^{\rm max}$ datum.  The quantities $\chi^{\rm max}J_1/Ng^2\mu_{\rm B}^2$, $k_{\rm B}T^{\rm max}/S(S+1)J_1$, and $\chi^{\rm max}T^{\rm max}/3C$ are listed in Table~\ref{ChiCalcData} for both the classical and quantum Monte Carlo simulations.  Using the values of $C$ obtained from Eq.~(\ref{CC}) using $g=2$, the predicted values of $\chi^{\rm max}T^{\rm max}$ for direct comparison to our experimental datum are listed in Table~\ref{ChiCalcData} for the quantum value of $S$ in the QMC simulations and for classical values $S=2$ and $S=5/2$ in the CMC simulations, respectively.    One sees from the table that the value of $\chi^{\rm max}T^{\rm max}$ is very sensitive to the ratio $J_2/J_1$ but that it hardly changes for a given $J_2/J_1$ as the interlayer coupling ratio $J_c/J_1$ is changed over the range from 0 to~0.1.

Plots of $\chi^{\rm max}T^{\rm max}$ versus $J_2/J_1$ are shown in Figs.~\ref{ChimaxTmax}(a) and~\ref{ChimaxTmax}(b) for $S=2$ and $S=5/2$, respectively.  The data were fitted by the second-order polynomials
\bea
\chi^{\rm max}T^{\rm max} &=& 1.012 -1.202\left(\frac{J_2}{J_1}\right) - 0.2712\left(\frac{J_2}{J_1}\right)^2\nonumber\\*
\label{Eq:ChimaxTmaxFcns}\\*
\chi^{\rm max}T^{\rm max} &=& 1.478 - 1.757\left(\frac{J_2}{J_1}\right) -0.4146\left(\frac{J_2}{J_1}\right)^2
\nonumber
\eea
for $S=2$ and $S=5/2$, respectively, as shown by the respective solid curves in Fig.~\ref{ChimaxTmax}, where the units of the fits are ${\rm cm^3\,K/mol}$.  The rms deviations of the fits from the data are 0.005 and 0.010 for $S=2$ and $S=5/2$, respectively.

\begin{figure}
\includegraphics [width=3.4in]{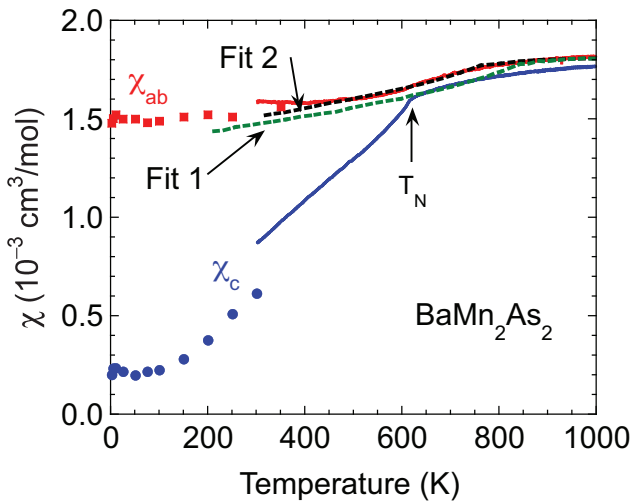}
\caption{(Color online) Two fits (Fit~1 and Fit~2, dashed curves) of the high-temperature magnetic susceptibility $\chi$ of ${\rm BaMn_2As_2}$ from Fig.~\ref{BaMn2As2_Hi_T_chi}(a) by CMC simulations.  The fits are only valid above $T_{\rm N}$ but are extrapolated to lower temperatures.  The fit parameters are given in Table~\ref{Tab:FitHighTChi}.  The temperatures of the breaks in slope of the fits are discernable and denote the predicted N\'eel temperatures in Eq.~(\ref{Eq:FitTN(JcJ2)}) for the respective parameters, which are somewhat above the observed value.}
\label{Fig:BaMn2As2_Hi_T_chi_All_Fit2}
\end{figure}

\begin{table}
\caption{\label{Tab:FitHighTChi}  Parameters determined from a fit of magnetic susceptibility data for ${\rm BaMn_2As_2}$ by classical Monte Carlo simulations.  The value $J_c/J_1 = 0.1$ was chosen to agree with the inelastic neutron scattering and band theoretical values which are both close to 0.1.  The values of $J_2/J_1$ were fixed at the listed values by comparing the predicted values of $\chi_{\rm spin}^{\rm max}T^{\rm max}$ with the experimental values for $S = 2$ and $S=5/2$.  Then the spin value $S$ was determined more precisely by fitting the respective CMC simulation to the experimental value of $\chi_{\rm spin}^{\rm max}T^{\rm max}$. Finally $J_1$ was found by fitting the experimental value of $\chi_{\rm spin}^{\rm max}$ to the theoretical value for the respective simulation.}
\begin{ruledtabular}
\begin{tabular}{lcc}
Quantity & Fit 1 & Fit 2 \\
\hline
$S$ & 2.06  & 2.64  \\
$J_1$ & 207~K = 17.8 meV & 210 K = 18.1 meV\\
$J_2$ & $\equiv41.4$ K = 3.6 meV & $\equiv85$ K = 7.3 meV\\
$J_2/J_1$ & $\equiv0.2$ & $\equiv 0.4$ \\
$J_c$ & $\equiv21$ K = 1.8 meV & $\equiv21$ K =1.8 meV\\
$J_c/J_1$ & $\equiv0.1$ & $\equiv0.1$ \\
\end{tabular}
\end{ruledtabular}
\end{table}

A comparison of the calculated values of $\chi^{\rm max}T^{\rm max}$ in Fig.~\ref{ChimaxTmax} and Eqs.~(\ref{Eq:ChimaxTmaxFcns}) with the observed value $\chi^{\rm max}T^{\rm max}=0.80~{\rm cm^3\,K/mol}$\,Mn in Eq.~(\ref{Eq:ChiMaxTMaxExp}) indicates that the local moment model can reproduce the observed $\chi^{\rm max}T^{\rm max}$ value with the following combinations of parameters
\begin{itemize}
\item{$g=2$, $S = 2$ and $J_2/J_1 \approx 0.17$}
\item{$g = 2$, $S=5/2$ and $J_2/J_1 \approx 0.36$}.
\end{itemize}
In the $J_1$-$J_2$ model, the G-type AF magnetic structure that is observed in ${\rm BaMn_2As_2}$ is stable against the stripe state as long as $J_2/J_1 < 1/2$ [Eq.~(\ref{Eq:JRestrictions})], which is satisfied by both of these estimates.  We do not have simulation data for precisely these two values of $J_2/J_1$.  Also, the parameter set $\{S, J_1, J_2, J_c\}$ is underdetermined by the experimental susceptibility data, so we have to make choices about some of the parameters when we fit the experimental data by the available classical Monte Carlo data.  We choose $J_c/J_1 = 0.1$ because this value is indicated both from the neutron scattering fit in Table~\ref{tbl2} above and from the theoretical results in Table~\ref{Tab:J1J2JcBandTheory} below.  For each of the two potential values $S=2$ and~5/2, we use the respective CMC $J_2/J_1$ simulation in Table~\ref{ChiCalcData} that shows the closest agreement with the experimental $\chi_{\rm spin}^{\rm max}T^{\rm max}$ for that spin value, namely
\begin{enumerate}
\item{Fit~1: $J_2/J_1= 0.2$, $J_c/J_1 = 0.1$, $S =  2$}
\item{Fit~2: $J_2/J_1 = 0.4$, $J_c/J_1 = 0.1$, $S = 5/2$.}
\end{enumerate}

Next, we have a choice of how to obtain a precise fit to the magnitude of the experimental data by the simulation data for $J_2/J_1= 0.2$ and 0.4, both with $J_c/J_1 = 0.1$.  We could adjust the $g$-factor, the orbital contribution to the susceptiblity, and/or the spin value.  At this stage such changes are just fitting parameters, so we arbitrarily choose to adjust the spin value slightly to obtain a good numerical fit of the particular simulation to the experimental value of $\chi_{\rm spin}^{\rm max}T^{\rm max}$ given in Eq.~(\ref{Eq:ChiMaxTMaxExp}).  Then we fix the value of $J_1$ by substituting the experimental value of $\chi_{\rm spin}^{\rm max}= 0.80\times 10^{-3}\ {\rm cm^3/mol\ Mn}$ and  $g=2$ into the expression $\chi^{\rm max}J_1/N_{\rm A}g^2\mu_{\rm B}^2$ and equating that with the $\chi^{\rm max}J_1/Ng^2\mu_{\rm B}^2$ value listed in Table~\ref{ChiCalcData}. The parameters obtained from the two fits are listed in Table~\ref{Tab:FitHighTChi}.  Remarkably, the value of $J_1$ is not sensitive to the values of $S$, $J_2$ or $J_c$, and a consistent value $J_1 \approx 210$~K = 18~meV is obtained for both fits.  The two fits are compared with the experimental data from Fig.~\ref{BaMn2As2_Hi_T_chi}(a) in Fig.~\ref{Fig:BaMn2As2_Hi_T_chi_All_Fit2}.  When plotting the fits, the calculated spin susceptibility per mole of spins has to be multipled by two (two atoms of Mn per formula unit) and then added to the orbital contribution given in Eq.~(\ref{Eq:ChiOrb}).  These fits are only valid in the paramagnetic regime above $T_{\rm N} = 625$~K, but they are extrapolated to lower temperatures.  The quality of the fits to the experimental data is reasonable for both fits.  Thus we cannot distinguish between the two possibilities $S= 2$ and $S = 5/2$ for the Mn spins on the basis of magnetic susceptibility measurements alone.

\section{\label{Sec:NMR} $^{75}$A\lowercase{s} NMR Measurements and Analysis}

\subsection{$^{75}$As NMR Spectrum}

\begin{figure}
\includegraphics [width=2.5in,viewport= 15 13 210 304,clip]{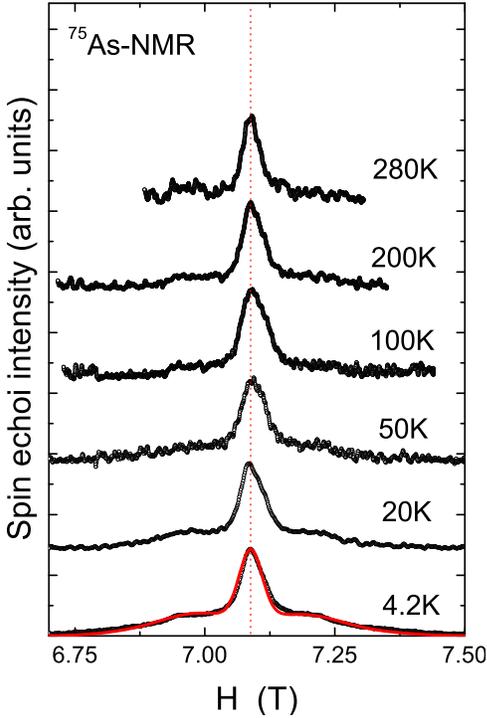}
\caption{(Color online) $^{75}$As NMR spectra for polycrystalline ${\rm BaMn_2As_2}$ at different temperatures. The solid red line is a fit to the spectrum at 4.2~K\@.}
\label{Fig:NMRSpectra}
\end{figure}

As shown in Fig.~\ref{Fig:BaMn2As2_Xtal_Mag_struct}, each As atom is coupled to four Mn atoms. Thus through $^{75}$As NMR one can probe the magnetism of the Mn sublattice in BaMn$_{2}$As$_{2}$. Figure~\ref{Fig:NMRSpectra} shows typical $^{75}$As NMR spectra in the magnetically ordered state at different temperatures $T < T_{\rm N}$ for a polycrystalline sample of ${\rm BaMn_2As_2}$. At low temperatures, along with the most intense central line the spectrum contains extra shoulder-like features on either
side. The broad linewidth is attributed to the random orientation of the internal field with respect to the
external field in the powder sample. $^{75}$As has an electric quadrupolar moment that interacts with the local electric field gradient (EFG) in the crystal giving rise to the splitting of the NMR line. Thus in principle one should see in the $^{75}$As spectra three allowed transitions: an $I_{z}= -\frac{1}{2} \leftrightarrow +\frac{1}{2}$ central transition, and two $I_{z}= \pm\frac{1}{2} \leftrightarrow \pm\frac{3}{2}$ satellite transitions. Therefore in an attempt to fit the experimental spectra taking into account both the EFG and the isotropic spin shift effects, we find that the spectrum at 4.2 K can be fitted reasonably well with iso-shift $K_{\rm iso} \simeq 0.38\%$, quadrupolar frequency $\nu_{Q} \simeq 2.1$ MHz, width of central peak $0.43$~kOe, width of satellite $1.13$~kOe, and EFG asymmetry parameter $\eta \simeq 0.0$.  The fit is shown as the solid red curve through the data at 4.2~K in Fig.~\ref{Fig:NMRSpectra}. The value of $\nu_{Q}$ is comparable to that reported for BaFe$_{2}$As$_{2}$ in the ordered state.\cite{kitagawa2008}

The linewidth and position were found to be almost temperature independent. As shown in Ref.~\onlinecite{YSingh2009} from magnetic neutron diffraction data, the sublattice magnetization is nearly saturated at 300~K\@. Since the NMR linewidth in the ordered state is a measure of the sublattice magnetization, the independence of the linewidth over our temperature range is consistent with the neutron diffraction results. 

The internal field at the $^{75}$As site can be analyzed by taking the crystal symmetry into consideration, which has been adopted in an analysis of the hyperfine field at the $^{75}$As site in ${\rm BaFe_2As_2}$ by Kitagawa et al.\cite{kitagawa2008}  According to their analysis, for a G-type antiferromagnetic spin structure the internal field at the $^{75}$As site is zero due to a perfect cancellation of the off-diagonal hyperfine field produced by four in-plane NN Mn spins when the spin moments are parallel to the $c$-axis. Thus the spin components along this axis do not produce any magnetic broadening in the $^{75}$As NMR spectra.  Only the $ab$-plane components of the ordered Mn spin can produce an internal field perpendicular to the $c$-axis at the $^{75}$As site.  On the other hand, for a stripe-type AF spin structure, a $c$-axis component of the spin moments produces an internal field $H_{\rm int}=z^\prime BS$ along the $a$-axis, where $z^\prime$ is the number of nearest neighbor Mn spins of the $^{75}$As site, $B$ is the off-diagonal hyperfine coupling constant and $S$ is the Mn spin.

Assuming that the broadening of the NMR spectra originates from $H_{\rm int}$ at the $^{75}$As site, $H_{\rm int}$ is estimated to be $\sim 215$~Oe from the spectral width.  Using the ordered moment $\mu=3.9~\mu_{\rm B}$/Mn and $z^\prime =4$, the off-diagonal hyperfine coupling constant $B$ is estimated to be $\sim{\rm 14\,Oe}/\mu_{\rm B}$ for the case of stripe-type AF order.  Such a small $B$ is of the order of the nuclear-nuclear dipolar field and is not likely due to transferred hyperfine couplings.  For the G-type AF structure, the $ab$-plane components can be produced by a canted component of the Mn spins when the magnetic field is applied perpendicular to the ordered moment axis, i.e., perpendicular to the $c$-axis. Using the perpendicular component of the spin susceptibility $\chi_\perp = {\rm 1.3\times 10^{-3}~cm^3/mol}$ from Fig.~\ref{BaMn2As2_Hi_T_chi}(a) and $H = 7.1$~T that we used for measurements of the spectra, the $ab$-component of the ordered Mn moment $\mu_{ab}$ is evaluated to be 0.0083~$\mu_{\rm B}$/Mn in this field. Now using $H_{\rm int} = 215$~Oe and $\mu_{ab} =0.0083~\mu_{\rm B}$, the off-diagonal hyperfine coupling constant is calculated to be $B = 6.5~{\rm kOe}/\mu_{\rm B}$. This value of $B$ is comparable to $B = 4.3~{\rm kOe}/\mu_{\rm B}$ reported in ${\rm BaFe_2As_2}$.\cite{kitagawa2008}  Thus our $^{75}$As NMR spectra observed in the AF ordered state are consistent with the G-type AF structure reported from the neutron experiment.\cite{YSingh2009}

\subsection{\label{Sec:NSLR} Nuclear Spin-Lattice Relaxation Rate}
\begin{figure}
\includegraphics [width=3.in,viewport= 13 13 223 215,clip]{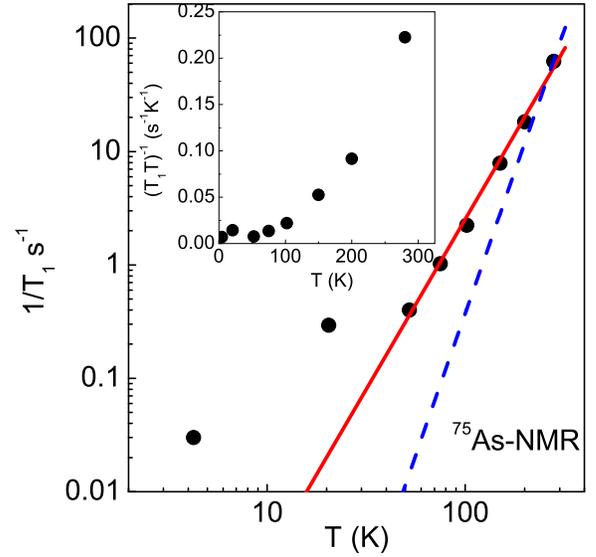}
\caption{(Color online) Nuclear spin-lattice relaxation rate ($1/T_{1}$) measured at the $^{75}$As site versus temperature $T$. The solid and dashed lines represent $T^{3}$ and $T^{5}$ behaviors, respectively. The slope of the former line fitted to the 50--300~K data is ${\rm 2.51\times 10^{-6}\,s^{-1}K^{-3}}$.  Inset: $(T_{1}T)^{-1}$ versus $T$.}
\label{Fig:NMR1T1}
\end{figure}

The longitudinal nuclear magnetization recovery curve following saturation was fitted by the double exponential function\cite{simmons1962}
\[
1-\frac{M(t)}{M(\infty)}=0.1\,e^{-t/T_{1}}+0.9\,e^{-6t/T_{1}},
\]
as expected for the center line of the spectrum of the $^{75}$As nuclear spin $I=\frac{3}{2}$, where $1/T_1$ is the $^{75}$As nuclear spin-lattice relaxation rate and $M(t)$ and $M(\infty)$ are the nuclear magnetization at time $t$ after saturation and the equilibrium nuclear magnetization at time $t = \infty$, respectively. The extracted $1/T_1$ as a function of temperature is shown in Fig.~\ref{Fig:NMR1T1}, where $1/T_{1}$ is seen to increase rapidly with increasing temperature. In the AF state, this rapid increase in $1/T_{1}$ with $T$ is a clear signature of relaxation due to scattering of magnons by the nuclear spins.  According to Beeman and Pincus,\cite{beeman1968} in the AF state for magnetic insulators, $1/T_{1}$ is mainly driven by such magnon processes, leading to a power law $T$-dependence.\cite{beeman1968, belesi2006, nath2009} For $T \gg \Delta/k_{\rm B}$, where $\Delta$ is the anisotropy gap in the spin-wave spectrum, it either follows a $T^{3}$ behavior due to a two-magnon Raman process or a $T^{5}$ behaviour due to a three-magnon process, while for $T \ll \Delta/k_{\rm B}$, it follows a thermally activated behaviour $1/T_{1} \propto T^{2}e^{-\Delta/k_{\rm B}T}$. As seen from Fig.~\ref{Fig:NMR1T1}, our $^{75}$As $1/T_{1}$ data in the $T$-range $50 \leq T \leq$ 300~K follow a $T^{3}$ behavior rather than a $T^{5}$ behavior indicating that the relaxation is mainly governed by the two-magnon Raman process. A $T^{3}$ fit over this $T$ range yields
\be
\frac{1}{T_1} = ({\rm 2.51 \times 10^{-6}~s^{-1}K^{-3}})\,T^3,
\label{Eq:1/T1fit}
\ee
as shown in Fig.~\ref{Fig:NMR1T1}.  The lack of activated behavior down to 50~K indicates that $\Delta/k_{\rm B}$ is smaller than 50~K\@.

For the two-magnon process, $1/T_1$ is determined by the slopes of the spin wave dispersion relations at $\omega \sim 0$ and thus by the spin wave velocities. The spin wave velocities within the $ab$-plane and along the $c$-axis in terms of the exchange constants in the $J_1$-$J_2$-$J_c$ Heisenberg model are given above in Eqs.~(\ref{Eq:SWvels}).  Since the spin wave velocity depends on the direction of propagation, $1/T_1$ should also depend on the spin wave direction.  Based on the $1/T_1$ expression for the two-magnon process reported by Beeman and Pincus,\cite{beeman1968} we have calculated the $1/T_1$ for ${\rm BaMn_2As_2}$ with the body-centered-tetragonal structure ($I$4/$mmm$) arising from the two spin wave velocities as
\bea
\left(\frac{1}{T_{1}}\right)^{-1}_i &=& \left(\frac{A}{\hbar}\right)^2 \frac{4z^\prime z\hbar\sin^{2}\theta}{(2\pi)^{3}}\left(k_{\rm B}T\right)^{3} \frac{(a^2c)^2}{(\hbar v_i)^4\alpha^2}\nonumber\\*
&& \hspace{0.6in}\times  \int_{\Delta/k_{\rm B}T}^{\infty} \frac{x}{e^{x}-1}\,dx,
\label{1/t1}
\eea
where $z^\prime = 4$ is the number of Mn nearest neighbors to a given $^{75}$As site (see Fig.~\ref{Fig:BaMn2As2_Xtal_Mag_struct}), $z=4$ is the number of nearest-neighbor Mn spins interacting with a given Mn spin, $i = ab$ with $\alpha=a$ or $i=c$ with $\alpha=c$,  and $\theta$ denotes the angle between the local hyperfine field at the $^{75}$As site and the anisotropy axis ($c$-axis) which is the Mn ordered moment axis.  We also have
\[
\frac{A}{\hbar}=\gamma_{\rm n}g\mu_{\rm B}A_{\rm hf},
\]
where $g$ is the electronic $g$-factor, $A_{\rm hf}$ is the hyperfine coupling constant and $\gamma_{\rm n}$ is the $^{75}$As nuclear gyromagnetic ratio given by $^{75}\gamma_{\rm n}/2\pi = 7.2919$~MHz/T\@.  Using $g=2$ and $A_{\rm hf}=6.5$~kOe/$\mu_{\rm B}$ that was estimated from the spectrum analysis, one obtains
\be
\frac{A}{\hbar} = 6.0 \times 10^7\ {\rm s^{-1}}.
\label{Eq:Ahbar}
\ee

For $T \gg \Delta/k_{\rm B}$ as in our temperature range 50--300~K where $1/T_1 \propto T^3$, the integral in Eq.~(\ref{1/t1}) approaches its maximum value $\pi^{2}/6$, so Eq.~(\ref{1/t1}) reduces to
\bea
\left(\frac{1}{T_{1}}\right)^{-1}_{ab} &=&  \left(\frac{A}{\hbar}\right)^2 \frac{z^\prime z\hbar a^2c^2 k_{\rm B}^3 \sin^{2}\theta}{12\pi(\hbar v_{ab})^4}\, T^3.
\label{1/t1fab}\\*
\left(\frac{1}{T_{1}}\right)^{-1}_c &=& \left(\frac{A}{\hbar}\right)^2 \frac{z^\prime z\hbar a^4k_{\rm B}^3 \sin^{2}\theta}{12\pi(\hbar v_c)^4}\, T^3
\label{1/t1fc}\\*
&\equiv& C_{ab\ {\rm or}\ c}\, T^3.\nonumber
\eea
The ratio $R_{ab/c}$ of the relaxation rates for $ab$-plane and $c$-axis spin waves should be independent of $T$.  If we assume that the hyperfine coupling $A$ of the electronic spins to the nuclear spins is isotropic, then $R_{ab/c}$ obtained using Eqs.~(\ref{Eq:SWvels}), (\ref{1/t1fab}) and~(\ref{1/t1fc}) is given by
\bea
R_{ab/c} &=& \frac{(1/T_1)_{ab}}{(1/T_1)_{c}} = \left(\frac{c}{a}\right)^2\left(\frac{v_c}{v_{ab}}\right)^4 \label{Eq:1/T1Ratio}\\*
&=& \frac{1}{4}\left(\frac{c}{a}\right)^{6}\left[\frac{(J_c/J_1)(1 + J_c/2J_1)}{(1-2J_2/J_1)(1+J_c/2J_1)}\right]^2.\nonumber
\eea
Taking $a = 4.15$ and $c =13.41$~\AA\ for the lattice parameters at 8~K,\cite{YSingh2009} and $J_2/J_1 = 0.29$ and $J_c/J_1=0.09$ from the neutron scattering fit in Table~\ref{tbl2}, Eqs.~(\ref{Eq:1/T1Ratio}) yield
\be
R_{ab/c} = 13.
\ee
Thus the nuclear spin-lattice relaxation rate due to spin waves traveling in the $ab$-plane is much larger than that due to $c$-axis spin waves and we will therefore assume that Eq.~(\ref{1/t1fab}) gives the observed $1/T_1$ to a good approximation.

Since $A_{\rm hf}$ was estimated in Eq.~(\ref{Eq:Ahbar}), one can obtain information on the exchange constants from the coefficient of the $T^{3}$ fit in Eq.~(\ref{Eq:1/T1fit}). Inserting $\hbar v_{ab}$ from Eqs.~(\ref{Eq:SWvels}) into~(\ref{1/t1fab}) gives
\bea
\left(\frac{J_1}{k_{\rm B}}\right)^4 &=& \frac{1}{C_{ab}}\left(\frac{A}{\hbar}\right)^2\left(\frac{c}{a}\right)^2 \frac{z^\prime z \hbar\langle\sin^2\theta\rangle}{192\pi k_{\rm B}S^4}\label{Eq:J1fromNMR}\\*
&&\times \left[\left(1-\frac{2J_2}{J_1}\right)\left(1+\frac{J_c}{2J_1}\right)\right]^{-2}.\nonumber
\eea
Our single fit parameter $C_{ab}$ in Eq.~(\ref{Eq:1/T1fit}) can only be used to determine a single exchange constant or a single combination of them.  We therefore estimate $J_1$ using the above values  $J_2/J_1 = 0.29$ and $J_c/J_1 = 0.09$ derived from our inelastic neutron scattering experiments.  In Eq.~(\ref{Eq:J1fromNMR}), we also use $z^\prime=4$, $z = 4$, we take $S$ to be the ordered spin $\langle S\rangle =2$ (from magnetic neutron diffraction experiments),\cite{YSingh2009} and $\langle\sin^{2}\theta\rangle=(1/2)\int_0^\pi \sin^3\theta\,d\theta = 2/3$ (i.e.\ considering an average over all angles).  Using Eq.~(\ref{Eq:J1fromNMR}), we then obtain
\bea
\frac{J_1}{k_{\rm B}} &=& {\rm 160~K}\label{Eq:J1Value}\hspace{0.2in}(S \equiv 2)\\*
J_1 &=& 14~{\rm meV}\nonumber.
\eea
This value is close to the value $J_1 = 16$~meV estimated in Table~\ref{tbl2} for $S=2$ from our neutron scattering data.  If we take the spin to be $S = 5/2$, the value of $J_1$ from Eq.~(\ref{Eq:J1fromNMR}) would be a factor of $(5/4)^4 = 2.4$ times smaller.

The overall temperature dependence of $1/T_{1}$ in BaMn$_{2}$As$_{2}$ in Fig.~\ref{Fig:NMR1T1} is similar to that reported for KMnF$_{3}$.\cite{mahler1967} In KMnF$_{3}$, a deviation from power law behavior was observed at low temperatures and $1/T_{1}$ shows a broad maximum. This broad feature at low temperature was attributed to the effects of defects or extrinsic impurities. Thus in BaMn$_{2}$As$_{2}$, the deviation of the data below 50~K  from the higher-temperature $T^{3}$ fit in Fig.~\ref{Fig:NMR1T1} is likely due to relaxation associated with defects and/or extrinsic impurities.

For a metallic system, one would expect a Korringa-like behaviour [$(T_{1}T)^{-1}$ = constant] as has been observed in (Ba,Ca)Fe$_{2}$As$_{2}$ (Refs.~\onlinecite{kitagawa2008}, \onlinecite{baek2008}, \onlinecite{baek2009}) and $R$FeAsO$_{1-x}$F$_{x}$ ($R$ = La, Pr) (Refs.~\onlinecite{nakai2008}, \onlinecite{matano2008}) in the paramagnetic state. In these compounds, $(T_{1}T)^{-1}$ is also constant at low temperature below $T_{\rm N}$ due to their metallic character and increases sharply near $T_{\rm N}$.  In contrast, $(T_{1}T)^{-1}$ in BaMn$_{2}$As$_{2}$ (inset of Fig.~\ref{Fig:NMR1T1}) shows a gradual increase with increasing temperature signifying the insulating ground state of the compound.

\section{\label{Sec:BandTheory} Band-Theoretical Estimates of the Exchange Couplings}

\begin{table*}
\caption{\label{Tab:J1J2JcBandTheory}  Parameters of ${\rm BaMn_2As_2}$ with NN ($J_{1}$), NNN ($J_{2}$) and interlayer ($J_{c}$) exchange interactions obtained from density functional theory.  Here $S$ is the calculated spin, $E$ is the total energy per Mn atom, FM means ferromagnetic structure and G-type AF structure is the N\'eel (checkerboard) antiferromagnetic structure with an in-plane structure shown in the top panel of Fig.~\ref{Fig:Magnetic_Structures}.  The two stripe structures have the in-plane antiferromangetic structure shown in the bottom panel of Fig.~\ref{Fig:Magnetic_Structures}, where the Stripe-AF structure has AF stacking and the Stripe-FM structure has FM stacking along the $c$-axis.  The estimated exchange constants $J$ in rows 5 and~6 are calculated from Eqs.~(\ref{Eq:JcfromSingh})--(\ref{Eq:FindJ2}) using the total energy values in column~4.  Our exchange constants using the LDA and GGA in the last two rows were calculated from the excitation energies from the magnetically ordered ground state.  In the reference (Ref.) column, ``PW'' means ``present work''.}
\begin{ruledtabular}
\begin{tabular}{lccc|ccccccc}
Magnetic  & $\mu$ & $S$ & $E$/Mn & $2J_1+J_c$ & $J_c$ & $J_1$ & $J_2$ & $J_c/J_1$ & $J_2/J_1$  & Ref. \\
Structure &  $(\mu_{\rm B}$/Mn) &  & (meV) & (meV) &(meV) & (meV)& (meV)\\
\hline
FM & 2.74        & 1.37 & $-330$ &  &  &&&&&     \onlinecite{an2009}\\
G-type AF & 3.20 & 1.60 & $-660$ & &&&&&&    \onlinecite{an2009}   \\
Stripe-AF & $\equiv 3.20$ &  $\equiv 1.60$ & $-515$ & &&&&&&    \onlinecite{an2009}   \\
Stripe-FM & $\equiv 3.20$ &  $\equiv 1.60$ & $-505$ & &&&&&&    \onlinecite{an2009}   \\
&&&& 41.0\footnotemark[1] & 2.0\footnotemark[1] & 19.5\footnotemark[1] & $-5.4$\footnotemark[1] & 0.10\footnotemark[1] & $-0.28$\footnotemark[1]    & \onlinecite{an2009}, PW\\
&&&& 28.5\footnotemark[2] & 1.2\footnotemark[2] & 13.7\footnotemark[2] & $-2.5$\footnotemark[2] & 0.09\footnotemark[2] & $-0.18$\footnotemark[2] &  \onlinecite{an2009}, PW\\
\hline
FM (LDA)& 2.8 & 1.4 &&& $\approx 0$ & $-9.1$ & $-2.2$ &&& PW\\
FM (GGA)& 3.0 & 1.5 &&&&&&&& PW\\
G-type AF (LDA)& 3.3 & $1.6_5$ && 27.2 & 1.03 & 13.1 & 2.8 & 0.08 & 0.21 & PW\\
G-type AF (GGA)& 3.6 & 1.8 && 26.2 & 1.0 & 12.6 & 2.7 & 0.08 & 0.21 & PW\\
\end{tabular}
\end{ruledtabular}
\footnotetext[1]{Calculated using Eqs.~(\ref{Eq:JcfromSingh})--(\ref{Eq:FindJ2}) as written.}
\footnotetext[2]{Calculated by replacing $S^2$ by $S(S+1)$ in Eqs.~(\ref{Eq:JcfromSingh})--(\ref{Eq:FindJ2}).}
\label{Tab:BandTheoryJs}
\end{table*}

\begin{figure}
\includegraphics [width=2in]{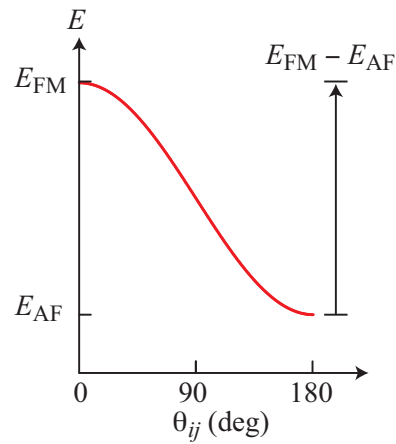}
\caption{(Color online) Total energy $E$ of a system when the polar angle $\theta_{ij}$ between two local magnetic moments $i$ and $j$ is varied in band theory.  The energies of the ferromagnetic $E_{\rm FM}$ and antiferromagnetic $E_{\rm AF}$ magnetic structures and their difference are indicated.  Here the antiferromagnetic state is the ground state.  For a ferromagnetic ground state, the minimum in energy would be at $\theta_{ij}=0$.}
\label{Fig:Total_Energy_Calcs}
\end{figure}

The quantitative analysis of the magnetic interactions in real magnets is based mostly on density functional theory.  To a large extent this theory is very similar to the Fermi-liquid theory of Landau, however, strictly speaking, it allows one to obtain only the total energy of the ground state, the distribution of charge and spin densities, and other quantities that can be directly determined by these.  Several notable exceptions (Mott insulators, rare earth systems, systems near quantum critical points) have been revealed but currently it is believed that for magnets of the Fe group the accuracy of the commonly used local density approximation (LDA) is acceptable for the description of the ground state properties including the equilibrium magnetic moments at $T=0$~K\@.

While the numerical agreement between experimental and theoretical magnetic moments is often very good, there are certain cases when the  local approximation numerically violates quantum mechanical laws.  For instance, even in insulating systems or magnetic molecules where the total magnetic moment is close to an integer number of Bohr magnetons, the value obtained from density functional theory is usually not an integer. This discrepancy is related to the fact that the wave function of the density functional method is often not an eigenfunction of the square of total spin (even without relativistic effects). This effect of `spin contamination' usually cannot be eliminated or easily resolved.  While non-integer values of the moment in metallic systems are traditionally explained by itineracy of the system and partial occupation numbers, the problematic issue of whether or not \textbf{S}$^{2}$ is an integral of the motion is usually ignored with the hope that such errors are small.  The relationship between the single-particle spectrum obtained in the density functional theory and the physical properties of the magnetic excitations is not clearly defined.  Nevertheless, the research of the last 20--25 years  revealed that LDA often provides good agreement between theory and experiment for the magnetic excitation spectra if the ground state is properly described.

The description of the intersite magnetic interactions represents a typical problem within the topic of magnetic excitations. By itself, the determination of a pairwise exchange parameter $J_{ij}$ between atoms $i$ and $j$ in an arbitrary magnetic material is not well-posed.  For instance, in very itinerant systems the effective spin Hamiltonian can have very non-Heisenberg behavior.  However, from the phenomenological theory of ferromagnetism\cite{Kittel1987,Moriya1985} the energy of any weak and smooth variation of spin density can be described by the effective classical Heisenberg Hamiltonian for equivalent classical spins $\mathbf{S}_{i}$ and $\mathbf{S}_{j}$ with magnitudes $S_i=S_j\equiv S$ given by
\be
E = \sum_{\langle ij \rangle} J_{ij}\mathbf{S}_{i}\cdot\mathbf{S}_{j} = S^2\sum_{\langle ij \rangle} J_{ij}\cos \theta _{ij}.
\label{Eq:ClassHeisThy}
\ee

Traditionally, a set $\{J_{ij}\}$ of exchange coupling constants in the density functional theory can be calculated using two approaches. In the first approach, using the effective Heisenberg model~(\ref{Eq:ClassHeisThy}) one can solve for the $J_{ij}$ from the set $\{F_{\alpha\beta}(J_{ij})\}$ of equations for the differences of the energies between different magnetic structures $\alpha$ and $\beta$ 
\be
F_{\alpha\beta}(J_{ij}) = \frac{E_{\alpha}\{J_{ij}\} -E_{\beta}\{J_{ij}\}}{2S^{2}}
\label{Eq:JfromDeltaEs}
\ee
obtained from band structure calculations as shown in Fig.~\ref{Fig:Total_Energy_Calcs}. This is the usual way to obtain $\{J_{ij}\}$ for highly localized magnetic insulators and is usually the most suitable method for the calculation of magnetic phase transition temperatures. 

Another approach is based on the definition of $J_{ij}$ as the second derivative of the total energy in Eq.~(\ref{Eq:ClassHeisThy}) with respect to rotation of moments from their magnetic alignment in a given magnetically ordered ground state
\be
J_{ij}=-\frac{1}{S^2}\frac{\partial ^{2}E}{\partial \theta _{ij}^{2}},
\label{Eq:Jijfrompartials}
\ee
which is proportional to the curvature of the total energy $E$ versus angle $\theta _{ij}$ near the minimum for an antiferromagnet at $\theta _{ij} = 180^\circ$ in Fig.~\ref{Fig:Total_Energy_Calcs}.  This definition of $J_{ij}$ corresponds to a linear response scheme and is usually the most suitable technique for the analysis of the excitations above the ground state (spin waves) and is directly related to the dynamical magnetic susceptibility measured in inelastic neutron scattering experiments.  The procedure for evaluating Eq.~(\ref{Eq:Jijfrompartials}) depends on the band structure
methods and the specifics of the linear response method employed. This technique has been used for many magnetic materials in the past.\cite{REF}  This approach can be understood as a static limit of the dynamic linear response technique which has been used for calculations of spin waves and Stoner excitation spectra in magnets.  One can analytically obtain an expression for the onsite stability parameter $J_{0}$ which should be the same as $\sum J_{ij}$. A comparison of $J_{0}$ and $\sum J_{ij}$ is a check on the consistency of the calculations and the reliability of the numerical scheme.  

We first consider the exchange constants obtained from total energy differences and then in the subsequent section from the energy of excitations from the magnetically ordered ground state.

\subsection{Exchange Interactions from Total Energy Calculations}

Using density functional theory in the LDA, An et al.\ correctly deduced from total energy calculations, prior to the availability of the experimental results, that the G-type antiferromagnetic structure of ${\rm BaMn_2As_2}$ has a lower energy than either the FM structure or of two types of stripe structure.\cite{an2009}  Their predicted ordered moment for the G-type AF structure was $\mu = 3.20~\mu_{\rm B}$/Mn, somewhat smaller than the value of $\mu = 3.9(1)~\mu_{\rm B}$/Mn observed later.\cite{YSingh2009}  Their LDA total energies and ordered moments for the FM and G-type AF structures are listed in Table~\ref{Tab:J1J2JcBandTheory}, together with their total energies of two commensurate collinear stripe states with the in-plane stripe structure shown in the bottom panel of Fig.~\ref{Fig:Magnetic_Structures}.\cite{an2009}  The Stripe-AF structure has AF alignment of the ordered moments along the $c$-axis, whereas the Stripe-FM structure has FM alignment along the $c$-axis.  As seen in  Table~\ref{Tab:BandTheoryJs}, the ordered moment $\mu$ of the Mn in the FM structure is not the same as the value of $\mu$ in the G-type AF structure.

From the LDA total energies and ordered moments calculated by An et al.\cite{an2009} for the magnetic structures listed in Table~\ref{Tab:J1J2JcBandTheory}, one can obtain estimates of the exchange couplings in ${\rm BaMn_2As_2}$ using the value for the spin $S$ obtained from the ordered moment $\mu$ as
\[
S=\frac{\mu}{g\mu_{\rm B}}=\frac{\mu}{2\mu_{\rm B}}
\]  
with $g=2$.  The classical energies per spin of the magnetic structures in Table~\ref{Tab:J1J2JcBandTheory} obtained using Eqs.~(\ref{Eq:StripevsNeel}) and Fig.~\ref{Fig:Magnetic_Structures} are
\bea
E_{\rm FM} &=& S_{\rm FM}^2(2J_1+J_c+2J_2)\label{Eq:solveJsfromEs}\\*
E_{\rm G} &=& S_{\rm G}^2(-2J_1-J_c+2J_2)\nonumber\\*
E_{\rm Stripe\,AF} &=& S_{\rm G}^2(J_c - 2J_2)\nonumber\\*
E_{\rm Stripe\,FM} &=& S_{\rm G}^2(-J_c - 2J_2).\nonumber
\eea
We have taken the Mn spin in the two stripe phases to be the same as in the G-type AF structure, since they were not given by An et al.  Because the total energy contains a constant term proportional to the square of the magnetization, we solve for the exchange constants using only differences between these total energies according to Eq.~(\ref{Eq:JfromDeltaEs}).  From the last two expressions in Eqs.~(\ref{Eq:solveJsfromEs}) we obtain
\be
J_c = \frac{1}{2S_{\rm G}^2}(E_{\rm Stripe\,FM} - E_{\rm Stripe\,AF}).
\label{Eq:JcfromSingh}
\ee
From the first two expressions in Eqs.~(\ref{Eq:solveJsfromEs}) we obtain
\be
2J_1 + J_c = \frac{1}{2}\left(\frac{E_{\rm FM}}{S_{\rm FM}^2} - \frac{E_{\rm G}}{S_{\rm G}^2}\right).
\label{Eq:J1JcfromSingh}
\ee
Thus Eqs.~(\ref{Eq:JcfromSingh}) and~(\ref{Eq:J1JcfromSingh}) determine the two exchange constants $J_1$ and $J_c$.  Then from the second and fourth of Eqs.~(\ref{Eq:solveJsfromEs}) we solve for $J_2$ according to
\be
J_2 = \frac{J_1}{2}-\frac{E_{\rm Stripe\,FM} - E_{\rm G}}{4S_{\rm G}^2}.
\label{Eq:FindJ2}
\ee

It is not clear whether to retain $S^2$ or to insert the quantum mechanical expectation value $S(S+1)$ of $\langle S^2\rangle$ in place of $S^2$ in Eqs.~(\ref{Eq:JcfromSingh})--(\ref{Eq:FindJ2}), so we calculate two sets of exchange parameters based on these two assumptions, which are given in Table~\ref{Tab:J1J2JcBandTheory}.  Using the second assumption, the values of $J_c$ and $J_1$ are respectively about the same as the values in Table~\ref{tbl2} deduced from our inelastic neutron scattering experiments, but $J_2$ has the opposite sign in the theory and experiment.

We studied the properties of BaMn$_{2}$As$_{2}$ using density functional calculations of the electronic structure and magnetic interactions in the FM and G-type AF structures. For consistency, we used the experimental values of the lattice parameters $a = 4.15~$\AA\ and $c = 13.47$~\AA\ and the theoretically optimized value of the internal As parameter $z_{\rm As}=0.3524$ utilized by An et al.\cite{an2009}    Our electronic structure calculations were performed using the recently developed full-potential linear muffin tin orbital program.\cite{Kotani2010}Ê The accuracy of the exchange couplings obtained is about 2--3\%.  The studies of the exchange couplings were done using the static linear response technique described in Refs.~\onlinecite{Antropov2003} and~\onlinecite{Antropov2006}.Ê 

Our results using LDA and the generalized gradient approximation (GGA) are very similar to those reported by An et al.\cite{an2009}  We find that that BaMn$_{2}$As$_{2}$ has a relatively large ordered moment in the G-type AF structure with $\mu = 3.3\,\mu_{B}$/Mn in LDA and $3.6\,\mu _{\rm B}$/Mn in GGA, as listed in Table~\ref{Tab:J1J2JcBandTheory}, with a small band gap (0.15~eV) in the electronic spectrum as observed, and with the G-type AF ordering having the lowest energy among all considered magnetic structures.  The ferromagnetic structure has no charge gap, i.e., the compound would be metallic.  Our total energy $E_{\rm FM}-E_{\rm G}$ differences were 350~meV/Mn (LDA) and 375~meV/Mn (GGA), which are similar to the values of 330~meV/Mn and 380~meV/Mn obtained in Ref.~\onlinecite{an2009}, respectively.

While the magnetic moments are relatively large, they show a significant dependence on the magnetic structure, in agreement with the results of Ref.~\onlinecite{an2009}.  For instance, the ferromagnetically ordered BaMn$_{2}$As$_{2}$ has an ordered moment of 2.8~$\mu_{\rm B}$/Mn in LDA and 3.0~$\mu _{\rm B}$/Mn in GGA, which deviate from the corresponding values for G-type AF ordering (Table~\ref{Tab:J1J2JcBandTheory}) by about 20\%. Due to this relatively strong dependence of the ordered magnetic moment on the magnetic structure, estimates of exchange couplings from total energy calculations should be used with caution.

\subsection{Exchange Interactions from Excitations from the Magnetically Ordered Ground State}
For comparison with the exchange constants deduced from inelastic magnetic neutron scattering experiments, calculations using the linear response technique\cite{REF} are preferable to the total energy technique, as noted above.  Our calculations of the parameters of the Heisenberg model are: $J_{1}=13.1$~meV, $J_{2}=2.8$~meV, $J_{c}=1.03$~meV (LDA) and $J_{1}=12.6$~meV, $J_{2}=2.7$~meV, $J_{c}=1.0$~meV (GGA), as summarized in Table~\ref{Tab:J1J2JcBandTheory}.  These values are quite comparable with the values deduced from our inelastic neutron scattering measurements in Table~\ref{tbl2}, and roughly similar to those in Table~\ref{Tab:J1J2JcMFT} obtained from molecular field analysis of our magnetic susceptibility data in Fig.~\ref{BaMn2As2_Hi_T_chi}.

The G-type AF ordering temperatures obtained from our spin value and exchange parameters in Table~\ref{Tab:BandTheoryJs} using the molecular field expression~(\ref{Eq:TNfromJ1J2Jc}) are 
\bea
T_{\rm N} &=& 730~{\rm K} \hspace{0.25in}({\rm G\ type\ AF,\ LDA})\\*
T_{\rm N} &=&  810~{\rm K}. \hspace{0.2in}({\rm G\ type\ AF,\ GGA})\nonumber
\eea
These mean-field N\'eel temperatures are somewhat larger than the observed value of 625~K, as expected, and indeed are approaching the temperature of the maximum of the measured susceptibility in Fig.~\ref{BaMn2As2_Hi_T_chi}(a) which from Table~\ref{ChiCalcData} is of order the mean-field transition temperature. 

The other longer-range pair exchange parameters appear to be much smaller, suggesting very short-ranged exchange interactions in this material.  In particular, the difference between the above parameter $J_{0}$ and $\sum J_{ij}$ over six NN and four NNN is only about 5\%, suggesting very short-ranged exchange interactions in BaMn$_{2}$As$_{2}$.  This is different from the corresponding results for many Fe pnictides, where the exchanges with further neighbors are not so small and definitely provide a finite contribution to the spin wave spectrum.\cite{pnictides}  We attribute this difference to the metallic character of the Fe pnictides and the semiconducting character of BaMn$_{2}$As$_{2}$.

To check the dependence of $\{J_{ij}\}$ on the type of magnetic order we also performed linear response calculations of the $\{J_{ij}\}$ for the FM phase. The stability parameter $J_0$ for this phase appears to be negative confirming the instability of such order with respect to the deviation of a single spin from the ordered moment direction (i.e., from $\theta_{ij} = 0)$. This directly supports the qualitative behavior of the total energy versus $\theta_{ij}$ in Fig.~\ref{Fig:Total_Energy_Calcs}. The pair exchanges in this phase are $J_1 = -9.1$~meV and $J_2 = -2.2$~meV with a very weak coupling along the $z$-direction, which are compared in Table~\ref{Tab:J1J2JcBandTheory} with the other exchange constant values discussed above. These results indicate that not only the ordered moments are different in the different magnetic phases, but the exchange coupling parameters depend on the type of magnetic order even in materials with a relatively large (3--$4\,\mu_{\rm B})$ magnetic moment.

Overall the localized Heisenberg model with four NN interactions $J_1$ and four NNN interactions $J_2$ in the $ab$-plane and two NN interactions $J_c$ along the $c$-axis is sufficient to theoretically describe the magnetic properties of BaMn$_{2}$As$_{2}$ quite well.

\section{\label{SecSWTMomentSuppr} Ordered Moment in the $J_1$-$J_2$-$J_c$ Heisenberg Model from Spin Wave Theory}

As previously noted, an ionic picture suggests that the spin of the Mn$^{2+}$ ion in ${\rm BaMn_2As_2}$
is $S = 5/2$, yielding for $g=2$ an ordered moment of
\be
\langle\mu\rangle = gS\mu_{\rm B} = 5~\mu_{\rm B} \ .
\ee
On the other hand, the observed ordered moment
\be
\langle\mu\rangle = g\langle S\rangle\mu_{\rm B}
\label{Eq:OrdMoment}
\ee
is only 3.9(1)~$\mu_{\rm B}$/Mn (Ref.~\onlinecite{YSingh2009}) implying a substantial spin reduction $\langle S\rangle$.  In view of the sizable frustrating AF next-nearest-neighbor exchange $J_2$ discussed above, it is natural to ascribe the moment reduction to enhanced quantum fluctuations. In the following, we shall use the conventional spin-wave theory to examine the quantum spin reduction for the layered $J_1$-$J_2$-$J_c$ square-lattice Heisenberg antiferromagnet.

The spin-wave theory provides an expansion of the
sublattice magnetization in powers of $1/S$:
\be
S - \langle S\rangle =   n_0 + \frac{n_1}{2S} + \frac{n_2}{(2S)^2} + \ldots \ ,
\label{SWE}
\ee
where the leading correction $n_0$ is determined by noninteracting spin waves,
while higher order corrections $n_{k\geq 1}$ come from magnon interactions.
For the nearest-neighbor square-lattice Heisenberg antiferromagnet
the two versions of the spin-wave expansion based either on the Dyson-Maleyev \cite{Hamer1992}
or the Holstein-Primakoff \cite{Igarashi1992}
representation of spin operators  yield identical results:
$n_0 = 0.19660$, $n_1 \equiv 0$,  and $n_2 = - 0.0035$, such that
the series (\ref{SWE}) rapidly converges and compares extremely well with  existing numerical
results.

Chandra and Doucot\cite{Chandra1988} used the harmonic spin-wave theory to investigate
the quantum renormalization of ordered moments for the next-nearest-neighbor $J_1$-$J_2$
square lattice Heisenberg antiferromagnet.  They found that the leading order correction
$n_0$ diverges as $J_2\to J_1/2$ due to a softening of the excitation spectrum seen in the first of Eqs.~(\ref{Eq:SWvelsPrimed}) above.
This fact is considered as an indication of a quantum spin-liquid state around the strongly 
frustrated point $J_2=J_1/2$. Subsequently, Chakravarty, Halperin,
and Nelson \cite{CHN89} calculated the next-order correction $n_1$,
which becomes finite for $J_2\neq 0$,  has an opposite sign compared to $n_0$,
and also diverges at $J_2=J_1/2$. Below, we extend the results of 
Chakravarty {\it et al.}\ to a finite
coupling between frustrated antiferromagnetic layers.
Comparison of two consecutive terms in the series (\ref{SWE}) is necessary to
judge the accuracy of the spin-wave expansion for large $J_2$.

In the spin-wave calculations for the $J_1$-$J_2$-$J_c$ model~(\ref{Eq:HamilJ1J2Jz}) with $H=0$ we use a single-rotating-sublattice
basis \cite{Chernyshev09} for the N\'eel structure with ${\bf Q}=(\pi,\pi,\pi)$ and apply
the Holstein-Primakoff transformation for spin operators expanded to first-order in $1/S$.
The quantum reduction of  ordered moments in the harmonic approximation is given
by an integral over the paramagnetic Brillouin zone
\begin{equation}
n_0 =  - \frac{1}{2} + \frac{1}{N}\sum_{\bf k} \frac{A_{\bf k}}{2\omega_{\bf k}} \ ,
\label{n0}
\end{equation}
where $N$ is the number of spins, 
\begin{eqnarray}
&& A_{\bf k} = 1 - j_2(1-\gamma_{2\bf k}) + \frac{1}{2}j_c \ , \label{Eq:AkBk} \\
&& B_{\bf k} = \gamma_{1\bf k} + \frac{1}{2}j_c \cos k_z \ , \quad
\omega_{\bf k} = \sqrt{A_{\bf k}^2-B_{\bf k}^2}
\nonumber
\end{eqnarray}
with $j_2=J_2/J_1$, $j_c=J_c/J_1$, where $\hbar\omega_{\bf k}$ is the magnon energy in units of $4J_1S$, and
\begin{equation}
\gamma_{1\bf k} = \frac{1}{2}(\cos k_x + \cos k_y) \ , \quad
\gamma_{2\bf k} = \cos k_x \cos k_y \ .
\end{equation}
Here the positive $x$ and $y$-directions are defined to be in the directions of the {\bf a} and {\bf b} primitive square-lattice translation vectors, respectively, the positive $z$-direction is perpendicular to the layers in the direction of the {\bf c} lattice translation vector, and we have set $a=b=c=1$.

To treat the effect of magnon interaction one needs to introduce various Hartree-Fock
averages of the bosonic operators compatible with the harmonic spectrum.
For the present model this procedure yields in addition to $n_0$ in Eq.~(\ref{n0})
three other integrals:
\begin{eqnarray}
&& m = \frac{1}{N}\sum_{\bf k} \frac{A_{\bf k}\gamma_{2\bf k}}{2\omega_{\bf k}} \ , \quad
\Delta_1 = \frac{1}{N}\sum_{\bf k} \frac{B_{\bf k}\gamma_{1\bf k}}{2\omega_{\bf k}} \ ,
\nonumber \\
&& \Delta_2 = \frac{1}{N}\sum_{\bf k} \frac{B_{\bf k}\cos k_z}{2\omega_{\bf k}} \ .
\label{Eq:Delta12m}
\end{eqnarray}
Then, the leading nonlinear correction
to the sublattice magnetization in Eq.~(\ref{SWE}) is expressed as
\begin{widetext}
\begin{equation}
n_1 = \frac{1}{N}\sum_{\bf k} \frac{B_{\bf k}}{\omega^3_{\bf k}} \Bigl[
j_2(m-\Delta_1)\gamma_{1\bf k}(1-\gamma_{2\bf k})
 + \frac{1}{2}j_2j_c(m-\Delta_2)
(1-\gamma_{2\bf k})\cos k_z 
+ \frac{1}{2}j_c(\Delta_1-\Delta_2)(\gamma_{1\bf k}-\cos k_z)\Bigr] \ .
\label{Eq:n1}
\end{equation}
\end{widetext}
For vanishing interlayer coupling $J_c = 0$, Eq.~(\ref{Eq:n1}) becomes
\bea
n_1 &=& j_2\Bigl[\frac{1}{N}\sum_{\bf k} \frac{\gamma_{1\bf k}^2(1-\gamma_{2\bf k})}{\omega^3_{\bf k}} \Bigr]\Bigl[\frac{1}{N}\sum_{\bf k} \frac{A_{\bf k}\gamma_{2\bf k}-\gamma_{1\bf k}^2}{2\omega_{\bf k}} \Bigr]\nonumber\\*
\omega_{\bf k} &=& \sqrt{A_{\bf k}^2-\gamma_{1\bf k}^2} \label{Eq:n1jc0}
\eea
as in Eq.~(A3) of Chakravarty {\it et al}.\cite{CHN89}  To obtain Eq.~(\ref{Eq:n1jc0}) we replaced  the factor ($m-\Delta_1$) in Eq.~(\ref{Eq:n1}) by its integral representation from Eqs.~(\ref{Eq:Delta12m}) and $B_{\bf k}$ by $\gamma_{1\bf k}$ from Eqs.~(\ref{Eq:AkBk}), and set $j_c=0$ in $A_{\bf _k}$.

\begin{table}
\caption{\label{Tab:SWTOrderedSpin}  The first-order and the second-order corrections, Eq.~(\ref{SWE}), to
the ordered moment in the $J_1$-$J_2$-$J_c$ stacked square-lattice Heisenberg antiferromagnet.}
\begin{ruledtabular}
\begin{tabular}{c|cr|cr|cr}
& \multicolumn{2}{c|}{$J_c=0$}  &  \multicolumn{2}{c|}{$J_c=0.05$} & \multicolumn{2}{c}{$J_c=0.1$}\\
$J_2/J_1$&  $n_0$& \multicolumn{1}{c|}{$n_1$}& $n_0$& \multicolumn{1}{c|}{$n_1$}     & $n_0$         & \multicolumn{1}{c}{$n_1$}   \\
\hline
0.00     &  0.1966  &   0.0000  &  0.1427        & 0.0087   &  0.1260 &   0.0088   \\
0.05     &  0.2124  & $-0.0047$ &  0.1529        & 0.0071   &  0.1347 &   0.0075   \\
0.10     &  0.2312  & $-0.0117$ &  0.1648        & 0.0045   &  0.1447 &   0.0055   \\
0.15     &  0.2542  & $-0.0227$ &  0.1790        & 0.0003   &  0.1566 &   0.0024   \\
0.20     &  0.2828  & $-0.0405$ &  0.1963        &$-0.0065$ &  0.1709 & $-0.0028$  \\
0.25     &  0.3198  & $-0.0710$ &  0.2177        &$-0.0182$ &  0.1885 & $-0.0115$  \\
0.30     &  0.3698  & $-0.1278$ &  0.2455        &$-0.0392$ &  0.2111 & $-0.0270$  \\
0.35     &  0.4423  & $-0.2485$ &  0.2832        &$-0.0809$ &  0.2414 & $-0.0573$  \\
0.40     &  0.5605  & $-0.5691$ &  0.3389        &$-0.1784$ &  0.2853 & $-0.1264$  \\
0.45     &  0.8074  & $-1.9842$ &  0.4354        &$-0.5045$ &  0.3594 & $-0.3483$  \\
0.49     &  1.6005  & $-23.899$ &  0.6257        &$-2.6777$ &  0.5008 & $-1.7482$  \\
\end{tabular}
\end{ruledtabular}
\end{table}

\begin{figure}
\centering
\includegraphics[width = 3.3in]{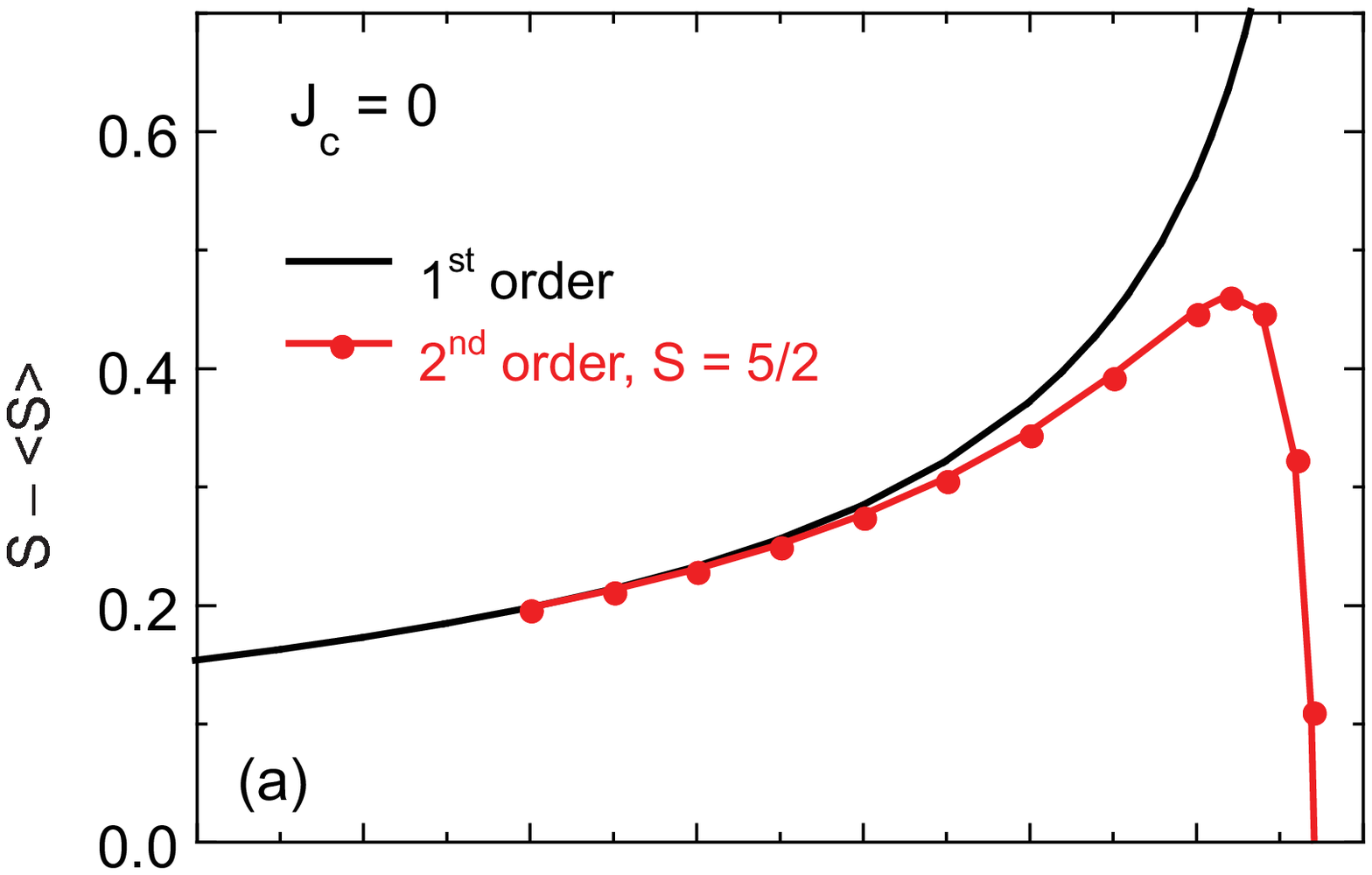}\vspace{-0.23in}
\includegraphics[width = 3.3in]{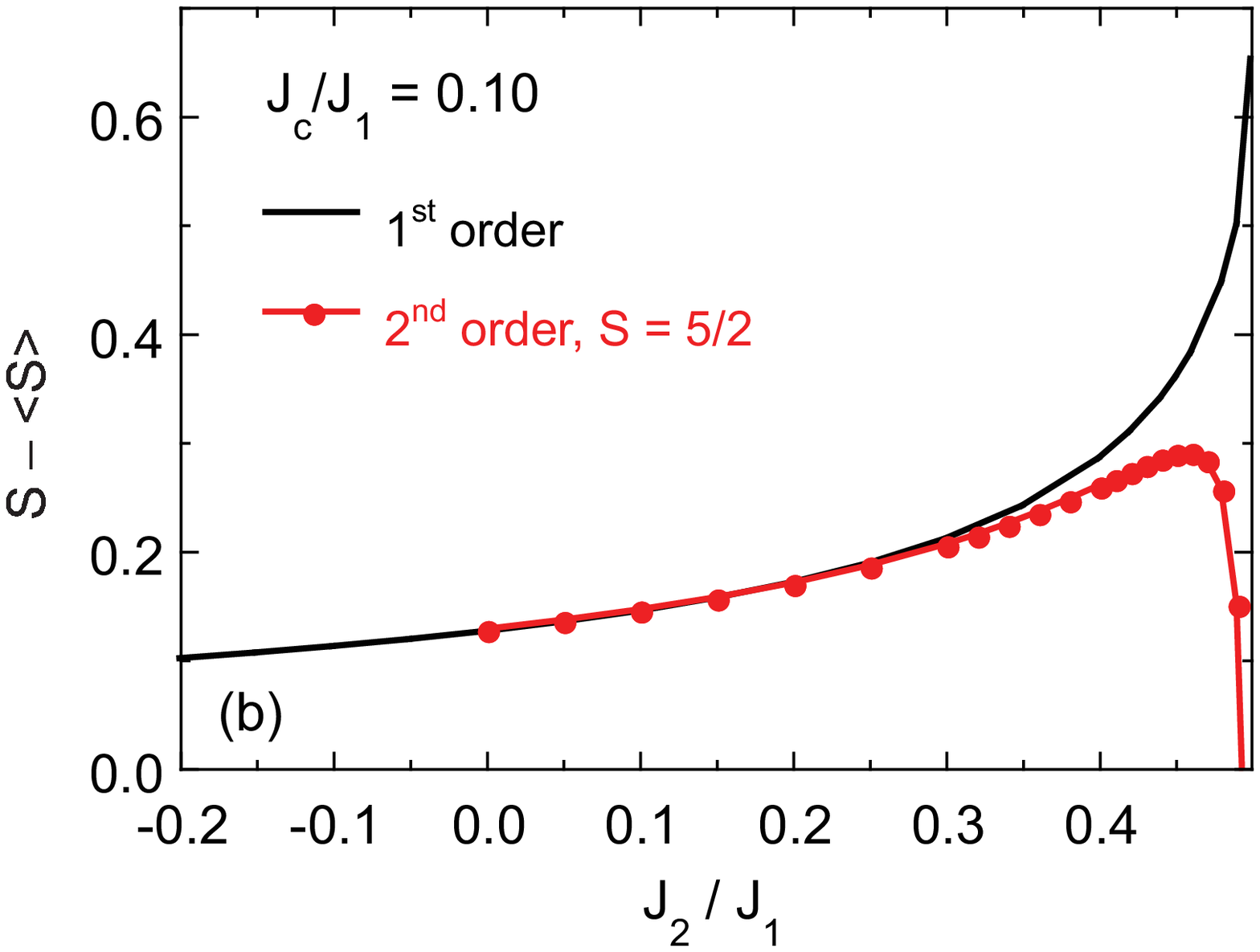}
\caption{(Color online) Reduction $S-\langle S\rangle$ in the ordered spin $\langle S\rangle$ from its value $S$ in the absence of quantum fluctuations versus $J_2/J_1$ according to linear spin wave theory to first order (solid black curves) and second order (filled circles) in $1/S$ for (a) $J_c = 0$ and (b) $J_c/J_1 = 0.1$.}
\label{Fig:MEZ_Ordered_Spin_J1-J2}
\end{figure}

Numerical values of $n_0$ and $n_1$ for a range of ratios $J_2/J_1 = 0$ to~0.49
and for $J_c/J_1 = 0$, 0.05 and 0.1 are listed
in Table~\ref{Tab:SWTOrderedSpin}. These allow one to compute the spin reduction
$S - \langle S\rangle$ for any value of $S$.
The spin reduction for  $S = 5/2$
is plotted in Fig.~\ref{Fig:MEZ_Ordered_Spin_J1-J2} for $J_c/J_2 = 0$ and~0.1.  Three-dimensional effects generally suppress quantum fluctuations, as seen in a comparison of Figs.~\ref{Fig:MEZ_Ordered_Spin_J1-J2}(a) and~\ref{Fig:MEZ_Ordered_Spin_J1-J2}(b), and extend the validity of the spin-wave expansion (\ref{SWE}) to somewhat larger values of $J_2/J_1\sim 0.44$, although the series remain divergent at $J_2/J_1\sim 0.5$.

The above fits to our experimental neutron scattering and magnetic susceptibility results for ${\rm BaMn_2As_2}$ and the band theoretical estimates of the exchange parameters in this compound yielded $J_c/J_1\approx 0.1$ and $J_2/J_1 = 0.2$--0.4.  According to the second-order calculations in Fig.~\ref{Fig:MEZ_Ordered_Spin_J1-J2}(b), together with Eq.~(\ref{Eq:OrdMoment}) with $g=2$, this parameter regime predicts an ordered moment reduction of $\approx 0.34$--0.52~$\mu_{\rm B}$/Mn due to quantum fluctuations.\cite{dS52Jc0.1}  This result appears to rule out the possibility that the spin of the Mn is $S = 2$ because the ordered moment would then be a maximum of $\sim 3.66~\mu_{\rm B}$/Mn for $g=2$, which is significantly smaller than the observed value\cite{YSingh2009} of 3.9(1)~$\mu_{\rm B}$/Mn.  On the other hand, if $S = 5/2$, then  the corresponding predicted ordered moment is $\lesssim 4.66~\mu_{\rm B}$/Mn, which is too large compared to the observed value.

It seems likely that charge and/or magnetic moment amplitude fluctuations which arise from both on-site and intersite interactions, can account for the additional reduction needed to reach agreement with the observed ordered moment for a Mn spin $S = 5/2$.  For example, the ordered moments of the Mn atoms are 3.50(4)~$\mu_{\rm B}$/Mn in ${\rm Sr_2Mn_3As_2O_2}$,\cite{Brock1996, Nath2010} 4.15(3)~$\mu_{\rm B}$/Mn in ${\rm La_2Mn_2Se_2O_3}$,\cite{Ni2010} and 4.04(8)~$\mu_{\rm B}$/Mn in ${\rm Ba_2MnMoO_6}$,\cite{Martinez-Lope2003} all containing Mn$^{+2}$ ions with nominal spin $S = 5/2$.  Such reductions are also often attributed to covalency.  On the other hand, for the more ionic compound MnF$_2$, the ordered moment of 4.82~$\mu_{\rm B}$/Mn$^{+2}$ is much closer to the  value of 5~$\mu_{\rm B}$/Mn expected for $S = 5/2$ with $g=2$,\cite{Noda2002} consistent with expectation.

\section{\label{Sec:Summary} Summary and Conclusions}

Our anisotropic magnetic susceptibility $\chi$ versus temperature $T$ measurements from 300 to 1000~K of single crystals of ${\rm BaMn_2As_2}$ yielded the N\'eel temperature $T_{\rm N} = 618$(3)~K, close to the value of 625(1)~K previously determined from neutron diffraction measurements on a polycrystalline sample.\cite{YSingh2009}  The $\chi(T)$ above $T_{\rm N}$ is nearly isotropic, indicating that single-ion anisotropy effects are small and that a Heisenberg model for the spin interactions is appropriate.  Below $T_{\rm N}$, the $\chi$ becomes strongly anisotropic, with $\chi_\perp$ nearly independent of $T$ and $\chi_\parallel$ dropping nearly to zero for $T\to0$, which corresponds qualitatively to the textbook behavior for collinear antiferromagnets in molecular field theory (MFT).  However, the temperature dependence of $\chi$ above $T_{\rm N}$ continues to increase, rather than decrease as expected from MFT, indicating the presence of strong short-range AF order above $T_{\rm N}$.  Such short-range AF order above $T_{\rm N}$ is expected for a quasi-two-dimensional spin lattice as in ${\rm BaMn_2As_2}$.  Magnetic inelastic neutron scattering measurements were carried out on a polycrystalline ${\rm BaMn_2As_2}$ sample at 8~K with momentum transfers up to 6~\AA$^{-1}$ and energy transfers up to 140~meV\@.  These data allow estimates of the magnetic exchange interactions in this compound to be made using appropriate models.  We also report $^{75}$As NMR measurements in the antiferromagnetically ordered state of a polycrystalline ${\rm BaMn_2As_2}$ sample from 4 to 300~K\@.  The nuclear spin-lattice relaxation rate is found to obey the power law dependence $1/T_1 \propto T^3$ from 50 to~300~K which we interpret in terms of the exchange interactions in this compound.

We developed various theories for the $J_1$-$J_2$-$J_c$ Heisenberg model in order to model our experimental data and extract values of the exchange constants between Mn spins and the value of the spin.  Our inelastic neutron scattering measurements indicate that this is the minimal model needed to understand these data.  For G-type antiferromagnetic ordering shown for the square spin lattice in the top panel of Fig.~\ref{Fig:Magnetic_Structures}, and where the $c$-axis alignment is also antiferromagnetic, linear spin wave theory  at $T\ll T_{\rm N}$ was used to calculate the magnon dispersion relations in Sec.~\ref{Sec:SWT}.  The in-plane spin waves soften as $J_2$ increases, and become unstable for $J_2\geq J_1/2$, signaling a phase transition to the in-plane stripe state shown in the bottom panel of Fig.~\ref{Fig:Magnetic_Structures}.  Thus the G-type AF ordered state requires $J_2<J_1/2$.  This theory is used in Sec.~\ref{Sec:NeutIntensity} to fit our inelastic magnetic neutron scattering data at 8~K, a temperature far below the N\'eel temperature of $\approx 625$~K, and obtain estimates of $SJ_1$, $SJ_2$ and $SJ_c$ for ${\rm BaMn_2As_2}$, where $S$ is the spin on the Mn atoms that is not determined separately in the spin wave fit to the data.  From the ordered moment $\mu =3.9(1)~\mu_{\rm B} = gS\mu_{\rm B}$, one would estimate $S=2$ for $g=2$.  On the other hand for the $d^5$ ion Mn$^{+2}$ one would estimate a high-spin $S = 5/2$.  In Sec.~\ref{Sec:SWTCmagTheory} we also calculated the spin wave contribution to the low-temperature heat capacity for comparison in Sec.~\ref{Sec:SWTCmagBaMn2As2} with our previously published\cite{singh2009} experimental heat capacity data for a single crystal of ${\rm BaMn_2As_2}$.  We also used spin wave theory to extend the nuclear spin-lattice relaxation rate $1/T_1$ calculations of Beeman and Pincus for the isotropic cubic Heisenberg spin lattice\cite{beeman1968} to the $J_1$-$J_2$-$J_c$ model.  We obtained Eqs.~(\ref{1/t1fab}) and~(\ref{1/t1fc}) that were used to analyze the fit to our $^{75}$As $1/T_1$ NMR data with $1/T_1 \propto T^3$ from 50 to 300~K for ${\rm BaMn_2As_2}$.

A molecular field theory (MFT) treatment of the $J_1$-$J_2$-$J_c$ Heisenberg model was described in Sec.~\ref{Sec:MFTJ1J2J3}.  In the paramagnetic state the system follows the Curie-Weiss law $C/(T+\theta)$ for $T\geq\theta$, which has the same form as described in many textbooks for the $J_1$-only model. The ratio $f$ of the Weiss temperature $\theta$ to $T_{\rm N}$, $f = \theta/T_{\rm N}$, is found to be $f=1$ for $J_2 = 0$, as expected for a bipartite spin lattice, but is $f>1$ for frustrating AF $J_2>0$ and is $f<1$ for nonfrustrating reinforcing FM $J_2 < 0 $, which are intrasublattice interactions so the spin lattice is no longer bipartite.  Thus for $J_2 > 0$, the Curie-Weiss law continues to be followed below $T = \theta$ down to $T = T_{\rm N}$, a characteristic already noted by Ramirez for geometrically frustrated antiferromagnets.\cite{Ramirez1994}  As shown in Fig.~\ref{Fig:Chi_ParaJ1J2MFT_S1_2}, we find that $\chi_\parallel(T)$ for $T<T_{\rm N}$ strongly depends on $J_2$, whereas $\chi_\perp$ is independent of $T$ and $J_2$ at $T<T_{\rm N}$, apart from the implicit influence of $J_2$ on $T_{\rm N}$.  We further find that the staggered moment and the magnetic heat capacity versus $T/T_{\rm N}$ at $T<T_{\rm N}$ are also independent of $J_2$, again apart from the implicit influence of $J_2$ on $T_{\rm N}$.  

We carried out quantum (QMC) and classical Monte Carlo (CMC) simulations of both $\chi(T)$ and the magnetic heat capacity $C_{\rm mag}(T)$ in $H=0$ versus $J_c/J_1$ and $J_2/J_1$.  Most of the QMC simulations were for $J_2=0$ due to severe negative sign problems when $J_2$ was taken to be positive, which is antiferromagnetic and frustrating for G-type AF order.  When we replaced the square of the spin, $S^2$, in the CMC simulations by the quantum mechanical expectation value $S(S+1)$, the QMC simulations for $J_2=0$ for increasing $S$ merged smoothly with the CMC simulation (which corresponds to $S\to\infty$) as shown in Fig.~\ref{square_chi_J2}, so we used the CMC simulations to fit the experimental $\chi(T)$ data for $T > T_{\rm N}$.  The CMC simulations of $C_{\rm mag}(T)$ as a function of $J_c$ showed AF phase transitions at temperatures $T_{\rm N}$ that increased with $J_c>0$ but decreased with $J_2>0$, as shown in Fig.~\ref{Fig:CMC_TN_vs_JcFit}.  The $T_{\rm N}(J_c,J_2)$ data are well-fitted by Eq.~(\ref{Eq:FitTN(JcJ2)}).

We also carried out band theoretical estimates of the exchange couplings in ${\rm BaMn_2As_2}$.  There are two generic ways to do this.  The first is to take the differences between the total energies for different spin configurations such as in Eqs.~(\ref{Eq:JcfromSingh}--\ref{Eq:FindJ2}), where the lowest energy spin configuration is the ground state.  This method is often used to determine exchange constants to be used in the calculation of magnetic transition temperatures and yields the exchange constants in rows~5 and~6 of Table~\ref{Tab:J1J2JcBandTheory} which will not be further discussed.  The second is to measure the change in the total energy due to small deviations of the spin directions from the magnetically ordered ground state via Eq.~(\ref{Eq:Jijfrompartials}), which gives the exchange values in rows~9 and~10 of Table~\ref{Tab:J1J2JcBandTheory}.  These values are considered to be more reliable for comparison with values extracted from inelastic neutron scattering experiments.

Our exchange constants from the MFT fit to our anisotropic $\chi(T)$ data for ${\rm BaMn_2As_2}$ below $T_{\rm N}$ in Fig.~\ref{Fig:BaMn2As2_Hi_T_chi_All_Fit2} and Table~\ref{Tab:FitHighTChi} are probably not reliable because that fit assumes that $\chi(T)$ follows the Curie-Weiss law above $T_{\rm N}$ (i.e., that there are no two-spin AF correlations stronger than $1/T$ above $T_{\rm N}$), which is strongly violated by the data in Fig.~\ref{BaMn2As2_Hi_T_chi}(a).  Similarly, although our fits by MFT to the ordered moment $\bar{\mu}_z(T)$ from neutron diffraction measurements\cite{YSingh2009} on ${\rm BaMn_2As_2}$ in Fig.~\ref{Fig:BaMn2As2_ord_mom_Fit} and to the heat capacity $C_{\rm p}(T)$ (Ref.~\onlinecite{singh2009}) in Fig.~\ref{Fig:C(T)-BaMn2As2-Xtal} are reasonable, they are not sufficient to distinguish between the possible spins $S = 2$ and $S = 5/2$ discussed above.  Furthermore, the fit to the $C_{\rm p}(T)$ data near room temperature by the sum of the Debye lattice heat capacity and the MFT prediction of the magnetic heat capacity indicated that the measured magnetic heat capacity is too low.  This discrepancy suggests the presence of strong AF fluctuations above $T_{\rm N}$ that reduce the magnetic entropy and magnetic heat capacity below $T_{\rm N}$, consistent with the behavior of the magnetic susceptibility above $T_{\rm N}$ in Fig.~\ref{Fig:BaMn2As2_Hi_T_chi_All_Fit2}.  The calculated $T^3$ contribution to the heat capacity at low temperatures from spin waves, without an anisotropy gap in the spin wave spectrum, is about 40\% of the measured value.  However, an anisotropy gap would suppress the spin wave contribution exponentially to zero at low temperatures.

\begin{table}
\caption{\label{Tab:J1J2JcFitSummary}  Summary of our most reliable exchange constants in ${\rm BaMn_2As_2}$ obtained from band theory and from fitting our experimental data by our predictions of the $J_1$-$J_2$-$J_c$ Heisenberg model.  The notation ``$\equiv$'' means that the value that follows it was assumed, not fitted.  Two spin values for the neutron fit are listed because the neutron fit gives the product of the spin $S$ and the respective exchange constant, not the two separately.  For the $\chi$ fit, a range of spin values can fit the data.  We chose two spin values close to 2 and 5/2, corresponding to $J_2/J_1 = 0.2$ and 0.4, respectively.   Two band theory estimates were obtained using the LDA and the GGA, as noted.}
\begin{ruledtabular}
\begin{tabular}{l|cccc}
Data & $S$ & $J_1$ & $J_2/J_1$ & $J_c/J_1$\\
    && (meV) \\
\hline
neutrons & 2 & $16.5\pm1.5$ & $0.29\pm0.05$ & $0.09\pm0.02$ \\
 & 5/2 & $13.2\pm1.2$ & $0.29\pm0.05$ & $0.09\pm0.02$ \\
$\chi(T>T_{\rm N})$ & 2.06 & 17.8 & $\equiv0.2$ & $\equiv0.1$  \\
 & 2.64 & 18.1 & $\equiv 0.4$ & $\equiv0.1$  \\
$^{75}$As NMR & $\equiv2$ & 14 & $\equiv0.29$ & $\equiv0.09$  \\
 \hspace{0.08in}  ($T\ll T_{\rm N}$)\\
band theory \\
\hspace{0.2in}LDA & $1.65$  & 13.1 & 0.21 & 0.08\\
\hspace{0.2in}GGA & 1.8 & 12.6 & 0.21 & 0.08\\
\end{tabular}
\end{ruledtabular}
\end{table}

We have gathered together in Table~\ref{Tab:J1J2JcFitSummary} our most reliable exchange constants in ${\rm BaMn_2As_2}$ from our band theory calculations and from the theoretical fits to our experimental data by the $J_1$-$J_2$-$J_c$ Heisenberg model.  Several features are noteworthy.  First, all three exchange constants are consistently positive (antiferromagnetic).  Second, the estimates give similar values of $J_1 \approx 13$--18~meV for $S$ values in the range from 2 to 5/2.  For the susceptibility fits, nearly the same $J_1$ was obtained from the two fits despite the significant differences between the respective $S$ and $J_2/J_1$ values.  Third, the estimates of $J_2/J_1$ from band theory and from analysis of the neutron and magnetic susceptibility measurements are in the range 0.2--0.4, which are below the value of 0.5 at which the in-plane G-type AF order would classically become unstable with respect to the stripe AF order [see Fig.~\ref{Fig:Magnetic_Structures} and Eq.~(\ref{Eq:JRestrictions})], and are therefore consistent with the observed G-type AF order.  From our classical Monte Carlo simulations of the heat capacity of stacked square lattice layers, the exchange parameters from the neutron scattering fit predict $T_{\rm N}\approx 640$~K if the Mn spin is $S = 5/2$, in close agreement with the experimental value of $\approx 625$~K\@.  

Finally, with the above range of exchange parameters, our second-order spin wave calculations in Sec.~\ref{SecSWTMomentSuppr} show that the ordered moment reduction due to quantum fluctuations alone is at least \mbox{$\sim 0.4~\mu_{\rm B}$/Mn}.  Because the measured ordered moment is $\langle\mu\rangle = 3.9(1)~\mu_{\rm B}$/Mn, this argues against assigning a spin $S=2$ to the Mn$^{+2}$ ions which gives $\langle\mu\rangle = gS\mu_{\rm B} = 4~\mu_{\rm B}$/Mn and favors $S = 5/2$ for which one would obtain $\langle\mu\rangle = 5~\mu_{\rm B}$/Mn for $g=2$ in the absence of quantum fluctuations.  The additional reduction needed to reach the experimental value is likely due to charge and/or magnetic moment amplitude fluctuations which arise from both on-site and intersite interactions, and/or from hybridization effects, consistent with the reduced ordered moment measured for other Mn$^{+2}$ compounds.\cite{Brock1996, Nath2010, Ni2010, Martinez-Lope2003}  This effect cannot be described in the Heisenberg model formalism used in this paper.  For instance, in Sec.~\ref{Sec:BandTheory} we discussed that despite the large moment of the Mn, the magnitude of the magnetic moment can vary by $\sim 20$\% depending on the specific magnetic configuration, suggesting that amplitude fluctuations of the magnetic moment may indeed be relevant.

As noted in the introduction, the view that ${\rm BaFe_2As_2}$ is an itinerant antiferromagnet is not universally held.  Furthermore, the results of many magnetic inelastic neutron scattering measurements on ${\rm BaFe_2As_2}$ have been analyzed in terms of local moment Heisenberg models, even when the authors of this modeling believe that the itinerant model is valid. The reason for this latter analysis, as has been stressed in the literature, is that the magnetism of itinerant models can often be parametrized by local moment Heisenberg models.  Further review and discussion of this issue is given in Ref.~\onlinecite{Johnston2010}.  

We therefore now compare the exchange constants in ${\rm BaMn_2As_2}$ with those in the isostructural (at room temperature) high-$T_{\rm c}$ $A{\rm Fe_2As_2}$ parent compounds ($A$ = Ca, Sr, Ba) within the context of the $J_1$-$J_2$-$J_c$ local moment Heisenberg model.  The $A{\rm Fe_2As_2}$ compounds order into an in-plane stripe-type antiferromagnetic structure below $\sim 200$~K (lower panel of Fig.~\ref{Fig:Magnetic_Structures}) and the lattice distorts to orthorhombic symmetry at or above $T_{\rm N}$.\cite{Johnston2010}  Within the orthorhombic structure and assuming $S = 1/2$ and $gS\mu_{\rm B} \approx 1~\mu_{\rm B}$, one defines the average $\langle J_1\rangle = (J_{1a} + J_{1b})/2$, yielding $\langle J_1\rangle/ J_2 = 0.7$--1.4 for a variety of $A{\rm Fe_2As_2}$ compounds,\cite{Johnston2010} which is in the regime $\langle J_1\rangle/J_2 < 2$ expected for in-plane stripe-type ordering [see Eq.~(\ref{Eq:JRestrictions})].  These values can be compared with those for ${\rm BaMn_2As_2}$ in Table~\ref{Tab:J1J2JcFitSummary} where $J_1/ J_2 \sim 3$.  In ${\rm BaMn_2As_2}$ and also in the $A{\rm Fe_2As_2}$ compounds, the interlayer coupling $J_c$ is weak compared to the in-plane couplings and the systems should be considered to have strongly spatially anisotropic exchange, but not necessarily two-dimensional.  Thus, the $AT_2$As$_2$ systems, where $T$ is a 3$d$-transition metal element, appear to be ideal systems to study the physics of the $J_1$-$J_2$-$J_c$ Heisenberg model, including the possibility of tuning the system to and through the quantum critical point $J_1/ J_2 = 2$ by doping.  Indeed, doping-dependent studies of Ba(Fe$_{1-x}$Cr$_x)_2$As$_2$ have recently revealed a transition from stripe-type AF order to G-type AF order at $x \approx 0.3$,\cite{Marty2011} and studies of Ba(Fe$_{1-x}$Mn$_x)_2$As$_2$ have revealed a transition to a new state, possibly due to competition between G-type and stripe-type AF ordering, at $x \approx 0.1$.\cite{Kim2010}

\acknowledgments

We are grateful to Andreas Kreyssig for insights about spin wave theory and to Ferenc Niedermayer for confirming the calculation of $\chi(T)$ (Ref.~\onlinecite{Hasenfratz1993}) in Eq.~(\ref{HN}).\cite{Hasenfratz2011}  Work at the Ames Laboratory was supported by the Department of Energy-Basic Energy Sciences under Contract No.~DE-AC02-07CH11358.  R.J.M. would like to thank F.~Trouw for assistance with Pharos.  The work has benefitted from the use of the Los Alamos Neutron Science Center (LANSE) at Los Alamos National Laboratory. LANSCE is funded by the U.S. Department of Energy under Contract No.~W-7405-ENG-36.  The work of A.H. was supported by the Deutsche Forschungsgemeinschaft through a Heisenberg Fellowship (Grant No.~HO~2325/4-2).  M.E.Z. acknowledges Grant No.~ANR-09-Blanc-0211 SupraTetrafer from the French National Research Agency.


\appendix

\section{\label{HTSE} High-Temperature Series Expansions}

From quite general considerations, one can show that the results of MFT at high temperatures for the magnetic susceptibility (the Curie-Weiss Law) is an exact result arising from a quantum mechanical treatment of local moment magnetism using a high-temperature series expansion HTSE\@.  We also discuss a complementary ``Curie-Weiss Law'' for the magnetic heat capacity at high temperatures, which is useful when discussing our Monte Carlo simulations of the magnetic heat capacity in Sec.~\ref{SecThy}.

\subsection{Magnetic Susceptibility and the Curie-Weiss Law}

Using the fluctuation-dissipation theorem, one can express the diagonal $\alpha\alpha$ components $\chi_{\alpha}\ (\alpha = x,y,z)$ of the magnetic susceptibility tensor in terms of the two-spin correlation functions 
\be
\Gamma_{\bf r}^\alpha \equiv \langle S_{\bf 0}^\alpha S_{\bf r}^\alpha \rangle
\label{CorrFcn}
\ee
where {\bf r} is the distance measured in the number $n$ of bonds, including zero, of spin $S_{\bf r}$ from a typical central spin $S_{\bf 0}$.\cite{Johnston1997, Fisher1962} In the isotropic Heisenberg model, one obtains\cite{Fisher1962}
\be
\chi = \frac{Ng^2\mu_{\rm B}^2}{k_{\rm B}T} \sum_{\bf r}\Gamma_{\bf r}^z.
\label{ChiFromCorr}
\ee
If one only considers the single-spin autocorrelation function ({\bf r} = 0), then one has $\Gamma_0^z = \langle S_z^2\rangle = \langle S^2\rangle/3 = S(S+1)/3$ which gives the Curie law
\be
\chi = \frac{Ng^2\mu_{\rm B}^2S(S+1)}{3k_{\rm B}T},
\label{ChiFromCorr2}
\ee
which in turn is the Curie-Weiss law~(\ref{EqCurieWeiss}) with $\theta = 0$ and Curie constant~(\ref{CC}).

Writing $r \equiv n = 0$, 1, 2, ..., in terms of the distance of a spin in number of bonds from the central spin at position~{\bf 0} (i.e., $n$ means the $n^{\rm th}$-nearest-neighbor of the spin at the origin in terms of the minimum number of bonds between them), one can express the temperature dependences of the two-spin correlation functions as high-temperature series expansions in $1/T$ with the general form
\be
\Gamma_n^z = \frac{\Gamma_{n,n}^z}{(k_{\rm B}T/J)^n} +  \frac{\Gamma_{n,n+1}^z}{(k_{\rm B}T/J)^{n+1}} + \cdots,
\label{GammaExpand}
\ee
where $\Gamma_0^z = S(S+1)/3$ as noted above and the lowest-order term for a given $\Gamma_n^z$ is $1/T^n$.\cite{Fisher1962}  Substituting the first three terms of Eq.~(\ref{GammaExpand}) into Eq.~(\ref{ChiFromCorr}) gives
\bea
\chi &=& \frac{Ng^2\mu_{\rm B}^2}{k_{\rm B}T} \left[\frac{S(S+1)}{3} + \frac{\Gamma_{1,1}^z}{k_{\rm B}T/J} + \frac{\Gamma_{1,2}^z + \Gamma_{2,2}^z}{(k_{\rm B}T/J)^2} + \cdots\right]\nonumber\\*
&=&\frac{C}{T} \Bigg[1 +\nonumber\\*
&&+ \frac{3}{S(S+1)}\left(\frac{\Gamma_{1,1}^z}{k_{\rm B}T/J} + \frac{\Gamma_{1,2}^z + \Gamma_{2,2}^z}{(k_{\rm B}T/J)^2} + \cdots\right)\Bigg],
\label{ChiFromCorr3}
\eea
where the Curie constant $C$ is the same as in Eq.~(\ref{CC}).  If one keeps only the first two terms in  the square brackets and uses the Taylor series expansion $1 + x \approx 1/(1 - x)$ for small $x$ to put the quantity in square brackets into the denominator, one gets the Curie-Weiss law~(\ref{EqCurieWeiss}) with Weiss temperature
\be
\theta = -\frac{3\Gamma_{1,1}^zJ}{S(S+1)k_{\rm B}  }.
\label{ThetaGamma}
\ee
Comparing Eqs.~(\ref{WT}) and~(\ref{ThetaGamma}) gives the coefficient
\be
\Gamma_{1,1}^z = -\frac{z[S(S+1)]^2}{9}.
\label{EqG11}
\ee

\subsection{The HTSE for the Magnetic Heat Capacity}

We discussed above that the Curie-Weiss law for the magnetic susceptibility of equivalent spins is rigorously derived from the  first ($1/T$) term in the HTSE of the nearest-neighbor two-spin correlation function and hence does not depend on the particular crystal structure or spin lattice dimensionality.  This suggests that there is an analogous term in the HTSE of the magnetic heat capacity $C_{\rm mag}$.  We show this to be the case [Eq.~(\ref{Eq:HTSECmag})] and apply the result in Sec.~\ref{SecThy}.

From Hamiltonian~(\ref{Eq:NNHeisHam}), the thermal-average magnetic configuration energy in zero field only depends on the nearest-neighbor two-spin correlation function as\cite{Fisher1962}
\be
E_{\rm mag}(T) = \frac{1}{2}NzJ\langle {\bf S}_0\cdot {\bf S}_1\rangle_{\rm T}(T),
\label{Eq:EmagCorrFcn}
\ee
where the factor of 1/2 is introduced to avoid double counting the distinct AF NN bonds and $\langle \cdots\rangle_{\rm T}$ denotes a thermal average of the quantum mechanical expectation value.  The magnetic specific heat $C_{\rm mag}(T)$ is obtained by differentiating Eq.~(\ref{Eq:EmagCorrFcn}) with respect to $T$.  Thus the first $1/T$ HTSE term of $E_{\rm mag}(T)$ gives the first term in the HTSE of $C_{\rm mag}(T)$ as a $1/T^2$ term.

Rushbrooke and Wood showed that the first six terms of the HTSE of $\chi(T)$ and $C_{\rm mag}(T)$ of a Heisenberg spin lattice containing equivalent spins $S$ can be expressed in terms of the identical NN exchange couplings $J$ and the bond connectivity (``lattice parameters'') of the specific spin lattice.\cite{Rushbrooke1958}  With respect to the present discussion, they found that the first ($1/T^2$) term of the HTSE for $C_{\rm mag}(T)$ is independent of the type of spin lattice and of the spin lattice dimensionality and only depends on $z$, $S$, and $J$ according to\cite{Rushbrooke1958}
\be
\frac{C_{\rm mag}}{R} = \frac{z}{6}\left[\frac{JS(S+1)}{k_{\rm B}T}\right]^2.
\label{Eq:HTSECmag}
\ee
This term is the same for FM and AF interactions because the exchange constant is squared, which gives a positive-definite result for $C_{\rm mag}(T)$.  Thus when comparing calculated $C_{\rm mag}(T)$ data for lattices with the same coordination number $z$ but different exchange constants and/or spins, a universal high-temperature behavior is obtained if the data are scaled according to
\be
\frac{C_{\rm mag}}{R} \ \ \ {\rm versus}\ \ \  \frac{k_{\rm B}T}{JS(S+1)}.
\label{Eq:CmagScaling}
\ee

The Curie law for the magnetic susceptibility arises because there is a nonzero susceptibility for isolated spins.  This is modified at high temperatures by the addition of a $1/T^2$ term in Eq.~(\ref{ChiFromCorr3}) arising from spin interactions, yielding a Curie-Weiss law with nonzero Weiss temperature $\theta$.  On the other hand, the magnetic heat capacity of isolated spins is zero, and hence there is no equivalent Curie law for $E_{\rm mag}(T)$ or $C_{\rm mag}(T)$: the values are just zero.  Equation~(\ref{Eq:HTSECmag}) can therefore be considered to be a ``Curie-Weiss law'' for $C_{\rm mag}(T)$.

\section{\label{Eq:AnisChi} Anisotropic Susceptibility below $T_{\rm N}$ for the $J_2$-$J_2$-$J_c$ Model in Molecular Field Theory}

\subsection{Perpendicular Susceptibility $\chi_\perp$ below $T_{\rm N}$}

\begin{figure}
\includegraphics [width=1.15in]{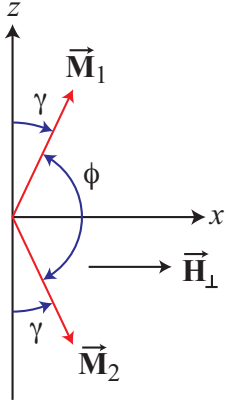}
\caption{(Color online) Influence of a perpendicular magnetic field ${\bf H}_\perp$ on the sublattice magnetizations of an ordered antiferromagnet.  The ${\bf H}_\perp$ tilts the ordered sublattice magnetizations ${\bf M}_1$ and ${\bf M}_2$ that are initially pointed along the $z$-axis by an angle $\gamma$ towards the applied field ${\bf H}_\perp$ along the $x$-axis.  The angle $\gamma$ is greatly exaggerated for clarity.}
\label{Fig:AF_Low_T_Perpendicular_Chi}
\end{figure}

Setting the external field $H$ to zero and using the exchange fields in Eq.~(\ref{Eq:H12exch}), the average exchange energy of the spin system is 
\be
E_{\rm exch} = -{\bf M}_1\cdot {\bf B_1} = -(\lambda_{\rm s}M_1^2 + \lambda_{\rm d}{\bf M}_1\cdot{\bf M}_2).
\label{Eq:Eexch}
\ee
We do not add a second term $-{\bf M}_2\cdot {\bf B_2}$ to this, because that would double-count the exchange interactions between the spins which occur \emph{pairwise}.  

We apply an external magnetic field ${\bf H}_\perp$ that is perpendicular to the ordered moment direction in the antiferromagnetically ordered state that induces a perpendicular magnetization ${\bf M}_\perp$ in the sample, where ${\bf M}_\perp = {\bf M}_1 + {\bf M}_2$ is the vector sum of the sublattice magnetizations.  From Fig.~\ref{Fig:AF_Low_T_Perpendicular_Chi}, the angle $\phi$ between ${\bf M}_1$ and ${\bf M}_2$ is
\[
\cos\phi = \cos(180^\circ -2\gamma) = -\cos(2\gamma) \approx -\left(1-2\gamma^2\right),
\] 
where we used $\cos(A-B)=\cos A\cos B + \sin A\sin B$ and on the right-hand side we used the small angle approximation $\cos x\approx 1-x^2/2$.  We assume that $\gamma$ is very small because $\chi_\perp = \lim_{H_\perp\to0}M_\perp/H_\perp$ by definition. Thus the exchange energy in Eq.~(\ref{Eq:Eexch}) becomes
\be
E_{\rm exch} = M_i^2[-\lambda_{\rm s} + \lambda_{\rm d}(1-2\gamma^2)],
\label{Eq:Eexch2}
\ee
where $M_1=M_2\equiv M_i$.

A perpendicular magnetic field ${\bf H}_\perp = H_\perp\hat{\bf i}$ causes the ordered AF spin sublattices to tilt towards the applied field direction, away from the ordered moment $z$-direction, as shown in Fig.~\ref{Fig:AF_Low_T_Perpendicular_Chi}.  The magnetic energy due to the perpendicular field is
\be
E_\perp = -2{\bf M}_i\cdot{\bf H}_\perp = -2M_iH_\perp\sin\gamma \approx -2M_iH_\perp\gamma,
\ee
where we have used the small-angle approximation $\sin x\approx x$.  Thus the total magnetic energy is
\be
E= E_{\rm exch}+E_\perp = M_i^2[-\lambda_{\rm s} + \lambda_{\rm d}(1-2\gamma^2)] -2M_iH_\perp\gamma.
\ee
The stable configuration minimizes the energy.  Taking the derivative of $E$ with respect to $\gamma$ and setting it to zero and setting $\lambda_{\rm d} = -|\lambda_{\rm d}|$ because $\lambda_{\rm d}$ is negative gives
\[
\gamma = \frac{H_\perp}{2|\lambda_{\rm d}|M_i}.
\]
Thus the interactions within the same sublattice ($\lambda_{\rm s}$, i.e., $J_2$) have no influence on this equilibrium condition.  The equilibrium value of the component $M_\perp$ of the total magnetization in the direction of ${\bf H}_\perp$, using the small-angle approximation $\sin\theta\approx\theta$,  is
\[
M_\perp = 2M_i\sin\gamma \approx 2M_i\gamma = \frac{H_\perp}{|\lambda_{\rm d}|},
\]
which gives the perpendicular susceptibility as 
\[
\chi_\perp = \frac{M_\perp}{H_\perp} = \frac{1}{|\lambda_{\rm d}|}.
\]
Note that $M_i$ and hence also its temperature dependence have dropped out, so that $\chi_\perp$ in this treatment is independent of $T$ below $T_{\rm N}$.  

From Eqs.~(\ref{Eq:lambdasdTN}) we have 
\be
|\lambda_{\rm d}| = \frac{T_{\rm N}(1+f)}{C}.
\ee
Thus we obtain
\[
\chi_\perp = \frac{1}{|\lambda_{\rm d}|} = \frac{C}{T_{\rm N}(1+f)}  = \frac{C}{T_{\rm N}+\theta}=\chi(T_{\rm N}), \hspace{0.1in}(T\leq T_{\rm N})
\]
which is Eq.~(\ref{Eq:ChiPerp}) in the text.  Although $J_2$ is not present explicitly, its influence is expressed through the implicit dependence of $T_{\rm N}$ on $J_2$.

\subsection{Parallel Susceptibility $\chi_\parallel$ below $T_{\rm N}$}

Here again we assume that the susceptibility in the absence of explicit exchange couplings is $\chi_0=C/T$, which is isotropic above $T_{\rm N}$.  We apply a small magnetic field $H$.  Below $T_{\rm N}$ a large exchange field develops as seen by each sublattice because of the ordered moments.  Therefore we must use a Brillouin function to describe the magnetization of each sublattice.  

The saturation magnetic moment of a spin $S$ is
\[\mu_{\rm sat} = g\mu_{\rm B}S.
\]
For $N$ spins, the saturation magnetization is therefore
\be
M_{\rm sat} = NgS\mu_{\rm B}.
\label{Eq:MsatGenS}
\ee
The magnetization of the $N$~spins is written
\be
M_z = M_{\rm sat}B_S(y)
\label{Eq:MzBrill}
\ee
where $B_S(y)$ is the Brillouin function given by
\be
B_S(y) = \frac{1}{2S} \left\{(2S+1)\coth\left[(2S+1)\frac{y}{2}\right]-\coth\left(\frac{y}{2}\right)\right\},
\label{Eq:BrillouinFunction}
\ee
where $0\leq B_S(y) \leq 1$ for $y\geq0$,
\be
y = \frac{g\mu_{\rm B}H}{k_{\rm B}T}
\ee
and the $g$-factor is usually set to the value $g=2$.

In MFT, we replace $H$ in the Brillouin function by the magnetic inductions $B_i$ in Eqs.~(\ref{Eq:B1,2}) and~(\ref{Eq:H12exch}), which include the exchange fields.  Thus we have
\bea
M_1 &=& \frac{1}{2}M_{\rm sat}B_S\left(\frac{g\mu_{\rm B}B_1}{k_{\rm B}T}\right) \nonumber\\*
&=& \frac{1}{2}M_{\rm sat}B_S\left[\frac{g\mu_{\rm B}(H + \lambda_{\rm s}M_1 + \lambda_{\rm d}M_2)}{k_{\rm B}T}\right]\nonumber\\*
\label{Eq:SublatticeMs}\\*
M_2 &=& \frac{1}{2}M_{\rm sat}B_S\left(\frac{g\mu_{\rm B}B_2}{k_{\rm B}T}\right) \nonumber\\*
&=& \frac{1}{2}M_{\rm sat}B_S\left[\frac{g\mu_{\rm B}(H + \lambda_{\rm d}M_1 + \lambda_{\rm s}M_2)}{k_{\rm B}T}\right]\nonumber
\eea
Substituting for $\lambda_{\rm s}$ and $\lambda_{\rm d}$ from Eqs.~(\ref{Eq:lambdasdTN}) gives
\bea
\frac{M_1}{M_{\rm sat}} &=& \frac{1}{2}B_S\left\{\frac{g\mu_{\rm B}\left[H -\frac{T_{\rm N}(f-1)}{C}M_1 -\frac{T_{\rm N}(f+1)}{C}M_2\right]}{k_{\rm B}T}\right\}\nonumber\\*
\label{Eq:SublatticeMs2}\\*
\frac{M_2}{M_{\rm sat}} &=& \frac{1}{2}B_S\left\{\frac{g\mu_{\rm B}\left[H -\frac{T_{\rm N}(f+1)}{C}M_1 -\frac{T_{\rm N}(f-1)}{C}M_2\right]}{k_{\rm B}T}\right\}.\nonumber
\eea

To simplify the notation and the solution to Eqs.~(\ref{Eq:SublatticeMs2}) we define reduced magnetic field, temperature and ordered moment variables, respectively, by
\bea
\tilde{h} &=& \frac{g\mu_{\rm B}H}{k_{\rm B}T_{\rm N}(H=0)}\nonumber\\*
t &=& \frac{T}{T_{\rm N}(H=0)}\label{Eq:RedVar}\\*
\bar{\mu}_{iz} &=& \frac{\mu_{iz}}{\mu_{\rm sat}}= \frac{\mu_{iz}}{gS\mu_{\rm B}}.\nonumber
\eea
Then using $M_i = N\mu_{iz}/2$ and $M_{\rm sat} = N\mu_{\rm sat} = NgS\mu_{\rm B}$ and the expression for the Curie constant $C$ in Eq.~(\ref{CC}), Eqs.~(\ref{Eq:SublatticeMs2}) become
\bea
\bar{\mu}_{1z} &=& B_S\left[\frac{\tilde{h}}{t}-\frac{3(f-1)}{2(S+1)}\frac{\bar{\mu}_{1z}}{t}-\frac{3(f+1)}{2(S+1)}\frac{\bar{\mu}_{2z}}{t}\right]\label{Eq:mu1zmu2zMFT}\\*
\bar{\mu}_{2z} &=& B_S\left[\frac{\tilde{h}}{t}-\frac{3(f+1)}{2(S+1)}\frac{\bar{\mu}_{1z}}{t}-\frac{3(f-1)}{2(S+1)}\frac{\bar{\mu}_{2z}}{t}\right].\nonumber
\eea
For specified $S$, $f$, $\tilde{h}$ and $t$, one can solve these two simultaneous equations numerically for $\bar{\mu}_{1z}(t,\tilde{h})$ and $\bar{\mu}_{2z}(t,\tilde{h})$.  The average reduced magnetization per spin is 
\be
\bar{\mu}_z(t,\tilde{h}) = \frac{1}{2}[\bar{\mu}_{1z}(t,\tilde{h}) + \bar{\mu}_{2z}(t,\tilde{h})].
\label{Eq:barmuparallel}
\ee
This solution is valid in both the paramagnetic and antiferromagnetic states.  The reduced parallel susceptibility per spin is 
\be
\tilde{\chi}(t) = \lim_{\tilde{h}\to0}\frac{\bar{\mu}_z(t,\tilde{h})}{\tilde{h}}.
\label{Eq:tildechi}
\ee
In the AF phase, the order parameter is the staggered ordered moment
\be
\bar{\mu}_z^\dagger = \frac{\bar{\mu}_{1z}-\bar{\mu}_{2z}}{2},
\label{Eq:muDagger}
\ee
which is one-half the difference between the $z$-components of the ordered moments of the two sublattices.  The term ``ordered moment'', when used in the context of a collinear antiferromagnet, is the staggered moment.  

The susceptibility is isotropic at $T_{\rm N}$ (at reduced temperature $t=1$).  Therefore setting $t=1$ and $\bar{\mu}_{1z} = \bar{\mu}_{2z}$ in Eqs.~(\ref{Eq:mu1zmu2zMFT}) gives
\[
\bar{\mu}_{iz} = B_S\left[\tilde{h}-\frac{3f}{S+1}\bar{\mu}_{iz}\right],
\]
where $i = 1,2$.  Using the expansion $B_S(y) = (S+1)y/3$ for $y\ll 1$ and solving for $\bar{\mu}_{iz}$ gives
\be
\bar{\mu}_{iz}(t=1) = \frac{S+1}{3(f+1)}\, \tilde{h}.
\ee
Then Eq.~(\ref{Eq:tildechi}) gives
\be
\frac{\chi_\parallel(T)}{\chi_\parallel(T_{\rm N})} = \frac{\tilde{\chi}_\parallel(t)}{\tilde{\chi}_\parallel(t=1)} =\frac{3(f+1)}{S+1}\,\lim_{\tilde{h}\to0} \frac{\bar{\mu}_z(t,\tilde{h})}{\tilde{h}}.
\label{Eq:MFTPredictionforChi||}
\ee

\section{\label{Sec:OrdMomentMFT} Ordered Moment versus Temperature below $T_{\rm N}$}

In the AF state, setting the applied magnetic field $\tilde{h}=0$ and the ordered moment $\bar{\mu}_{2z}(t) = -\bar{\mu}_{1z}(t)$ in the first of Eqs.~(\ref{Eq:mu1zmu2zMFT}) gives the simple result
\be
{\bar{\mu}_{z}^\dagger}(t) = B_S\left[\left(\frac{3}{S+1}\right)\frac{\bar{\mu}_{z}^\dagger(t)}{t}\right]
\label{Eq:OrderedMomentMFT}.
\ee
Thus the reduced exchange field as in Eqs.~(\ref{Eq:B1,2}) and~(\ref{Eq:H12exch}) is
\be
\tilde{h}_{\rm exch}(t) = \frac{3}{S+1}{\bar{\mu}_{z}^\dagger}(t)
\label{Eq:AFExchField}.
\ee

\section{\label{Sec:CpMFT} Zero-Field Magnetic Heat Capacity $C_{\rm mag}$ below $T_{\rm N}$}

In the presence of the staggered exchange field with $z$-components $H_{1\,{\rm exch}} = -H_{2\,{\rm exch}}$ and nonzero $\mu_{1z} = -\mu_{2z}$, the energy of a collinear G-type AF system in zero applied field is
\be
E_{\rm ave} = -\frac{N}{2}\mu_{1z} H_{1\,{\rm exch}} = -\frac{1}{2}M_{1z}H_{1\,{\rm exch}},
\label{Eq:EaveMagExch}
\ee
where the factor of 1/2 is included to avoid counting each magnetic moment twice (once in $\mu_{1z}$ or $M_{1z}$ and again in $H_{1\,{\rm exch}}$).  The exchange field seen by sublattice~1 can be written using Eqs.~(\ref{Eq:H12exch}) and~(\ref{Eq:lmdas-lmdad2}) as
\be
H_{1\,\rm exch} = (\lambda_{\rm s}-\lambda_{\rm d})M_{1z} = \frac{3k_{\rm B}T_{\rm N}}{g^2\mu_{\rm B}^2S(S+1)}\mu_{1z}.
\label{Eq:Hexch2}
\ee
Then using the expression for the saturation moment $\mu_{\rm sat}=gS\mu_{\rm B}$, one can rewrite this as
\be
H_{1\,\rm exch} = \left(\frac{3k_{\rm B}T_{\rm N}S}{S+1}\right)\frac{\mu_{1z}}{\mu_{\rm sat}^2}.
\label{Eq:Hexch3}
\ee
Inserting Eq.~(\ref{Eq:Hexch3}) into~(\ref{Eq:EaveMagExch}) gives
\be
E_{\rm ave}(T) = -\frac{3Nk_{\rm B}T_{\rm N}S}{2(S+1)}\left[\frac{\mu_{1z}(T)}{\mu_{\rm sat}}\right]^2.
\label{Eq:EaveMag2}
\ee
Using the dimensionless reduced variables introduced in Eqs.~(\ref{Eq:RedVar}) and the definition~(\ref{Eq:muDagger}) of the staggered moment, we obtain
\be
E_{\rm ave}(t) = -\frac{3Nk_{\rm B}T_{\rm N}S}{2(S+1)}\,(\bar{\mu}_z^\dagger)^2(t).
\label{Eq:EaveMag3}
\ee

By setting $N$ equal to Avogadro's number $N_{\rm A}$ and using the definition of the molar gas constant $R=N_{\rm A}k_{\rm B}$ one obtains from Eq.~(\ref{Eq:EaveMag3}) the molar magnetic energy
\be
E_{\rm ave}(t) = -\frac{3RT_{\rm N}S}{2(S+1)}(\bar{\mu}_z^\dagger)^2(t).
\label{Eq:EaveMag4}
\ee
The molar magnetic heat capacity is then
\[
\frac{C_{\rm mag}(t)}{R} = \frac{1}{R\, T_{\rm N}}\frac{dE_{\rm ave}(t)}{dt} = -\frac{3S}{S+1}\bar{\mu}_z^\dagger (t)\frac{d{\bar{\mu}_z^\dagger}(t)}{dt}.\ \ \ (\ref{Eq:CmagQuantum})
\]

\section{\label{App:CmagSW} Low-Temperature Heat Capacity of Spin Waves}

At low temperatures, only the lowest energy spin waves contribute to the heat capacity, so we can use generic Eq.~(\ref{Eq:SWEnergies0}) for the dispersion relation.  To evaluate the integrals in Eq.~(\ref{Eq:EaveSWB}), we change variables in the integrals from wave vector {\bf q} to the vector $\vec{\epsilon}$ with dimensions of energy and with components
\bea
\epsilon_x &=& \hbar v_xq_x, \hspace{0.2in}\epsilon_a = \hbar v_x\pi/a\nonumber\\*
\epsilon_y &=& \hbar v_yq_y, \hspace{0.2in}\epsilon_b = \hbar v_y\pi/b\label{Eq:epsilonDef}\\*
\epsilon_z &=& \hbar v_zq_z, \hspace{0.2in}\epsilon_c = \hbar v_z\pi/c.\nonumber
\eea
Now the dispersion relation~(\ref{Eq:SWEnergies0}) is written symmetrically as
\be
E_{\vec{\epsilon}} = \hbar \omega_{\vec{\epsilon}} = \sqrt{\epsilon_x^2 + \epsilon_y^2 + \epsilon_z^2} \equiv \epsilon
\label{Eq:EvsEpsiloni}
\ee
and Eq.~(\ref{Eq:EaveSWB}) becomes
\bea
E_{\rm ave} &=& \frac{NV_{\rm spin}}{(2\pi)^3\hbar^3v_xv_yv_z}\label{Eq:EaveSW2}\\*
&& \times\int_{-\epsilon_a}^{\epsilon_a}d\epsilon_x\int_{-\epsilon_b}^{\epsilon_b}d\epsilon_y \int_{-\epsilon_c}^{\epsilon_c}d\epsilon_z\ \frac{E_{\vec {\epsilon}}}{e^{E_{\vec {\epsilon}}/k_{\rm B}T}-1},\nonumber
\eea
in which the anisotropy in the dispersion relation~(\ref{Eq:SWEnergies0}) has been moved to anisotropies in the limits of integration and in the prefactor.

At low temperatures, only the lowest energy spin wave states are populated, so we can take the limits of each integral to be $-\infty$ to $\infty$, which also eliminates the anisotropy between the limits of integration of the three integrals.  We can then convert the integrals over the three Cartesian coordinates to an integral over radius in spherical coordinates according to $E_{\vec{\epsilon}}\to\epsilon$ and
\[
\int_{-\infty}^{\infty}d\epsilon_x\int_{-\infty}^{\infty}d\epsilon_y \int_{-\infty}^{\infty}d\epsilon_z \to 4\pi \int_0^\infty d\epsilon\ \epsilon^2.
\]
Now we will integrate only about the $\Gamma$ point, so we must multiply by two to take into account the low-energy spin wave branches at the corners of the Brillouin zone of the primitive tetragonal direct lattice as discussed in the text.  We then obtain
\be
E_{\rm ave} = \frac{NV_{\rm spin}}{\pi^2\hbar^3v_xv_yv_z}\int_0^\infty \frac{\epsilon^3}{e^{\epsilon/k_{\rm B}T} - 1}\,d\epsilon. \hspace{0.2in} ({\rm low}\ T)
\label{Eq:EaveSW3}
\ee
Changing variables in the integral to $x = \epsilon/k_{\rm B}T$ gives
\be
E_{\rm ave} = \frac{NV_{\rm spin}(k_{\rm B}T)^4}{\pi^2\hbar^3v_xv_yv_z}\int_0^\infty \frac{x^3}{e^x - 1}\,dx. \hspace{0.2in} ({\rm low}\ T)
\label{Eq:EaveSW3A}
\ee
The integral is $\pi^4/15$, yielding
\be
E_{\rm ave} = \frac{\pi^2NV_{\rm spin}(k_{\rm B}T)^4}{15\hbar^3v_xv_yv_z}. \hspace{0.2in} ({\rm low}\ T)
\label{Eq:EaveSW4}
\ee
Then setting $N = N_{\rm A}$ (Avogadro's number), the magnetic heat capacity due to the spin waves per mole of spins is
\[
\frac{C_{\rm mag}}{R} = \frac{1}{R}\frac{dE_{\rm ave}}{dT} = \left(\frac{4\pi^2k_{\rm B}^3V_{\rm spin}}{15\hbar^3v_xv_yv_z}\right)\,T^3, \hspace{0.1in} (T \ll T_{\rm N})\ (\ref{Eq:CmagSW})
\]
where $R=N_{\rm A}k_{\rm B}$ is the molar gas constant.

We now calculate the two-dimensional spin wave theory prediction of $C_{\rm mag}$ to check consistency with Eq.~(\ref{Eq:CmagHN}) that was derived using chiral perturbation theory.  In two dimensions (i.e., $J_c=0$), the area of the sample is $A = Na^2$, where $N$ is the number of spins and $ab = a^2$ is the area of the square unit cell which contains one spin in its basis.  Then Eq.~(\ref{Eq:EaveSWB}) is replaced by
\be
E_{\rm ave} = \frac{1}{\frac{(2\pi)^2}{Na^2}}\int_{-\pi/a}^{\pi/a}dq_x\int_{-\pi/b}^{\pi/b}dq_y \frac{\hbar\omega_{\bf q}}{e^{\hbar\omega_{\bf q}/k_{\rm B}T}-1}.
\label{Eq:EaveSW2D}
\ee
Following the same steps as for the three-dimensional case above, converting the two-dimensional integrals to polar coordinates according to
\[
\int_{-\infty}^{\infty}d\epsilon_x\int_{-\infty}^{\infty}d\epsilon_y \to 2\pi \int_0^\infty d\epsilon\ \epsilon,
\]
and multiplying by two to take into account the spin wave excitations at the corners of the Brillouin zone, gives
\be
\frac{C_{\rm mag}}{R} = \frac{6\zeta(3)}{\pi(\hbar v/a)^2 }\, (k_{\rm B}T)^2,
\label{Eq:Cmag2DSW}
\ee
where $v$ is the isotropic spin wave velocity in the $ab$-plane and we have used $\int_0^\infty\frac{x^2}{e^x-1}dx = 2\zeta(3).$  Equation~(\ref{Eq:Cmag2DSW}) is identical to Eq.~(\ref{Eq:CmagHN}) obtained for the isotropic Heisenberg square lattice from chiral perturbation theory.\cite{Hasenfratz1993}



\begin{thebibliography}{999}
\bibitem{dS52Jc0.1} The second order data in Fig.~\ref{Fig:MEZ_Ordered_Spin_J1-J2}(b) for $J_c/J_1=0.1$ and $j_2\equiv J_2/J_1 = 0$ to~0.42 are described by $S-\langle S\rangle = 0.12776 + 0.17474 j_2 - 0.026127 j_2^2 + 1.0365 j_2^3 $ with an rms deviation of $2.1\times10^{-4}$.
\end{thebibliography}
\end{document}